\documentclass[aps, prd, twocolumn, lengthcheck, superscriptaddress, %
nofootinbib]{revtex4-2}
\usepackage{amsmath}
\usepackage{amssymb}
\usepackage{graphicx}
\usepackage{xcolor}
\usepackage[colorlinks=true,
urlcolor=blue,
anchorcolor=blue,
citecolor=blue,
filecolor=blue,
linkcolor=red,
menucolor=blue,
linktocpage=true,
pdfproducer=medialab,
pdfa=true]{hyperref}

\usepackage{float}

\graphicspath{{figures/}}

\usepackage{bookmark}
\usepackage{ulem}


\begin{document}
\title{Probing soft signals of gravitational-wave memory with space-based interferometers}
\author{Yan Cao}
\email{yancao@smail.nju.edu.cn}
\affiliation{School of Physics, Nanjing University, Nanjing 210093, China}

\author{Yong-Liang Ma}
\email{ylma@nju.edu.cn}
\affiliation{School of Physics, Nanjing University, Nanjing 210093, China}
\affiliation{School of Frontier Sciences, Nanjing University, Suzhou 215163, China}

\author{Yong Tang}
\email{tangy@ucas.ac.cn}
\affiliation{School of Astronomy and Space Sciences, University of Chinese Academy of Sciences (UCAS), Beijing 100049, China}

\begin{abstract}
Gravitational-wave displacement memory is a remarkable, ubiquitous prediction of general relativity that has not yet been detected. Unlike the oscillatory components of gravitational waveforms, displacement memory is associated with soft gravitons and provides the leading low-frequency observable of its parent event. Analogous soft signals may also be associated with velocity and integrated-displacement memory. The simple, universal spectral shapes of soft waveforms provide effective templates for matched filtering and parameter estimation. In this paper, we investigate the detection prospects for such soft memory signals with future space-based laser interferometers. As realistic examples, we examine the infrared spectral features of gravitational waves from moderately relativistic nonspinning compact binary scattering and comparable-mass quasicircular, nonprecessing black hole mergers. In both cases, the low-frequency spectrum can be described by a soft displacement-memory waveform with finite-frequency corrections. Using the adopted instrumental-noise models, simulated Bayesian parameter-estimation analyses show that a single LISA-like detector can constrain the memory amplitude and arrival time for signals with matched-filter signal-to-noise ratios of order 10. Big Bang Observer (BBO) Stage 1 could measure null displacement memory from sufficiently loud stellar-mass compact binary mergers. We also evaluate the detectability of an idealized stochastic background of soft displacement-memory signals. These results establish gravitational-wave bursts with memory as concrete targets for space-based interferometers.
\end{abstract}
	
\maketitle
	
\section{Introduction}
Besides the oscillatory components, gravitational radiation in general relativity generally contains a nonoscillatory component that results in a \textit{displacement memory}~\cite{Braginsky:1987kwo,PhysRevLett.67.1486,PhysRevD.45.520,PhysRevD.44.R2945,Blanchet:1992br,Favata_2010}. It may have an ordinary or null origin~\cite{Bieri_2014}, including both the traditional linear memory~\cite{Braginsky:1987kwo} and the nonlinear Christodoulou memory~\cite{PhysRevLett.67.1486}, and is associated with the supertranslation 
charge-balance laws of asymptotically flat spacetimes as well as with Weinberg's soft graviton theorem~\cite{Weinberg_1965,strominger2014gravitationalmemorybmssupertranslations,He_2015}. Displacement memory can therefore serve as an important test of general relativity. In addition to their theoretical interest, gravitational wave (GW) bursts with displacement memory are expected to be produced by a variety of astrophysical sources, including compact binaries~\cite{Favata_2009,Favata_2009-2,Favata_2011,PhysRevD.98.064031,Yang_2018,Ebersold_2019,Tiwari_2021,Liu_2021,Zhao_2021,Islam_2023,PhysRevD.108.104016,Yoo_2023,Dandapat_2023,Hait_2024,dutta2025curiouscaseparabolicencounters,Lopez_2024,ngfw-dvwz,cpl_42_10_101101,Cunningham_2025,pqcz-tqnn,bamber2025gravitationalwavememorybinary,PhysRevD.111.044052,wzqk-62wc,caldwell2025}, gamma-ray bursts~\cite{Sago_2004,Sakai_2025}, and core-collapse supernovae~\cite{Mukhopadhyay_2021,Richardson_2022}, as well as other hypothetical sources such as cosmic-string cusps and kinks~\cite{Jenkins:2021ckm}. Searches for displacement memory in the GWs from compact binary mergers have been considered for ground-based detectors~\cite{Lasky_2016,McNeill:2017uvq,Yang_2018,Ebersold_2020,H_bner_2020,H_bner_2021,Tiwari_2021,Grant_2023,Cheung_2024,Goncharov_2024}, space-based detectors~\cite{islo2019prospectsmemorydetectionlowfrequency,Ghosh_2023,Gasparotto_2023,Inchauspe:2024ibs,Sun:2024nut,Goncharov_2024,cogez2026detectabilitygravitationalwavememorylisa,zosso2026claimingdetectiongravitationalmemory}, pulsar timing arrays~\cite{van_Haasteren_2010,Wang_2014,islo2019prospectsmemorydetectionlowfrequency,Aggarwal_2020,Sun_2023,Dandapat_2024,PhysRevD.111.102002, agazie2025nanograv15yeardataset,Tomson:2025smbhbpta}, and astrometry~\cite{Madison_2020}. Potentially observable imprints of displacement memory may also appear in the cosmic microwave background~\cite{Zwick:2024hfn}.
	
Although displacement memory ubiquitously occurs in gravitational radiation, it has so far remained undetected. For example, the null memory of a compact binary merger produces only a small modification of the waveform and is not easily disentangled from the oscillatory ``parent'' signal in the time domain. Resolving such modifications through sensitive detectors and refined waveform modeling would enable the detection of memory and the improvement of parameter estimation~\cite{Grant_2023,Gasparotto_2023,Inchauspe:2024ibs,Sun:2024nut}. Accurately modeling waveforms that include memory effects becomes increasingly challenging for more complex sources. For a compact binary scattering event, even with a sufficiently accurate waveform model, parameter estimation is still challenging due to degeneracies~\cite{PhysRevD.111.044024}. Waveform modeling is further complicated by possible environmental influences~\cite{Cheng:2024mgl,singh2025gravitationalmemorymeetsastrophysical,yyv5-3y1c}. However, because of its soft nature, displacement memory produces a signal that persists at low frequencies, with a spectrum of \textit{universal} shape. Although much information about the sources is lost in such ``orphan'' signals~\cite{van_Haasteren_2010,McNeill:2017uvq,Ghosh_2023}, they remain valuable targets for probing the memory components, either by searching for them or by measuring them independently, thanks to the ubiquity of the sources and the simplicity of the templates. It is also possible that some yet-unknown processes could produce observable GW bursts with displacement memory, making soft memory signals a potential probe of these processes.
	
In addition to the soft waveform associated with displacement memory (which corresponds to a steplike strain in the time domain), other types of soft waveforms may also be observed, including those associated with \textit{velocity memory}~\cite{GrishchukPolnarev1989,Bieri_2024} (corresponding to a ramplike strain in the time domain) and integrated-displacement memory (discussed in Sec.~\ref{Sec:Integrated-displacement memory}). Although these additional templates are not fixed by the leading soft theorem, they may arise over sufficiently long timescales and serve as phenomenological low-frequency descriptions.
	
In this paper, we study the prospects for detecting and measuring soft memory signals with future space-based interferometers, including Laser Interferometer Space Antenna (LISA)~\cite{amaroseoane2017laser,babak2021lisasensitivitysnrcalculations}, Taiji~\cite{10.1093/nsr/nwx116}, TianQin~\cite{Luo_2016}, and Big Bang Observer (BBO) Stage 1~\cite{Corbin_2006}. We investigate detector responses and parameter-estimation precision for both a single LISA-like detector and the proposed LISA-Taiji network~\cite{Ruan_2020}. In optimal configurations, the signal-to-noise ratio in LISA, Taiji, and TianQin can reach 10 for a displacement memory of amplitude $\sim 10^{-20}$, or a velocity memory of amplitude $\sim 10^{-22}\,\text{Hz}$. To examine the measurement precision, we perform numerical simulations for Bayesian parameter estimation. We find that with a single detector, the memory amplitude and arrival time remain measurable for signal-to-noise ratios $\gtrsim 10$, while the angular parameters are limited by low-frequency degeneracies. These angular and polarization degeneracies are mitigated either by the LISA-Taiji network or by external sky localization. For the adopted BBO instrumental-noise model in the sub-Hz to Hz band, a displacement memory of amplitude $\sim 10^{-23}$ gives a signal-to-noise ratio of 10, as does a velocity memory of amplitude $\sim 10^{-23}\,\text{Hz}$. Under this model, sufficiently loud stellar-mass compact-binary mergers are promising targets for measuring null displacement memory with BBO Stage 1.
	
The soft waveform is an idealized zero-frequency limit. We examine finite-frequency corrections to this limit in two realistic examples, compact-binary hyperbolic encounters and comparable-mass quasicircular binary black hole (BBH) mergers, and assess their impact on displacement memory measurements using a soft template. In the parameter-estimation tests, we inject realistic waveforms and use a soft-template likelihood that does not require knowledge of the correction factor. The resulting posterior shifts quantify the waveform-shape systematics of this strategy. We find that the soft-template parameters remain measurable in these tests, while finite-frequency waveform-shape corrections can produce measurable posterior shifts, especially at high SNR. Soft signals may also contribute to the stochastic gravitational-wave background~\cite{Zhao:2021zlr}. We simulate an idealized background of soft displacement-memory signals, and evaluate its detectability at the considered detectors. Our results identify soft signals from bursts with memory as concrete targets for space-based gravitational-wave detectors.
	
This paper is organized as follows. In Sec.~\ref{sec:2}, we introduce the soft waveforms and discuss their main properties. Section~\ref{sec:3} examines the low-frequency GW spectrum of compact binary scattering and merger events, as concrete examples of realistic waveforms exhibiting displacement memory. In Sec.~\ref{sec:4}, we investigate the responses of space-based detectors to soft memory signals. Section~\ref{sec:5} presents the results of simulated parameter estimation. The stochastic background of soft displacement-memory signals and its detectability in space-based detectors are discussed in Sec.~\ref{sec:6}. We provide a brief summary and discussion in Sec.~\ref{sec:7}. Details of waveform propagation and coordinate transformations are collected in the Appendixes. Throughout this paper, we use the natural units $c = G = \hbar = 1$ and the mostly-plus metric signature.
	
\section{Soft memory waveforms}\label{sec:2}
We refer to a specific form of the zero-frequency limit of the spectrum that depends only on the asymptotic difference in a waveform observable in the time domain as a \textit{soft waveform}. This is a useful notion, since such a limit is shared by all waveforms with the same ``memory'' and becomes a good approximation to the actual signal at sufficiently low frequencies. In this section, we introduce three types of soft waveforms, corresponding, respectively, to displacement memory, velocity memory, and integrated-displacement memory. In the subsequent sections on detectability, however, we focus primarily on displacement memory.
	
The GWs emitted by distant sources are well approximated by plane waves when they reach our detectors. Consider a flat spacetime perturbed by a weak plane GW\footnote{The weak plane-wave approximation used here for radiation from distant
sources should not be confused with exact plane-wave spacetimes, where related but distinct displacement and velocity memory
effects have been studied~\cite{Zhang_2017_PLB,Zhang_2017_PRD,Achour_2024}.} in the transverse-traceless (TT) gauge:
\begin{equation}
ds^2=g_{\mu\nu}dx^\mu dx^\nu=-dt^2+(\delta_{ij}+h_{ij})\,dx^idx^j, \label{GW_spacetime}
\end{equation}
where $h_{ij}(t,\mathbf{x})$ is the strain tensor field satisfying $\delta_{ij}h_{ij}=\partial_i h_{ij}=0$. The strain tensor can be decomposed into two linearly polarized states, and the waveform at $\mathbf{x}=0$ reads
\begin{equation}
h_{ij}(t)\equiv h_{ij}(t,\mathbf{0}) =\sum_{\lambda=+,\times}h_\lambda (t)\,e_{ij}^\lambda.
\end{equation}
Here the orthonormal polarization tensors are constructed as
\begin{equation}\label{TT_polarization}
e_{ij}^+=a_ia_j-b_ib_j,
\quad
e_{ij}^\times=a_ib_j+b_ia_j,
\end{equation}
where the three unit vectors $\{\mathbf{a,b,\hat{k}}\}$ form a right-handed triad such that the GW propagates along the direction of $\hat{\mathbf{k}}$, and $e_{ij}^{\lambda}e_{ij}^{\lambda'*}=2\delta_{\lambda,\lambda'}$. We now consider the waveform of a given linearly polarized state, denoted temporarily by $h(t)$ in this section. For the Fourier transform, we use the convention $	h(t) =\int_{-\infty}^\infty df\,\text{e}^{-2\pi i f t}\,\tilde h_\infty (f)$ and $\tilde h_\infty(f)\equiv \int_{-\infty}^\infty dt\,\text{e}^{2\pi i ft}\,h(t)$, where the subscript $\infty$ is a reminder of the idealized assumption of an infinite time duration. Since $h(t)\in\mathbb{R}$, $\tilde h_\infty(-f)=\tilde h_\infty^*(f)$.
	
To illustrate the observable effects of displacement and velocity memory, we introduce a free-falling timelike test particle at $\mathbf{x}=0$ in the spacetime \eqref{GW_spacetime} and the Fermi normal coordinates (FNC) $\bar x^a=(\bar t,\bar{\mathbf{x}})$ centered at it, with $\frac{\partial x^\mu(\bar x)}{\partial \bar x^a}\approx \delta^\mu_a$ and the particle's proper time $\bar t\approx t$. In this FNC, the equation of motion for a nearby free-falling classical test particle with $\bar x^j(\bar t=0)= X^j$ and negligible initial velocity is
\begin{equation}\label{EOM_in_FNC}
\frac{d^2\bar x^i}{d\bar t^2}\approx \frac{1}{2}\,\ddot h_{ij}(\bar t)\,\bar x^j\approx \frac{1}{2}\,\ddot h_{ij}(\bar t)\,X^j.
\end{equation}
The response of an interferometer to low-frequency GWs can be understood as arising from such relative displacements between test masses in the FNC, where the effects of GWs on light propagation are negligible. However, this approximation is not adequate for space-based detectors, and (as discussed in Sec.~\ref{sec:TDI response}) it is convenient to analyze the signals in the TT coordinates.
	
\subsection{Displacement memory}\label{subsec:Displacement memory}
	
Displacement memory refers to a DC (direct current) shift of $h(t)$ between the asymptotic past and future, i.e., $\Delta h\equiv h(\infty)-h(-\infty)\ne 0$. In Eq.~\eqref{EOM_in_FNC}, this leads to a ``permanent'' displacement: $\Delta \bar x^i \approx \frac{1}{2}\Delta h_{ij}X^j$. In practice, $\infty$ refers to a moment that is sufficiently far from the observer's time of interest. The soft waveform of displacement memory depends solely on the amplitude $\Delta h$ of the memory jump.
	
To make the limiting procedure concrete, we consider the arctan model:
\begin{align}
h(t) &=\frac{\Delta h}{\pi} \arctan\left(f_*(t-t_*)\right), \label{arctan_model}
\\
\tilde h_\infty(f)&=\frac{i\Delta h}{2\pi f}\text{e}^{-2\pi |f|/f_*}\text{e}^{2\pi i f t_*}, \label{arctan_model_2}
\end{align}
with $f_*>0$. Here the Fourier transform is understood in the sense of tempered distributions, and a global DC shift of the time-domain waveform is unobservable in the spectrum at $f>0$. The asymptotic transition is carried by the zero-frequency mode, and $1/f_*$ measures the time scale of the jump. In the limit $f_*\to\infty$, the waveform approaches a step function at $t=t_*$ with the step size $\Delta h$, corresponding to $\tilde h_\infty(f)=i\Delta h/(2\pi f)\text{e}^{2\pi  i ft_*}$ at $f>0$. For a finite $f_*$, the spectrum approaches the zero-frequency limit at $f/f_*\to 0^+$,
\begin{equation}\label{displacement_memory_waveform}
\tilde h_\infty(f)\to\frac{i\Delta h}{2\pi f}\,\text{e}^{2\pi i f t_*},
\end{equation}
which is the soft waveform of displacement memory. This form is shared by waveforms with the same displacement memory $\Delta h$ when the detailed transition profile is unresolvable. Note that the spectral density of the graviton number fluence, $dN/(dAdf)=[(\pi/2)f^2|\tilde h_\infty|^2]/(2\pi f)$, associated with this waveform diverges as $f\to 0$, whereas the spectral density of the energy fluence remains finite.
	
For $f\ll 1/|t_*|$, the phase factor $\text{e}^{2\pi i f t_*}$ can be further approximated by unity. This approximation is generally poor for the GW signal, since the frequency spectrum of the waveform can only be measured at \textit{finite} frequency, with the lowest resolvable frequency limited by the signal duration $[t_\text{min}, t_\text{max}]$ and instrumental noise. For the same reason, detection of the soft waveform \eqref{displacement_memory_waveform} would provide strong evidence for a low-frequency memory component, but cannot by itself demonstrate the persistence of the memory jump at arbitrarily low frequencies. Likewise, even if a detector observes the spectrum in the form \eqref{displacement_memory_waveform} at $f>f_\text{min}$, this does not determine the spectrum at lower frequencies. A more complete measurement of the waveform therefore requires detectors with better sensitivity in the low-frequency band.
	
This finite-window issue is important when relating the idealized spectrum to the actual detector response. For a waveform $h(t)$ with displacement memory satisfying $h(t<t_\text{min})= h(-\infty)$ and $h(t>t_\text{max})= h(\infty)$, $\tilde h_\infty(f)$ is drastically different from the Fourier transform of $h(t)$ in the finite interval $[t_\text{min},t_\text{max}]$:
\[
\begin{aligned}
\tilde h(f) &\equiv \int_{t_\text{min}}^{t_\text{max}} dt\,\text{e}^{2\pi i ft}\,h(t)
\\
&= \tilde h_\infty-\frac{i}{2\pi f}
\left[h(\infty)\,\text{e}^{2\pi i f t_\text{max}}
-h(-\infty)\,\text{e}^{2\pi i f t_\text{min}}\right],
\end{aligned}
\]
which, in particular, is dominated by the second term at sufficiently high frequencies. For a GW interferometer, however, the observable depends not on $h(t)$, but on temporal differences such as $h(t-D)-h(t)$, with $D$ approximately constant within sufficiently short time intervals. Assuming $0<D\ll |t_\text{max}-t_\text{min}|$ and $h(t\ge t_\text{max}-D)= h(\infty)$, one finds
\[
\begin{aligned}
&\int_{t_\text{min}}^{t_\text{max}} dt\,\text{e}^{2\pi if t}[h(t-D)-h(t)]
=\left(\text{e}^{2\pi i fD}-1\right)\tilde h_\infty(f).
\end{aligned}
\]
Thus the frequency spectrum of such an interferometer signal well below the sampling frequency can be approximately derived with $\tilde h_\infty(f)$, but not with $\tilde h(f)$. Numerically, $\tilde h_\infty(f)$ can be obtained from $\dot h(t)$ as $\tilde{\dot h}/(-2\pi i f)$, with $\tilde{\dot h}$ well approximated by its discrete Fourier transform (DFT); note that a windowed finite-duration waveform can reproduce the same spectrum only over an intermediate frequency range. Even if the memory jump does not occur strictly within $[t_\text{min}, t_\text{max}]$, the approximation typically remains valid for $f>1/|t_\text{max}-t_\text{min}|$, provided that $h(t)$ varies slowly outside this interval, in which case $\tilde{\dot h}/(-2\pi i f)$ is close to $\tilde h_\infty(f)$. In subsequent sections, we compute the numerical Fourier spectrum of a time-domain waveform $h(t)$ over a finite interval $[t_\text{min},t_\text{max}]$ as $\int_{t_\text{min}}^{t_\text{max}} dt\,\text{e}^{2\pi i ft}\,\dot h(t)\,/(-2\pi i f)$.
	
If the interval $[t_\text{min},t_\text{max}]$ is too short to capture most of the jump, the signal spectrum at $f\gtrsim 1/|t_\text{max}-t_\text{min}|$ can differ significantly from that of the soft waveform. As an extreme example, consider a linearly growing waveform $\dot h(t)=\text{const}$. In this paper, we focus on measurements using space-based interferometers. As explained in Sec.~\ref{sec:TDI response}, the relevant time-domain signals [e.g., Eq.~\eqref{X_1(t)}] are built up by observables such as $[h(t-D_1)-h(t)]-[h(t-D_1-D_2)-h(t-D_2)]$, with $D_{1,2}$ constant over sufficiently short times. A linearly growing waveform thus has a vanishing signal in the static-detector approximation. In fact, we find that the signal remains extremely weak compared with that of a soft waveform with $\Delta h = \dot h\,T$ when the signal duration $T$ is not short, so that the static-detector approximation no longer holds.
	
\subsection{Velocity memory}\label{Sec:Velocity memory}
	
In analogy with displacement memory, velocity memory is a DC shift in $\dot h(t)$:
$\Delta \dot h\equiv \dot h(\infty)-\dot h(-\infty)\ne 0$, which can arise in certain astrophysical scenarios~\cite{GrishchukPolnarev1989,Bieri_2024}. In Eq.~\eqref{EOM_in_FNC}, this leads to a ``permanent'' velocity change: $\Delta \frac{d\bar x^i}{d\bar t}\approx \frac{1}{2}\Delta \dot h_{ij}X^j$. Since $\lim_{f\to 0}\tilde{\dot h}_\infty (f)\to\frac{i\Delta \dot h}{2\pi f}\text{e}^{2\pi i f t_*}$, the soft waveform is given by
\begin{equation}\label{velocity_memory_waveform}
\tilde h_\infty (f)\to - \frac{\tilde{\dot h}_\infty}{2\pi i f}=-\frac{\Delta \dot h}{(2\pi f)^2}\text{e}^{2\pi i f t_*}.
\end{equation}
As mentioned above, we are interested in observables such as $[h(t-D_1)-h(t)]
-[h(t-D_1-D_2)-h(t-D_2)]$, which tends to zero at $t\to\pm \infty$. The frequency spectrum of the signal can also be obtained approximately from $\tilde h_\infty(f)$ if the observation time is sufficiently long.
	
A simple velocity-memory waveform is the ramp model: $h(t)=\Delta\dot h_+\,(t-t_*)\Theta(t-t_*)+\Delta\dot h_-\,(t_*-t)\Theta(t_*-t)$, where $\Theta(t)$ is the Heaviside unit step function. The frequency domain waveform at $f>0$ is exactly given by Eq.~\eqref{velocity_memory_waveform}, with $\Delta \dot h =\Delta\dot h_++\Delta\dot h_-$. As discussed above, a strictly indefinite ramp should be understood as an idealization over the observation time; a pure linear ramp is strongly suppressed in the static TDI response.
	
\subsection{Integrated-displacement memory}\label{Sec:Integrated-displacement memory}
	
It is tempting to extend this hierarchy and consider a waveform with $\frac{d^nh}{dt^n}(\infty)-\frac{d^nh}{dt^n}(-\infty)\ne0$, with $n\ge2$. This would be more problematic than velocity memory. A DC shift in $\dot h$ gives an idealized ramplike strain, but the radiative tidal field $R_{0i0j}\approx-\frac{1}{2}\partial_t^2h_{ij}$ can still vanish away from the transition. By contrast, a DC shift in $\frac{d^nh}{dt^n}$ with $n\ge2$ would make $R_{0i0j}$ approach a nonzero constant for $n=2$, or grow polynomially for $n>2$, on at least one side of the waveform. We therefore do not consider such higher-derivative memories as soft templates~\cite{GrishchukPolnarev1989}. Soft spectra of the form $\frac{O}{|f|}$ and $\frac{iO}{f|f|}$ with $O\in\mathbb{R}$ are likewise unphysical, since they correspond to time-domain waveforms that grow as $\ln|t|$ and $t\ln|t|$, respectively, as $t\to\pm\infty$.
	
If hereditary observables are taken into account, additional types of soft waveforms can be defined. For example, the waveform $h(t)$ with a finite zeroth time moment is accompanied by an asymptotic change in the observable $O(t)\equiv \int_{-\infty}^t dt'\,h(t')$, and has the zero-frequency limit
\begin{equation}\label{integrated-displacement_memory_waveform}
\tilde h_\infty (f)\to O\,\text{e}^{2\pi i f t_*},
\end{equation}
with $O=O(\infty)$. We therefore call this the soft waveform of an ``integrated-displacement memory.''
It can naturally arise as the soft signal associated with subleading GW memory, for which the persistent observable is a change in the time-integrated strain~\cite{Pasterski:2015tva,Nichols:2017rqr,Nichols:2018qac,Hamada:2018cjj,Flanagan:2019ezo}. For a localized GW source, the spin and center-of-mass memories are collectively referred to as drift memory~\cite{Grant:2021hza,Grant:2023xyz}; the next level in this hierarchy is ballistic memory, associated with the second moment of the Bondi news (which is obtained by projecting $\dot h^{\rm TT}_{ij}$ onto the sphere centered on the source)~\cite{Grant:2023xyz,Siddhant:2026}. More generally, Eq.~\eqref{integrated-displacement_memory_waveform} may simply be regarded as a phenomenological low-frequency template. Note that the imaginary version of \eqref{integrated-displacement_memory_waveform}: $\lim_{f\to 0}\tilde h_\infty (f)\to iO\,\text{sgn}(f)\,\text{e}^{2\pi i f t_*}$ with $O\in\mathbb{R}$, corresponds to an asymptotic time-domain waveform $\lim_{|t|\to\infty}h(t)\to\frac{O}{\pi}\frac{1}{t-t_*}$.
	
\subsection{Correction to the soft waveform}\label{sec:Correction to the soft waveform}
A soft waveform is the exact zero-frequency limit of the spectrum in the presence of memory. In reality, we observe the spectrum at a finite frequency and therefore observe a deviation from the soft waveform. There are no generic constraints on such corrections other than that they vanish at $f=0$, since an arbitrary waveform $\mathtt{h}(t)$ with zero memory can be superposed on the original waveform without affecting the memory. The correction therefore contains information from the bulk of the time interval, and the leading-order (LO) nonanalytic correction encodes the next-to-leading order term in the asymptotic time-domain waveform.
	
We focus on displacement memory (setting $t_*=0$), and define the exact correction factor to be
\begin{equation}
\mathcal{C}(f)= \left(\frac{2\pi f}{i\Delta h}\right)\tilde h_\infty(f)-1=[\mathcal{C}(-f)]^*.
\end{equation}
$\mathcal{C}(f)$ is generally not analytic at $f=0$ for realistic waveforms. For example, $\lim_{|t|\to\infty}\mathtt{h}(t)\to -1/(2|t|)$ corresponds to $\tilde {\mathtt{h}}_\infty(f\to 0)\to \ln |f|$; $\lim_{|t|\to \infty}\mathtt{h}(t)\propto |t|^{-\alpha}$ with $\alpha\in (0,1)$ corresponds to $\tilde{\mathtt{h}}_\infty(f\to 0)\propto |f|^{\alpha-1}$. Note that $\mathcal{C}(f)$ cannot generally be obtained from the Taylor expansion about a reference frequency $f_\text{ref}$, which may not converge away from $f_\text{ref}$. If $\dot h/\Delta h\ge0$, as is relevant in some realistic cases, $\mathcal{C}(f)$ satisfies $|1+\mathcal C|\le 1$. If the memory growth is further confined to a time interval of duration $T$, then $|1+\mathcal C|\ge\cos(\pi |f|T)$ for $|f|T\le1/2$.
	
We are interested in situations where both linear polarization components can be approximately described by soft waveforms of the same type, and their arrival times $t_*^\lambda\approx t_*$ satisfy $f|(t_*^+-t_*^\times)|\ll 1$. In the zero-frequency limit, the GW is then linearly polarized (i.e., $\tilde h_+/\tilde h_\times\in\mathbb{R}$). This may not hold in general due to finite-frequency corrections. Nevertheless, we can define
\begin{equation}
\begin{aligned}
\tilde h_+ -i \tilde h_\times \equiv \frac{iH\text{e}^{-2i\psi}}{2\pi f}\text{e}^{2\pi i f t_*}[1+C(f)],
\end{aligned}
\end{equation}
where $\psi\in\mathbb{R}$, $H>0$, and $C(f)$ is defined on the two-sided frequency domain. For example, for $C(0)=0$, $H\text{e}^{-2i\psi}=\Delta h_+-i\Delta h_\times$.
	
\section{Memory of compact binaries}\label{sec:3}
	
In this section, we examine examples of soft waveforms of displacement memory produced by compact binaries. Before considering the concrete waveforms, we briefly discuss the effects of GW propagation on the observed waveforms in the eikonal approximation.
	
In an asymptotically flat spacetime, the gravitational waveform of a localized source can be described at future null infinity in radiative coordinates, which fix a particular Bondi frame, as
\begin{equation}
h(u)=h_+-ih_\times\equiv\sum_{l\ge 2,m} h_{l,m}(u)\,_{-2}Y_{lm}(\Theta,\Phi),
\label{source_frame_waveform}
\end{equation}
where $u=t-R$ is the retarded time, $R$ is the observer's distance to the source, $(\Theta,\Phi)$ are the spherical polar coordinates of the observer, and $_{s}Y_{lm}$ are the spin-weighted spherical harmonics with spin weight $s$. Here the TT polarization is defined by $\mathbf{a}=\mathbf{e}_\Theta$ and $\mathbf{b}=\mathbf{e}_\Phi$ in Eq.~\eqref{TT_polarization}. We also use the Cartesian coordinates $(X,Y,Z)=R(s_\Theta c_\Phi, s_\Theta s_\Phi, c_\Theta)$, where $c_z\equiv \cos z$ and $s_z\equiv \sin z$. The one-sided GW energy spectrum is given by ${dE_\text{gw}}/{df}=\frac{R^2}{8\pi}\sum_{l,m}|\tilde h_{l,m}(f)|^2(2\pi f)^2$, with $\tilde h_{l,m}(f)= \int dt\,\text{e}^{2\pi i ft}\,h_{l,m}(t)$.
	
The waveform modes $h_{l,m}(u)$ are expressed in terms of the radiative mass and current multipole moments $U_{l,m}$ and $V_{l,m}$ as $h_{l,m}=\left(U_{l,m}-iV_{l,m}\right)/(\sqrt{2}\,R)$~\cite{Favata_2009}. At leading post-Newtonian (PN) order, the $l=2$ modes correspond to the Newtonian-order (0PN) quadrupole waveform as given by the formula:
\begin{equation}
h_{ij}=\frac{2}{R}\,\Lambda_{ij,kl}\,\ddot M_{kl}, \label{quad_formula}
\end{equation}
where $M_{ij}\equiv \int d^3x\,\rho\,x_ix_j$ is the Newtonian mass quadrupole moment tensor with $\rho$ denoting the Newtonian mass density, and $\Lambda_{ij,kl}\equiv \frac{1}{2}(P_{ik}P_{jl}+P_{il}P_{jk}-P_{ij}P_{kl})$ is the TT projector, with $P_{ij}\equiv \delta_{ij}-\hat k_i\hat k_j$.
	
\subsection{Waveform propagation}\label{Propagation of waveforms}
	
Equation~\eqref{source_frame_waveform} is an idealized description, since in reality the spacetime is not asymptotically flat and the observer is not located at infinity. For a locally plane GW with sufficiently high frequency, its propagation in a curved spacetime background can be described by the eikonal approximation. If the observed frequency of a Fourier mode with original frequency $f$ is shifted to $f'= \varrho f$, the observed frequency spectrum $\tilde h_\infty'(f)$ is related to the original spectrum $\tilde h_\infty (f)$ via
\begin{equation}\label{frequency_shift}
\tilde h_\infty'(f)=(F/\varrho)\,\tilde h_\infty(f/\varrho),
\end{equation}
where $F$ is the amplitude modulation factor (which becomes frequency dependent in the wave-optics regime~\cite{Takahashi_2003,Harte_2019,Emma2026}). Applying this to the soft waveform of displacement memory, $\tilde h_\infty=i\Delta h/(2\pi f)$, the observed memory amplitude is given by $\Delta h'=F\Delta h$. Here we are considering a frequency band that is low enough for the soft waveform to be a good approximation, while still high enough for the eikonal approximation to be valid. More details of the eikonal approximation are collected in Appendix~\ref{appendix_propagation_details}.
	
For a spherical wave, the amplitude decay $\propto 1/(a_\text{o}R)$ needs to be included, where $R$ is the distance in comoving coordinates $\mathbf{x}$, related to the source's luminosity distance $d_L=z/H_0+\mathcal{O}(z^2)$ via $a_\text{o}R=d_L/(1+z)$, with $H_0$ denoting the Hubble parameter today.\footnote{For the numerical calculations, we adopt a spatially flat $\Lambda$CDM cosmology with $H_0=67.4\,\mathrm{km\,s^{-1}\,Mpc^{-1}}$ and $\Omega_m=0.315$, neglecting radiation.} Using Eq.~\eqref{frequency_shift} with $F=1$, a waveform $\tilde h(f,R)$ in an asymptotically flat spacetime is converted to $\tilde h'(f,d_L)=(1+z)\tilde h((1+z)f,d_L/(1+z))=(1+z)^2\tilde h((1+z)f,d_L)$. For displacement memory, this implies that $d_L\Delta h'(d_L)=(1+z)R\Delta h(R)$, where $\Delta h(R)$ is observed at a distance $R$ sufficiently close to the source where the approximation of an asymptotically flat spacetime is good. This is consistent with the analysis of \cite{PhysRevD.94.064040,PhysRevD.94.044009,Bieri_2017}. In summary, the effect of cosmic expansion on the soft waveform of displacement memory is a memory amplitude $\Delta h'(d_L)=(1+z)\Delta h(R=d_L)$, where $\Delta h$ is computed in an asymptotically flat spacetime with a source distance $R=d_L$; meanwhile, any finite-frequency correction is evaluated at the higher source-frame frequency $(1+z)f$, so the deviation from the soft waveform at a fixed observed frequency increases when $|\mathcal{C}(f)|$ grows with frequency. Additional Doppler shifts from source or observer motion, as well as perturbative metric-induced shifts, can be treated by the same frequency-rescaling factor (see Appendix~\ref{appendix_propagation_details}).
	
\subsection{Displacement memory from compact binary scattering}\label{sec:Displacement memory from compact binary scattering}
	
Displacement and velocity memory can appear at 0PN in linearized gravitational radiation, since according to the quadrupole formula \eqref{quad_formula}, they can be produced, respectively, by asymptotic changes in the second and third time derivatives of $M_{ij}$, namely $\ddot M_{ij} =\int d^3x\,\rho\, (2v_iv_j+a_i x_j+x_ia_j)$ and $\dddot M_{ij}=\int d^3x\,\rho\, (3a_iv_j+3v_ia_j+\dot a_i x_j+x_i\dot a_j)$, where $v_i={dx^i}/{dt}$ is the velocity and $a_i={dv^i}/{dt}$ is the acceleration of the nonrelativistic matter. Velocity memory can thus be produced by a change of acceleration that persists over the relevant timescale while the velocity remains nearly constant~\cite{GrishchukPolnarev1989,Bieri_2024}, whereas displacement memory is produced in generic unbound scattering processes.\footnote{In an asymptotically flat spacetime, the displacement memory at future null infinity can be decomposed into ordinary and null components~\cite{Bieri_2014}, sourced by the energy-momentum of massive matter and null energy flux, respectively.} In this section, we examine the low-frequency waveforms of nonspinning compact binary scattering events in the moderately relativistic regime, where the PN approximation is applicable. We describe the scattering in the two-body center-of-mass (CM) frame, and denote by $x_I^\mu=(t,\mathbf{x}_I(t))$ the four-position of the $I$th body with mass $m_I$. We also define $\mathbf{x}\equiv \mathbf{x}_2-\mathbf{x}_1$, $\mathbf{v}\equiv \dot{\mathbf{x}}$, and $r=|\mathbf{x}|$. In the CM frame, the initial state can be specified by the relative velocity $\mathbf{v}_0=\mathbf{v}(t\to-\infty)$ and the impact parameter vector $\mathbf{b}=[\mathbf{x}-(\mathbf{x}\cdot\hat{\mathbf{v}})\,\hat{\mathbf{v}}]_{t\to-\infty}$, with $\hat{\mathbf{v}}=\mathbf{v}/|\mathbf{v}|$; we denote $v_0=|\mathbf{v}_0|$ and $b=|\mathbf{b}|$. Covariantly, the initial state can be defined by the four-velocities $u_I^\mu=(1-|\mathbf{v}_I|^2)^{-1/2}(1,\mathbf{v}_I)$ with an impact parameter four-vector $b^\mu$ satisfying $b_\mu u_1^\mu=b_\mu u_2^\mu=0$. The CM frame is defined by $m_1 u_1^i + m_2 u_2^i=0$, in which $b^\mu=(0,\mathbf{b})$. Note that $\mathbf{v}$ is not the initial velocity of body 2 in the initial rest frame of body 1, whose magnitude is instead given by $\sqrt{
1-
{(1-|\mathbf v_1|^2)(1-|\mathbf v_2|^2)}/
{(1-\mathbf v_1\cdot\mathbf v_2)^2}
}$.
	
We first consider the 0PN quadrupole waveform of a Newtonian hyperbolic encounter. The orbit of $\mathbf{x}(t)$ is taken to be a Keplerian orbit with semimajor axis $a$ and eccentricity $e$. Correspondingly, $b=a\sqrt{e^2-1}$, $v_0=\sqrt{M/a}$; so $e=\sqrt{1+(b/M)^2v_0^4}$, $a=M/v_0^2$. This approximation is reasonable if the incident velocity $v_0$ (for a given impact parameter $b$) is neither too large nor too small, since in the former case the conservative PN corrections become important, while in the latter case the gravitational radiation reaction matters.\footnote{For example, a Keplerian parabolic orbit produces only null displacement memory~\cite{Favata_2011}. However, when gravitational radiation reaction is included, a final state with vanishing relative velocity arises from an initial state with a nonzero relative velocity, thereby producing nonzero (and likely non-negligible) ordinary displacement memory.} The quadrupole moment in the CM frame is given by $M_{ij}=\nu M x_ix_j$, where $M=m_1+m_2$ is the total mass and $\nu=m_1m_2/M^2 \in (0,1/4]$ is the symmetric mass ratio.
	
\begin{figure}[hbt]
\centering
\includegraphics[width=0.47\textwidth]{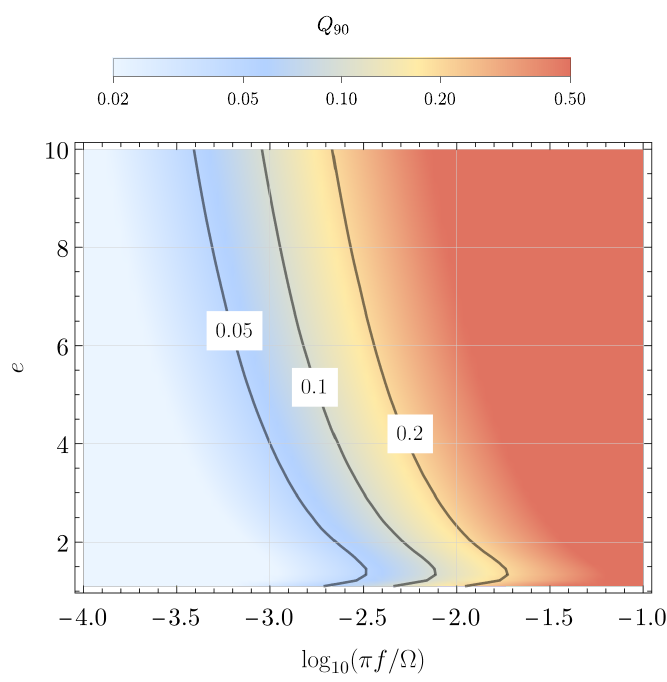}
\caption{
Directional 90th-percentile deviation from the soft limit for 0PN compact-binary scattering waveforms. Here $Q_{90}$ is the 90th percentile of $|C|$ over directions satisfying $|\Delta h|\geq 0.1|\Delta h|_{\max}$, excluding near-node directions where the relative correction is ill conditioned. The contours show $Q_{90}\in\{0.05,0.1,0.2\}$. Varying the exclusion threshold between $0.05|\Delta h|_{\max}$ and $0.2|\Delta h|_{\max}$ shifts the contours by at most about $0.06$ in $\log_{10}(\pi f/\Omega)$.
}\label{fig:scattering_soft_validity}
\end{figure}

We choose a source frame in which the orbital angular momentum is aligned with the $Z$-axis and the periastron lies on the $X$-axis; the time coordinate is shifted such that $t=0$ corresponds to the periastron crossing. With this choice, only $h_{2,\pm 2}$ and $h_{2,0}$ are nonzero. Using the eccentric-anomaly parametrization of the hyperbolic orbit (see, e.g., \cite{Cheng:2024mgl}) and the integral
representation of the Hankel function of the first kind: $H_m^{(1)}(z)=\int_{-\infty}^\infty \frac{d\xi}{i\pi}\text{e}^{z\sinh\xi-m\xi}$, the idealized positive-frequency spectrum of the 0PN waveform can be derived as
\begin{equation}\label{hyperbolic_encounter_1}
\begin{aligned}
\frac{f\tilde h_{2,\pm 2}(f)}{\pi h_0}&=\left(
\frac{\pm\sqrt{e^2-1}}{2e}+\frac{e^2-1}{2e} n
\right)
nH_{1+in}
\\
&\quad +\left[\left(\frac{i\mp\sqrt{e^2-1}}{2e^2}-\frac{i}{4}\right)\right.
\\
&\left.\quad-\frac{e^2-1}{2e^2}\left(1\pm i\,\sqrt{e^2-1}\right) n\right]n H_{in}
,
\end{aligned}
\end{equation}
\begin{equation}\label{hyperbolic_encounter_2}
\frac{f\tilde h_{2,0}(f)}{\pi h_0}=
-\frac{i}{2\sqrt{6}}n H_{in},
\end{equation}
with $h_0\equiv \sqrt{{64 }/{(5\pi)}}\,{\nu M^2}/{(aR)}$, $n \equiv 2\pi f/\Omega$, $\Omega\equiv\sqrt{M/a^3}$,  and $H_m\equiv H_{m}^{(1)}(ine)$. Note that $\tilde h_{2,2}(-f)=[\tilde h_{2,-2}(f)]^*$, $\tilde h_{2,0}(-f)=[\tilde h_{2,0}(f)]^*$. Using the small-$n$ asymptotic forms $H_{in} \sim \frac{2i}{\pi}\left(\ln \frac{ne}{2}+\gamma_E\right)$ and $H_{1+in} \sim \left(-\frac{2}{\pi e n}-\frac{1}{e}\right)+\frac{2i}{\pi e}\left(\ln \frac{ne}{2}+\gamma_E\right)$, where $\gamma_E$ is the Euler-Mascheroni constant, we obtain from Eqs.~\eqref{hyperbolic_encounter_1} and \eqref{hyperbolic_encounter_2} the soft waveform $\tilde h(f)\to{i(\Delta h_+-i\Delta h_\times)}/{(2\pi f)}$ (since the periastron crossing happens at $t=t_*=0$) with its leading low-frequency correction:
\begin{equation}
\begin{aligned}\label{soft_waveform_0PN_scattering}
\frac{f\tilde h_{2,\pm 2}(f)}{h_0} \approx &\mp\frac{\sqrt{e^2-1}}{e^2}\left(1+\frac{\pi}{2}n\right)
+\left(\frac{1}{e^2}-1\right)n
\\
&+\left(\frac{1}{2}-\frac{1}{e^2}\right) n\,\left(\ln \frac{ne}{2} +\gamma_E\right)
,
\\
\frac{f\tilde h_{2,0}(f)}{h_0} \approx&
\frac{1}{\sqrt{6}}\,n \left(\ln \frac{ne}{2} +\gamma_E\right)
.
\end{aligned}
\end{equation}
As $f\to0$, $f\,\operatorname{Im}\tilde h_\lambda
\propto1+(\pi^2/\Omega)|f|$, whereas $\text{Re}\,\tilde h_\lambda$ contains a logarithmic subleading term.\footnote{The mass-monopole tail gives an additional logarithmic correction~\cite{Laddha:2018myi} $\tilde h(f)\supset -2M\Delta h\left[\ln|f|-\frac{i\pi}{2}\operatorname{sgn}(f)\right]$, with $\Delta h$ the displacement memory. This contribution is not included below since it starts at $1.5$PN order~\cite{Blanchet:1992br}.} These corrections are associated with the $1/t$ falloff of the waveform as $t\to \pm \infty$.\footnote{Generically, if $h_\lambda(t)\to A_\pm+B_\pm/t+\mathcal{O}(|t|^{-1-\epsilon})$ (with $\epsilon>0$ and $A_+\ne A_-$) as $t\to\pm\infty$, then for $f\to0$, $\tilde h_\lambda(f)\to\frac{i(A_+-A_-)}{2\pi f} -(B_+-B_-)\ln|f| +\frac{i\pi}{2}(B_++B_-)\operatorname{sgn}(f)+C_\text{finite}$, where $C_\text{finite}\in\mathbb{R}$ is a generally nonuniversal term. Relative to the soft pole, the logarithmic term gives an imaginary correction $\mathcal{O}(f\ln |f|)$ and therefore dominates the \textit{phase} correction, whereas the $(B_++B_-)$ term gives the leading \textit{amplitude} correction. Hence the leading correction to the energy spectrum comes from the latter, unless its coefficient vanishes.} From Eq.~\eqref{soft_waveform_0PN_scattering}, the memory amplitude $|\Delta h|=\sqrt{\Delta h_+^2+\Delta h_\times^2}=\frac{\nu M^2}{R}\frac{\sqrt{e^2-1}}{ae^2}\sqrt{-8 s^4_\Theta c_{4\Phi}+28 c_{2\Theta}+c_{4\Theta}+35}$. Thus $|\Delta h|\propto M\sqrt{e^2-1}\,/(ae^2)=(b/M)/[(b/M)^2+1/v_0^4]$, which increases with $v_0$ for given $b$, and peaks at $b=M/v_0^2$ for given $v_0$. For given $b$ and $v_0$, $|\Delta h|$ takes the maximum value $8\frac{\nu M^2}{R}\frac{\sqrt{e^2-1}}{ae^2}$ at $\Theta\in\{0,\pi\}$, and vanishes at $\Theta=\pi/2 \,\land\, \Phi=0\mod\pi/2$. To quantify the source-side range of validity of the soft approximation, we examine $C(f)=\frac{2\pi f}{i\Delta h(\Theta,\Phi)}\tilde h(f;\Theta,\Phi)-1$, and compute $Q_{90}(e,\pi f/\Omega)$, the 90th percentile of $|C|$ over viewing directions with uniform solid-angle measure, excluding near memory-node directions with $|\Delta h|<0.1|\Delta h|_{\max}$. The result is shown in Fig.~\ref{fig:scattering_soft_validity}; to the left of a contour $Q_{90}=\epsilon$, 90\% of the retained directions have a relative deviation below $\epsilon$.
	
In the high-frequency limit, a ``hard'' waveform can be derived from Eqs.~\eqref{hyperbolic_encounter_1} and \eqref{hyperbolic_encounter_2} by using the asymptotic forms~\cite{Garc_a_Bellido_2018}:\footnote{The exponential factor is set by the nearest complex-time singularity of the Keplerian hyperbola, $e\cosh \xi=1$, where $\xi$ is the eccentric anomaly, for which $\Omega\,\operatorname{Im} t(\xi)=\sqrt{e^2-1}-\operatorname{arcsec} e$.}
\begin{equation}\label{high-freq}
\begin{aligned}
H_{in} & \sim
-\frac{i\,\text{e}^{-n \left(\sqrt{e^2-1}-\text{arcsec}\,e\right)}}{ (e^2-1)^{1/4} \sqrt{n\pi/2}}
,
\\
H_{1+in} &\sim
-\frac{\left(\sqrt{e^2-1}+i\right) \text{e}^{-n  \left(\sqrt{e^2-1}-\text{arcsec}\,e\right)}}{e (e^2-1)^{1/4} \sqrt{n\pi/2}}
.
\end{aligned}
\end{equation}
	
\begin{figure}[hbt]
\centering
\includegraphics[width=0.48\textwidth]{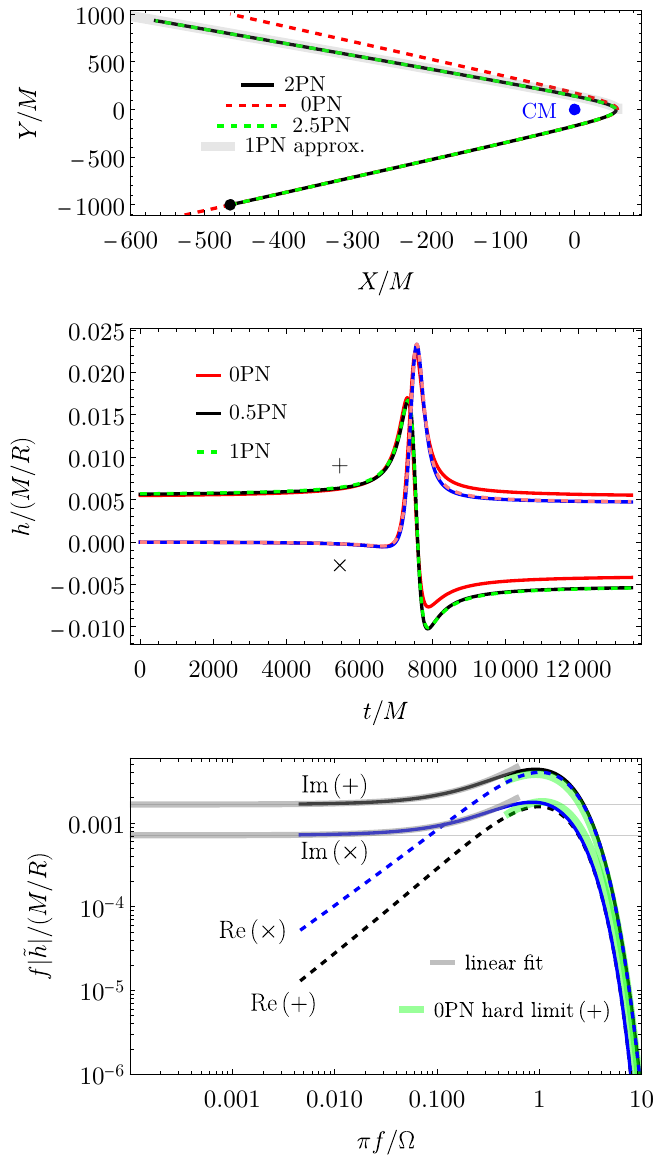}
\caption{Relative orbits in the CM frame (top panel) and gravitational waveforms in the time domain (middle panel) and frequency domain (bottom panel) for nonspinning compact-binary scattering, with $\nu=0.2$, $b/M\approx 100$, $v_0\approx 0.13$ (corresponding to $e\approx 2$), and $\{\Theta,\Phi\}=\{\pi/6,\pi/3\}$. In the middle panel, the 0PN waveform is computed using the 0PN orbit, whereas the 0.5PN and 1PN waveforms are computed using the 2PN orbit. The bottom panel shows the spectrum of the 1PN waveform, with $\Omega=\sqrt{M/a^3} = v_0^3/M \approx \pi (v_0/0.13)^3(100\,M_\odot/M)\times 1.4\,\text{Hz}$. The spectrum is computed with the time origin set at periastron crossing. The horizontal line in the bottom panel corresponds to $\Delta h_\lambda$ of the time-domain waveform.}\label{fig:scattering}
\end{figure}

Several features outlined above persist once PN corrections are included, at least when those corrections are not too large. We include PN corrections to the CM-frame equation of motion in harmonic coordinates and to the gravitational waveform, following \cite{PhysRevD.52.821}. The initial condition at $t=-\infty$ (as defined by $b$ and $v_0$) can be specified by an osculating Keplerian hyperbolic orbit about a central mass $M$ located at the CM, with orbital elements $\{a,e\}$ (see, e.g., \cite{Cheng:2024mgl}), coinciding with the one considered above. For simplicity, we neglect the spins of the two bodies, thus $\ddot{\mathbf{x}}=-(M/r^3)\,\mathbf{x}+\mathbf{F}$, with $\mathbf{F}=\mathbf{F}_\text{1PN}+\mathbf{F}_\text{2PN}+\mathbf{F}_\text{2.5PN}+\mathcal{O}(\text{3PN})$; the PN orbit also lies in the $XY$ plane, but its periastron deviates from that of the Keplerian orbit. At 1PN order, the change of the argument of periastron $\varphi_0$ of the osculating orbit during the scattering is given by~\cite{Cheng:2024mgl,Brax:2024myc}
\begin{equation}\label{1PN_periastron shift}
\frac{\Delta\varphi_0}{M/a}=\frac{(5\nu+2)e^2-5\nu+4+\frac{12e^2\arctan\frac{e+1}{\sqrt{e^2-1}}}{\sqrt{e^2-1}}}{e^2\sqrt{e^2-1}}.
\end{equation}
Numerically, a finite-radius initialization is more convenient, with the
initial condition set at $r_0$, $\{a_K,e_K\}$ denoting the initial osculating
elements, $v_{0K}=\sqrt{M/a_K}$, and $b_K=a_K\sqrt{e_K^2-1}$. The results for $\nu=0.2$, $e_K=2$, $b_K/M=100$, and $r_0/M=1105$ are shown in Fig.~\ref{fig:scattering}. To verify the convergence of PN corrections, we perform the computation at different PN orders. The approximation \eqref{1PN_periastron shift} (shown as the thick gray line in the top panel of Fig.~\ref{fig:scattering}) is found to agree well with the 2.5PN orbit, indicating that higher-order PN corrections are relatively small. The orbit in this example is in fact well described by the 1PN-accurate quasi-Keplerian parametrization~\cite{1985AIHPA..43..107D}, and the extrapolated values of $\{v_0,b\}$ are close to $\{v_{0K},b_K\}$. The 1PN correction to the waveform is already small; note that the null memory contribution to the waveform in the present case enters at 2.5PN order~\cite{Favata_2011}, and is thus negligible.
	
The spectrum of the 0PN waveform computed using the 0PN orbit agrees well with Eqs.~\eqref{hyperbolic_encounter_1} and \eqref{hyperbolic_encounter_2}. We also find that when $|C(f_\text{ref})|$ is small, the linear fit performed near the lowest sampled frequency $f_\text{ref}$ provides a reasonable extrapolation to frequencies $0<f<f_\text{ref}$. This extrapolation improves as $f_\text{ref}$ decreases. The spectra of $|\text{Re}\,\tilde h_\lambda(f)|$ and $|\text{Im}\,\tilde h_\lambda(f)|$ of the 1PN waveforms are shown in the bottom panel of Fig.~\ref{fig:scattering}. The numerical spectrum at the lowest frequency is close to the soft waveform with $\Delta h_\lambda$ computed in the time domain. In the high-frequency limit, the radiation is elliptically polarized.\footnote{Using \eqref{high-freq}, the high-frequency limit of the 0PN waveform is dominated by $|m|=2$ modes and becomes elliptically polarized, since $\lim_{f\to\infty}(\tilde h_+,\tilde h_\times)\propto \text{e}^{2i\Phi}(3+c_{2\Theta}, 4ic_\Theta)$, similar to that of a circular orbit. This is also reflected in the 0PN spectra of the $Z$-component angular momentum and graviton number fluxes, whose ratio approaches 2 at $f\to\infty$ (as for an elliptical orbit).} The decaying behavior of the spectrum at high frequencies is similar to the 0PN approximation given by Eq.~\eqref{high-freq}, as depicted by the green lines in the bottom panel of Fig.~\ref{fig:scattering}.
	
\begin{figure*}[hbt]
\centering
\includegraphics[width=0.325\textwidth]{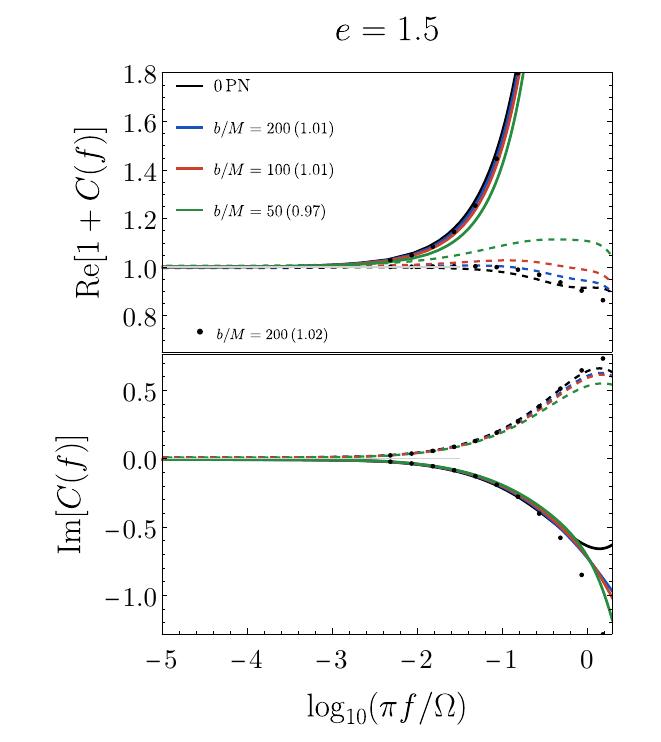}
\includegraphics[width=0.325\textwidth]{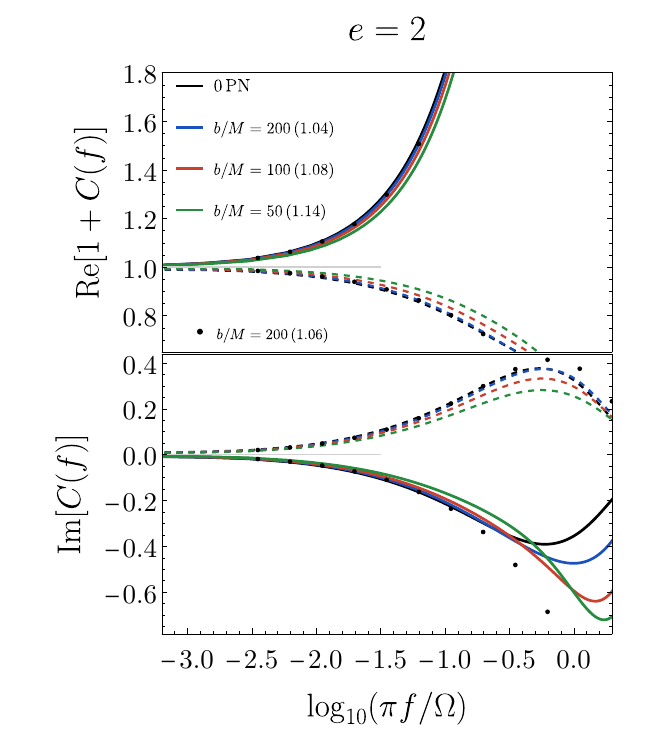}
\includegraphics[width=0.325\textwidth]{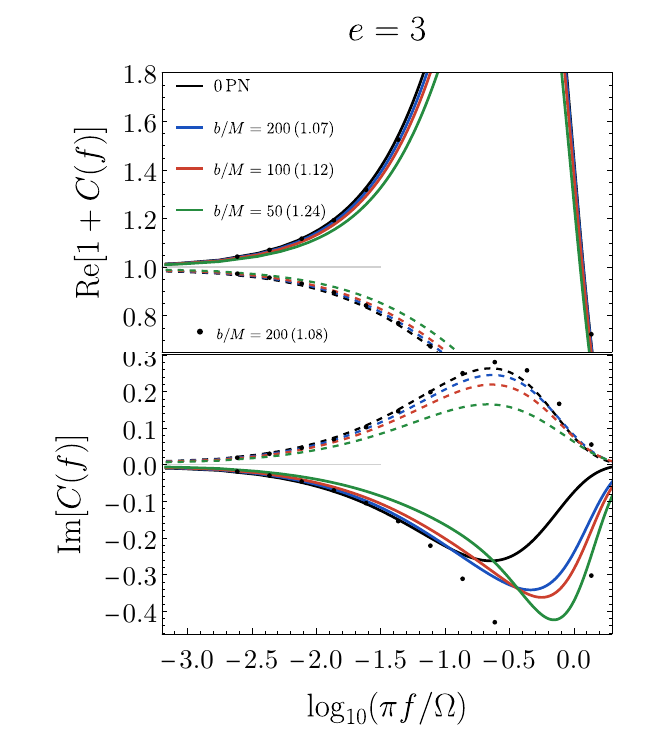}
\caption{Examples of the finite-frequency correction factor for nonspinning compact-binary scattering, with $C(0)=0$. Black lines denote the Newtonian result, while colored lines denote the 1PN waveform computed using the 2PN orbit for different impact parameters $b/M$; the numbers in parentheses give the corresponding ratio $H/H_{\rm 0PN}$ of the memory amplitudes. Solid and dashed lines represent the positive- and negative-frequency spectra, respectively. The gray horizontal lines mark the soft limits. The black points denote the frequency-domain results for nonspinning extreme-mass-ratio scattering obtained by solving the Regge-Wheeler-Zerilli master equations in the test-body approximation, including spherical-harmonic modes with $l\le 6$, for $b/M=200$, following Refs.~\cite{HopperCardoso2018Scattering,Hopper2018UnboundRWZ}; the corresponding soft limits are given by Eq.~\eqref{EMR_memory}.}\label{fig:scattering_2}
\end{figure*}

Figure~\ref{fig:scattering_2} shows the correction factor $C(f)=\frac{2\pi f}{i\Delta h}\tilde h(f)-1$, with the time origin set at periastron crossing and the orbital periastron on the $X$ axis, for $\nu=0.2$, $\{\Theta,\Phi\}=\{1,0.7\}$, and various pairs $\{b,e\}$ with $e=\sqrt{1+(b/M)^2v_0^4}$. The reference frequency is $\Omega=v_0^3/M$. The solid and dashed lines show the positive- and negative-frequency spectra, respectively. For fixed $e$, the 0PN correction factor is independent of $\nu$ and
$b$; the associated directional deviations are summarized statistically
in Fig.~\ref{fig:scattering_soft_validity}. At higher frequencies, the
PN result shows a clearer $b$ dependence, and the memory amplitude
is also modified by PN corrections.
	
For an extreme-mass-ratio scattering involving a large black hole (BH), the orbit of the small compact body can be approximated by an unbound geodesic in the background BH spacetime. Self-force effects are expected to be negligible at sufficiently high incident velocities. We compute the ordinary displacement memory produced by such a system in the direction $\hat{\mathbf{k}}(\Theta,\Phi)$ using the general formula~\cite{Braginsky:1987kwo}:
\begin{equation}\label{scattering_memory}
\Delta h_{ij}=\Lambda_{ij,kl}\sum_I \frac{4}{R}\Delta\left[\frac{m_Iv_I^kv_I^l}{\sqrt{1-|\mathbf{v}_I|^2}\,(1-\mathbf{v}_I\cdot\hat{\mathbf{k}})}\right].
\end{equation}
The corresponding spin-weighted spherical-harmonic decomposition is $\Delta h =\sum_{l,m}(\Delta h)_{l,m}{}_{-2}Y_{lm}(\Theta,\Phi)$,
\begin{equation}\label{scattering_memory_spherical}
(\Delta h)_{l,m}=\sum_I \frac{4}{R}\Delta\left[\frac{m_I|\mathbf{v}_I|^2F_l(|\mathbf{v}_I|)}{\sqrt{1-|\mathbf{v}_I|^2}}\,Y_{lm}^*(\theta_I,\phi_I)\right],
\end{equation}
where $(\theta_I,\phi_I)$ is the direction of $\mathbf v_I$, and
\begin{equation}
F_l(v)=4\pi\sqrt{\frac{(l-2)!}{(l+2)!}}\frac{(1-v^2)^2}{2v^2}\int_{-1}^1dz\,\frac{P_l(z)}{(1-vz)^3},
\end{equation}
where $P_l(z)$ denotes the Legendre polynomial of degree $l$.  In the null limit, $F_l(1)=4\pi\sqrt{\frac{(l-2)!}{(l+2)!}}$. Equation~\eqref{scattering_memory} can be derived from the linearized Einstein equation in a flat background, but holds also for asymptotically flat spacetimes when the incoming and outgoing objects approach inertial asymptotic states. The same result follows from the classical limit of the leading soft graviton theorem~\cite{strominger2014gravitationalmemorybmssupertranslations}, while the underlying infrared structure is also closely related to the dressing of asymptotic states and to decoherence from unresolved soft gravitons~\cite{Carney2017InfraredQuantumInformation,Carney_2018}. The low-velocity expansion of Eq.~\eqref{scattering_memory} up to $\mathcal{O}(v^4)$ (using the 1PN approximation for $\mathbf{v}_{1,2}$ in the CM frame) agrees with the 1PN approximation:
\begin{equation}\label{1PN_memory}
\begin{aligned}
\Delta h_{ij}&=\Lambda_{ij,kl}\,\Delta\Big[2v^kv^l\Big(1+\frac{m_1-m_2}{M}(\mathbf{v}\cdot \hat{\mathbf{k}})
\\
&\quad
+(1-3\nu)\left[(\mathbf{v}\cdot \hat{\mathbf{k}})^2+v^2/2\right]\Big)\Big]\frac{2\nu M}{R}.
\end{aligned}
\end{equation}
For simplicity, we consider a nonspinning central BH, and thus compute the deflection angle $\chi$ of a test particle with mass $\nu M$ in a Schwarzschild spacetime with mass parameter $M$. The source frame is again chosen such that the initial orbit lies in the $XY$ plane, the initial and final three-velocities are along the direction $\mathbf{N}_\text{in}=c_\vartheta \mathbf{e}_X+s_\vartheta\mathbf{e}_Y$ and $\mathbf{N}_\text{out}=-c_\vartheta \mathbf{e}_X+s_\vartheta\mathbf{e}_Y$, respectively, with $\vartheta = (\pi-\chi)/2$. The displacement memory from Eq.~\eqref{scattering_memory} is given by
\begin{equation}\label{EMR_memory}
\begin{aligned}
\frac{\Delta h_+}{F/R} &=
\frac{c_{\chi +2 \Phi}+v_0^{-2}}{1-v_0 s_\Theta s_{\chi/2+\Phi}}-\frac{c_{\chi -2 \Phi}+v_0^{-2}}{1+ v_0 s_\Theta s_{\chi/2 - \Phi}}-\frac{s_\Theta s_{\chi/2} c_\Phi}{v_0/2},
\\
\frac{\Delta h_\times}{F/R} &=
\frac{v_0 s_{2 \Theta} s_{\chi/2} s_\Phi (c_\chi-c_{2 \Phi})+2c_\Theta s_\chi c_{2 \Phi}}{\left(v_0 s_\Theta s_{\chi/2 - \Phi}+1\right) \left(v_0 s_\Theta s_{\chi/2+\Phi}-1\right)},
\end{aligned}
\end{equation}
where $F=2\omega $ and $v_0=1$ for a null particle with energy $\omega$, and $F=\frac{2m v_0^2}{\sqrt{1-v_0^2}}$ for a timelike particle with mass $m$.
	
\begin{figure}[hbt]
\centering
\includegraphics[width=0.48\textwidth]{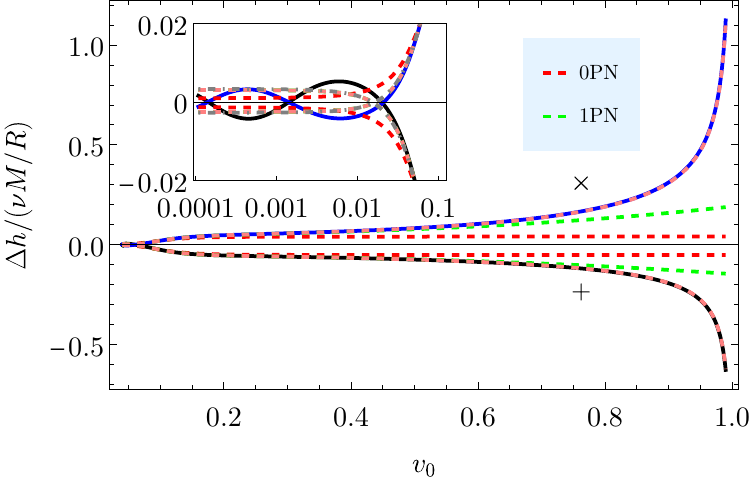}
\caption{Ordinary displacement memory from the scattering of a test body with mass $\nu M$ by a nonspinning BH with mass $M$, for $b/M=100$ and $\{\Theta,\Phi\}=\{\pi/5,\pi/7\}$. The blue and black solid lines show the test-body approximation. The red (green) dashed lines show the 0PN (1PN) approximations. The pink dashed lines show the results computed from Eq.~\eqref{EMR_memory}, using the 1PN approximation to the deflection angle. The critical incident velocity is $v_\text{plunge}\approx 0.04$. The inset shows the results for small incident velocities as a function of $v_0-v_\text{plunge}$. The gray dot-dashed lines show the results for a central BH with dimensionless spin $\chi_\bullet=1$ along the $Z$-axis, for which $v_\text{plunge}=0.02$.
}\label{fig:EMR}
\end{figure}
	
The deflection angle depends on the initial velocity $v_0$ and the dimensionless impact parameter $b/M$. For the small body to be scattered rather than plunge into the BH,\footnote{
Such a plunge will also produce displacement memory. If the incident velocity at infinity is sufficiently large, the ordinary memory is expected to dominate and is approximately given by Eq.~\eqref{scattering_memory}, with only the incoming particle contributing. If the incident velocity at infinity is strictly zero, the ordinary contribution vanishes at leading order in the mass ratio (neglecting the recoil of the final remnant), but there is a null memory contribution.
} the impact parameter needs to exceed a critical value~\cite{Unruh:1976fm}: ${b_\text{plunge}}/{M}=\sqrt{4+\frac{\sqrt{8 v_0^2+1}-1}{2v_0^4}+\frac{4\sqrt{8 v_0^2+1}+10}{v_0^2}}$. For a null particle, $b_\text{plunge}(v_0=1)=3\sqrt{3}M$; in the nonrelativistic limit, $b_\text{plunge}(v_0\to 0)\to 4M/v_0$. It is possible for high-energy massless or ultrarelativistic particles to be scattered by a BH, but this cannot produce an observable memory, since even a photon with $\omega=10\,\text{TeV}$ amounts to a mass of only $9\times 10^{-54}\,M_\odot$.
	
For the scattering of timelike particles, a concrete example is shown in Fig.~\ref{fig:EMR}, where the results are compared with the 0PN approximation in Eq.~\eqref{soft_waveform_0PN_scattering} (red dashed lines) and the results computed using Eq.~\eqref{1PN_memory} and the 1PN approximation to the deflection angle: $\chi=2\arccos(-1/e)-\pi+\Delta\varphi_0$ (green dashed lines), with $\Delta\varphi_0$ given by Eq.~\eqref{1PN_periastron shift}. For comparison, the pink dashed lines show Eq.~\eqref{EMR_memory} evaluated with the 1PN deflection angle, which agrees well with the test-body result even at high velocities for this large impact parameter. For $v_0\ll1$ and $b\gg M/v_0$, the finite-frequency correction factor of the waveform is close to the 0PN result (see Fig.~\ref{fig:scattering_2} for an example). The 0PN approximation is accurate only for intermediate velocities $v_0$, and the near-$v_\text{plunge}$ regime is not captured by the 1PN deflection angle either. For $v_0\gtrsim v_\text{plunge}$, the periastron occurs at $r\gtrsim 4M$; the result is therefore sensitive to the spin of the central BH. Equation~\eqref{EMR_memory} can also be applied to equatorial scattering in Kerr spacetime; the gray dot-dashed lines in Fig.~\ref{fig:EMR} show the result for a prograde equatorial orbit around a central BH with dimensionless spin $\chi_\bullet=1$ aligned with the $Z$-axis; they lie closer to the pink dashed lines because the prograde spin lowers the plunge threshold, making the orbit less close to the strong-field regime. At very high velocities, the memory amplitudes are considerably larger than the 0PN prediction.
	
\subsection{Displacement memory from BBH merger}\label{Sec:Displacement memory from quasi-circular BBH merger}
	
Because gravity is nonlinear, displacement memory can also be produced by gravitational radiation itself~\cite{PhysRevLett.67.1486}. Perturbatively, the spherical-harmonic modes of the produced secondary GW (memory modes) are given by the time integral~\cite{PhysRevD.45.520,PhysRevD.44.R2945,Blanchet:1992br,Favata_2009}:\footnote{This can be obtained, e.g., from Eq.~\eqref{scattering_memory_spherical} by identifying the outgoing particles with a flux of gravitons~\cite{PhysRevD.45.520}, for which $|\mathbf{v}_I(\theta,\phi)|=1$, and ${m_I}/{\sqrt{1-|\mathbf{v}_I|^2}}\to \int dt'd\Omega_2(\theta,\phi)\frac{dE_\text{gw}}{dt'\,d\Omega_2}$.}
	
\begin{equation}
h_{l,m}^\text{(mem)}(u)
\approx\frac{16\pi}{R} \sqrt{\frac{(l-2)!}{(l+2)!}} \int^u dt'\int d\Omega_2 \,\frac{dE_\text{gw}}{dt'\,d\Omega_2}\,Y_{lm}^*,
\label{null_memory_formula}
\end{equation}
where $d\Omega_2(\theta,\phi)\equiv\sin\theta\,d\theta\,d\phi$. A similar result holds for other forms of null radiation, but the gravitational case is special since it arises from the self-interaction of GWs. The angular distribution of the instantaneous energy flux of the primary GW at null infinity can be written as
\[
\frac{\frac{dE_\text{gw}}{dt\,d\Omega_2}}{R^2}=\frac{|\dot h|^2}{16\pi}
=\frac{1}{16\pi}\sum_{g}\dot h_{l_1,m_1} \dot h^*_{l_2,m_2}\,_{-2}Y_{l_1m_1} \,_{-2}Y^*_{l_2m_2},
\]
summing over $g=\{l_1,m_1,l_2,m_2\}$. For a quasiperiodic GW source, Eq.~\eqref{null_memory_formula} contains both nonoscillatory and oscillatory contributions; the nonoscillatory component grows approximately linearly over a time interval short compared with the secular evolution time. As discussed in Sec.~\ref{subsec:Displacement memory}, the detectability of such a waveform is poor. For searches for null displacement memory, one is therefore interested in systems capable of producing strong and predictable GW bursts with negligible ordinary displacement memory. Binary black hole mergers are ideal targets. In this section, we discuss some features of the potentially observable low-frequency spectra of such systems. We consider quasicircular nonprecessing BBH systems and employ the numerical relativity surrogate model \texttt{NRHybSur3dq8\_CCE}~\cite{Yoo_2023} (constructed for the mass ratio $q=m_1/m_2\in [1,8]$) to compute their gravitational waveforms.
	
\begin{figure}[hbt!]
\centering
\includegraphics[width=0.48\textwidth]{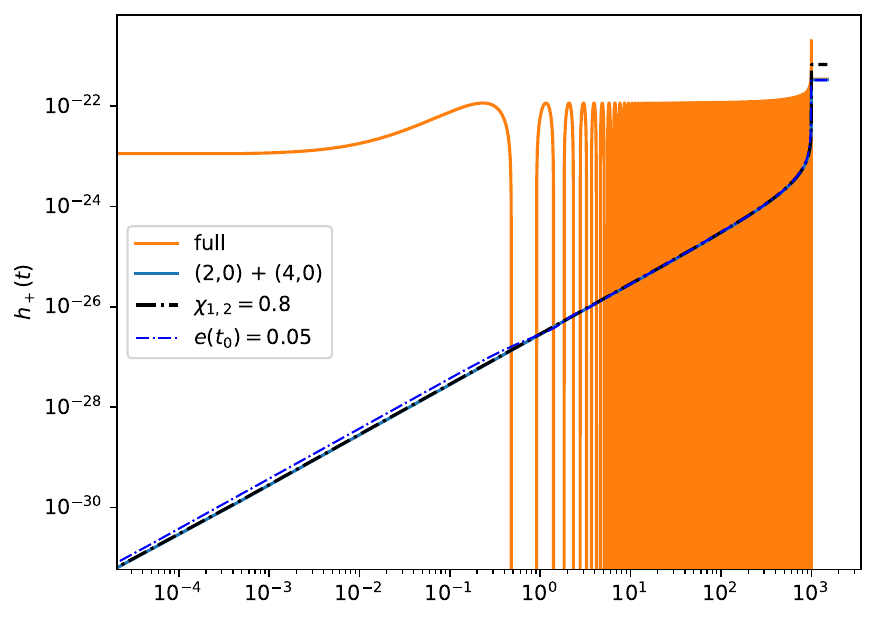}
\\
\includegraphics[width=0.48\textwidth]{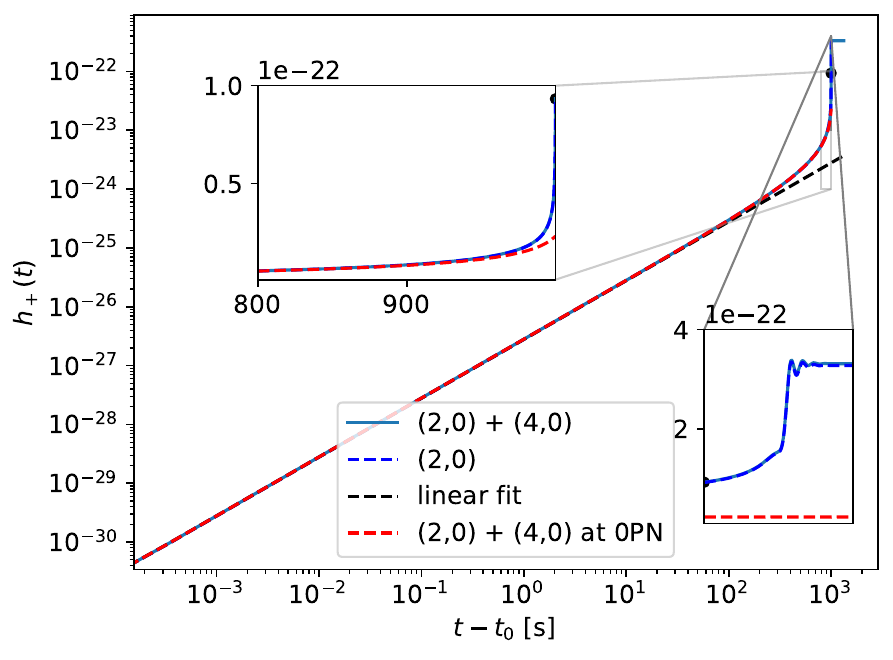}
\caption{Gravitational waveform of a quasicircular BBH merger from \texttt{NRHybSur3dq8\_CCE}, with $q=1$, $M=100\,M_\odot$, $\Theta=\pi/3$, $R=400\,\text{Mpc}$ (the effect of cosmic expansion is not included), and $t_0=-1000\,\text{s}$. The top panel compares the full waveform (orange) with the memory-mode waveform from the $h_{2,0}$ and $h_{4,0}$ modes (purple) for a nonspinning system. The black dot-dashed line shows the memory-mode waveform for aligned dimensionless spins $\chi_{1,2}=0.8$. The blue dot-dashed line shows the $h_{2,0}$ mode for an eccentric BBH waveform with $e(t_0)=0.05$ and the same initial orbit-averaged orbital angular frequency, reconstructed from the \texttt{SEOBNRv5EHM} oscillatory modes using Eq.~\eqref{null_memory_formula} (the resulting $h_{2,0}(t)$ lines for $e(t_0)=0.05$ and $e(t_0)=0$ nearly overlap after a relative time shift of about $8.8\,\mathrm{s}$). The bottom panel compares the $h_{2,0}$ mode (blue dashed) with its 0PN approximation (red dashed) and linear approximation (black dashed) for a nonspinning system.}\label{fig:circular_merger12}
\end{figure}

\begin{figure}[hbt!]
\centering
\includegraphics[width=0.48\textwidth]{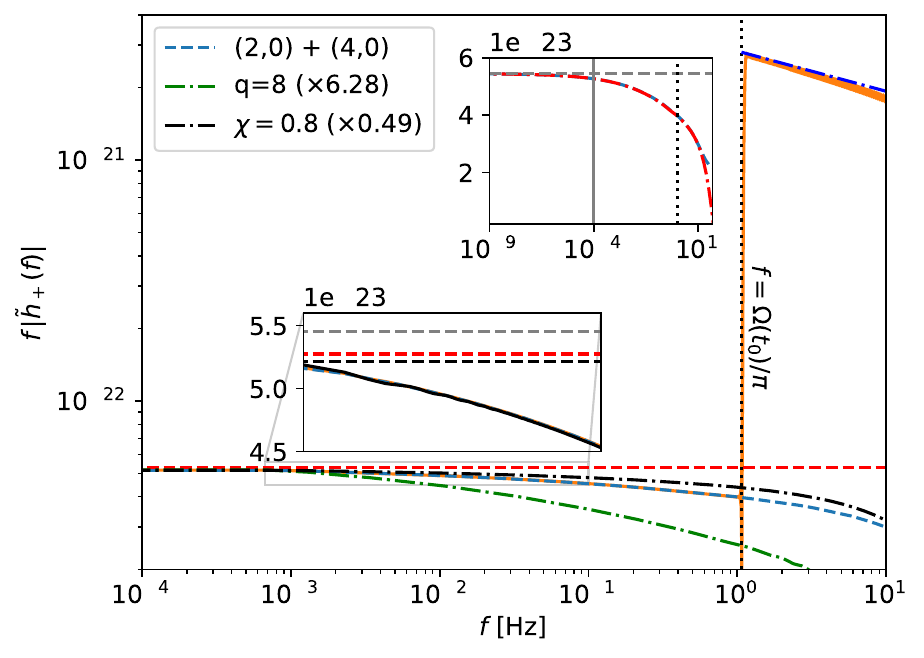}
\caption{Frequency spectra of the waveforms in Fig.~\ref{fig:circular_merger12}. The horizontal red (black) dashed line corresponds to $\Delta h_+$ of the full (shortened) time-domain waveform in $[t_0,t_\text{final}]$; the horizontal gray dashed line corresponds to $(\Delta h_+)_\text{max}$. The black solid line is the spectrum of the shortened waveform. The green dot-dashed line shows the memory-mode spectrum for $q=8$, multiplied by 6.28, and the black dot-dashed line shows the memory-mode spectrum for $q=1$ and aligned spins $\chi_{1,2}=0.8$, multiplied by 0.49. The blue dot-dashed line shows the 0PN oscillatory-mode spectrum ($f|\tilde h_+|=\frac{1+\cos^2 \Theta}{2}\,\frac{M^{5/6}}{\pi^{2/3}R}\left(\frac{5\nu}{24}\right)^{1/2}f^{-1/6}$) in the stationary phase approximation. The lower inset shows a zoomed-in view of the low-frequency region. The left inset shows a fit to the full spectrum over $[10^{-4},1]\,\text{Hz}$. The upper inset compares $f\,|\tilde h_+(f)|$ with $f\,\text{Im}\,\tilde h_+(f)$ (red dot-dashed line) of the memory-mode waveform. $f_\text{ISCO}\equiv\sqrt{M/(6M)^3}/\pi\approx 44\,\text{Hz}$, and the maximum oscillation frequency $\sim 100\,\text{Hz}$.}\label{fig:circular_merger3}
\end{figure}

Long before the merger, the 0PN approximation to the waveform is adequate. In this regime, the orbital evolution is given by $\dot x=J x^5$, where $J\equiv 64\nu/(5M)$, and $x\equiv (M\Omega)^{2/3}$ is the PN parameter, with $\Omega$ being the orbital frequency. Meanwhile, the gravitational radiation in a source frame with $Z$-axis aligned with the orbital angular momentum is dominated by the $h_{2,\pm 2}$ modes, corresponding to the quadrupole waveform. Because the orbital evolution is adiabatic, the steadily accumulated memory contribution also appears at 0PN order and is dominated by the $h_{2,0}$ mode.\footnote{For precessing binaries, there could be non-negligible contributions from $h_{2,m\ne 0}$ modes~\cite{PhysRevD.98.064031,pqcz-tqnn}.} The LO memory-mode waveform with $h_{l,m}^\text{(mem)}(x=0)=0$ is given by\footnote{For a primary GW given by $h_{2,\pm 2}(t)=A\,\text{e}^{\mp 2i\Omega t}$ with $A\in\mathbb{R}$, Eq.~\eqref{null_memory_formula} yields $\dot h_\times^{\text{(mem)}}=0$ and $\dot h_+^{\text{(mem)}}=R A^2  \Omega^2  s^2_\Theta\,(17+c^2_\Theta)/(48\pi)$. For the 0PN oscillatory waveform of a circular binary, $A=8\sqrt{\pi/5}\,\nu Mx(t)/R$. Note that the oscillatory waveform of a scalar or vector gravitational atom in the saturated superradiant ground state~\cite{Siemonsen:2022yyf,Cao:2024wby} is similar, with $A\approx A(t_0)/[1+(t-t_0)/\tau_\text{gw}]$, so that $\Delta h_+^{\text{(mem)}}\propto 1-\tau_\text{gw}/(t-t_0+\tau_\text{gw})$.}~\cite{PhysRevD.44.R2945} $h_+^\text{(mem)}(t)=Kx$ and $h_\times^\text{(mem)}(t)=0$, where $K\equiv{\nu M s^2_\Theta\,(17+c^2_\Theta)}/(48 R)$. The change of $h_+^\text{(mem)}$ with respect to its initial value at $t_0$ is given by $\Delta h_+^\text{(mem)}(t)\equiv h_+^\text{(mem)}(t)-h_+^\text{(mem)}(t_0)=Kx_0\left\{[1-4Jx_0^4(t-t_0)]^{-1/4}-1\right\}$, where $x_0=x(t=t_0)$. A concrete example with $M=100\,M_\odot$ and $q=1$ is shown in Fig.~\ref{fig:circular_merger12}, where the merger occurs near $t=0$. As shown in the bottom panel, the initial memory-mode waveform is well described by an \textit{effective} 0PN extrapolation: $h_{+,\text{eff}}^\text{(mem)}=K x_\text{eff}$, with $\dot x_\text{eff}=Jx_\text{eff}^5$. Here $x_\text{eff}(t_0)$ is fixed by matching the memory growth rate to $\dot h_+^\text{(mem)}$ at $t_0$. In this example we choose $t_0=-1000\,\text{s}$, for which $x(t_0)\approx 0.0141$ and $x_\text{eff}(t_0)\approx 0.0140$. The Newtonian inspiral is followed by the late inspiral and merger, during which the memory can deviate significantly from the effective 0PN extrapolation, before finally settling to a nearly saturated value in the ringdown phase. Throughout this process, $h_{2,0}$ is the dominant memory mode. The top panel of Fig.~\ref{fig:circular_merger12} compares the memory-mode waveform with the full (memory + oscillatory) waveform including all modes available in the surrogate model. Although the oscillatory component dominates the waveform before merger, it rapidly decays during the ringdown, leaving only the memory component. The frequency spectrum of $h_+(t)$ is shown by the orange line in Fig.~\ref{fig:circular_merger3}, whose low-frequency part is dominated by the memory modes. The low-frequency spectrum of $h_\times(t)$ is dominated by the spin memory mode $h_{3,0}$~\cite{Nichols:2017rqr} (a magnetic-parity memory component), but its amplitude is much smaller than that of $h_+(t)$. In practice, the low-frequency signal from spin memory would therefore be masked by that of displacement memory.
	
\begin{figure}[hbt!]
\centering
\includegraphics[width=0.48\textwidth]{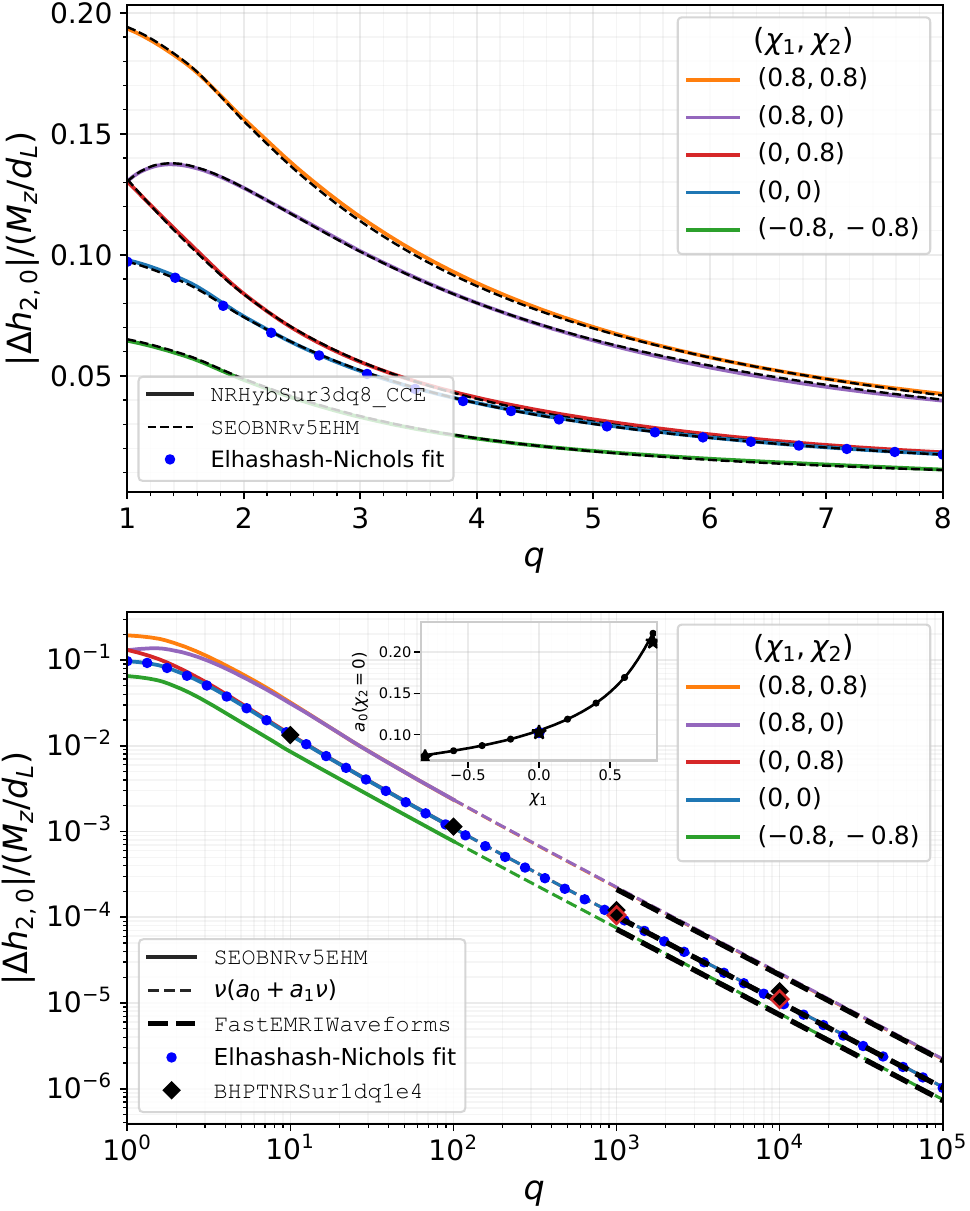}
\caption{
Dimensionless amplitude $\mathcal{A}_{2,0}\equiv |\Delta h_{2,0}|/(M_z/d_L)$ as a function of the mass ratio $q$. In the top panel, the solid lines show the \texttt{NRHybSur3dq8\_CCE} results for different aligned-spin configurations $(\chi_1,\chi_2)$, while the black dashed lines show the corresponding \texttt{SEOBNRv5EHM}-based results. The blue points show the nonspinning numerical relativity (NR) surrogate fit from Ref.~\cite{wzqk-62wc}. The bottom panel shows $\mathcal{A}_{2,0}$ over a wider mass-ratio range. The colored solid lines show the \texttt{SEOBNRv5EHM}-based results. The black diamonds and the thick black dashed lines show the results computed from the \texttt{BHPTNRSur1dq1e4} and \texttt{FastEMRIWaveforms} oscillatory modes, respectively. Diamonds with red edges mark \texttt{BHPTNRSur1dq1e4} results obtained with an effective extrapolation using the 3.5PN expression for $h_{2,0}(x)$~\cite{Cunningham_2025} and the 3PN expression for $\dot x$~\cite{Blanchet:2001aw}. The thin dashed extensions continue the high-$q$ fit $\mathcal{A}_{2,0}=\nu(a_0+a_1\nu)$, with $\nu=q/(1+q)^2$. The inset shows the fitted leading coefficient $a_0$ as a function of the primary spin $\chi_1$ for $\chi_2=0$; the \texttt{FastEMRIWaveforms} results are marked by stars.
}
\label{fig:delta_h20_scaled_vs_q}
\end{figure}
	
Because memory accumulates slowly during the early inspiral, the low-frequency spectrum in Fig.~\ref{fig:circular_merger3} does not faithfully reflect the memory amplitude. The horizontal red dashed line in Fig.~\ref{fig:circular_merger3} indicates the value $|\Delta h_+ ^\text{(mem)}(t_\text{final})|/(2\pi)$, while the black dashed line indicates the value $|\Delta h_+ ^\text{(mem)}(t_\text{final})-\Delta h_+ ^\text{(mem)}(t_\text{mid})|/(2\pi)$, which is the memory amplitude of the shortened waveform (with zero padding for $t<t_\text{mid}$): $h_+'(t)=[h_+(t)-h_+(t_\text{mid})]\,\Theta(t-t_\text{mid})$, with\footnote{This is chosen so that $h_{2,0}(t<t_\text{mid})$ can be described by the effective 0PN extrapolation. We verified that the numerical spectrum of the waveform $h_+(t)-h_+'(t)$, describing the growth before $t_\text{mid}$, agrees well with the analytic spectrum derived from the 0PN memory-mode waveform.} $t_\text{mid}=-300\,\text{s}$. Neither horizontal line precisely matches the low-frequency end of the corresponding spectrum, although the differences are small. The difference between the spectra of the shortened and full waveforms is modest and appears primarily below $f= 10^{-2}\,\text{Hz}$, reflecting the orbital evolution in $[t_0,t_\text{mid}]$. Assuming that the memory is nearly saturated at $t_\text{final}$, we can approximately construct $\tilde h^\text{(mem)}_{+,\infty}(f)$ using $h_+^\text{(mem)}(t)$ in $[t_\text{mid},t_\text{final}]$ and the analytic spectrum of the 0PN memory-mode waveform, which is almost indistinguishable from the blue dashed line. At sufficiently low frequencies, $\tilde h^\text{(mem)}_{+,\infty}(f)$ approaches the soft waveform corresponding to $(\Delta h_+)_\text{max}\equiv h_+^\text{(mem)}(\infty)-h_+^\text{(mem)}(-\infty)\approx h_+ ^\text{(mem)}(t_\text{final})-h_+ ^\text{(mem)}(t_0)+Kx_\text{eff}(t_0)$; see the upper inset of Fig.~\ref{fig:circular_merger3}. Since late-time ringdown and Price-law tails should be subleading in the $f\to 0$ expansion, the exact LO correction factor $C(f)$ for $\tilde h^{(\mathrm{mem})}_{+,\infty}(f)$ is set by the early-time Newtonian behavior $h_+^{(\mathrm{mem})}\propto(-t)^{-1/4}$, and therefore scales as $f^{1/4}$. For a given total mass $M$, as the mass ratio $q>1$ increases, $\Delta h_+^\text{(mem)}(t_\text{final})$ decreases, while the relative correction $|C(f)|$ increases. This is illustrated in Fig.~\ref{fig:circular_merger3} by comparing the normalized spectra of systems with $q=1$ and $q=8$. Figures~\ref{fig:circular_merger12} and \ref{fig:circular_merger3} also illustrate that $\Delta h_+^\text{(mem)}(t_\text{final})$ is insensitive to a small initial orbital eccentricity but can be significantly affected by the BH spins. As the observation time increases, lower frequency components of the memory mode can be resolved, while the dominant quadrupolar oscillatory mode starts at $f\approx \Omega(t_0)/\pi$. In the Newtonian regime, we have ${\Omega(t_0-\Delta t)}/{\Omega(t_0)}=\ell$, with
\[
\ell\approx\left[1+26\left(\frac{\Omega(t_0)}{\pi\text{Hz}}\right)^{\frac{8}{3}}\left(\frac{M/M_\odot}{100}\right)^{\frac{5}{3}}\left(\frac{\Delta t/\text{yr}}{10^{-3}}\right)\left(\frac{\nu}{\frac{1}{4}}\right)\right]^{-\frac{3}{8}}.
\]
Assuming that $\Omega(t_0)/\pi=1\,\text{Hz}$, for $M=100\,M_\odot$ and $\nu=1/4$, $\ell(\Delta t=10^{-4}\,\text{yr})\approx 0.62$, $\ell(\Delta t=10^{-3}\,\text{yr})\approx 0.3$, while $\ell(\Delta t=1\,\text{yr})\approx 0.02$. For smaller $M$, $\Omega(t_0)/\pi=1\,\text{Hz}$ corresponds to an earlier stage of the evolution. Therefore, for quasicircular stellar-mass BBH mergers and $\Delta t<10^{-3}\,\text{yr}$, $\Omega(t_0-\Delta t)/\pi$ typically remains at the sub-Hz to Hz scale.
	
Because the orbital dynamics and gravitational waveforms obey a mass-scaling symmetry at fixed dimensionless spins and mass ratio, the memory-mode spectra in an expanding universe (see Sec.~\ref{Propagation of waveforms}) exhibit the usual mass-redshift degeneracy:
\[
\tilde h_\infty'(f,M',d_L)=\left(\frac{M'_z}{M}\right)^2\tilde h_\infty\left(\frac{M'_z}{M}f,M,d_L\right),
\]
where $M_z'\equiv(1+z)M'$. Considering nonspinning circular binaries and $\Theta=\pi/2$, for
$(q,M/M_\odot,d_L/\text{pc})=(1,10^2,4\times 10^8)$,
$(\Delta h_+)_\text{max}\sim 5\times 10^{-22}$.
Subsolar-mass compact binary mergers, such as primordial-BH binaries, may occur
at much smaller distances and thus be detectable~\cite{Ebersold_2020}. For
example, $(q,M/M_\odot,d_L/\text{pc})=(1,2,10^4)$ gives
$(\Delta h_+)_\text{max}\sim 4\times 10^{-19}$, while
$(q,M/M_\odot,d_L/\text{pc})=(1,4\times10^{-4},10^{-2})$~\cite{PhysRevLett.127.071102}
gives $(\Delta h_+)_\text{max}\sim 7\times10^{-17}$.
In the latter case, the memory-mode spectrum below $0.1\,\text{Hz}$ may closely
resemble a soft waveform.
	
\begin{figure*}[hbt]
\centering
\includegraphics[width=\textwidth]{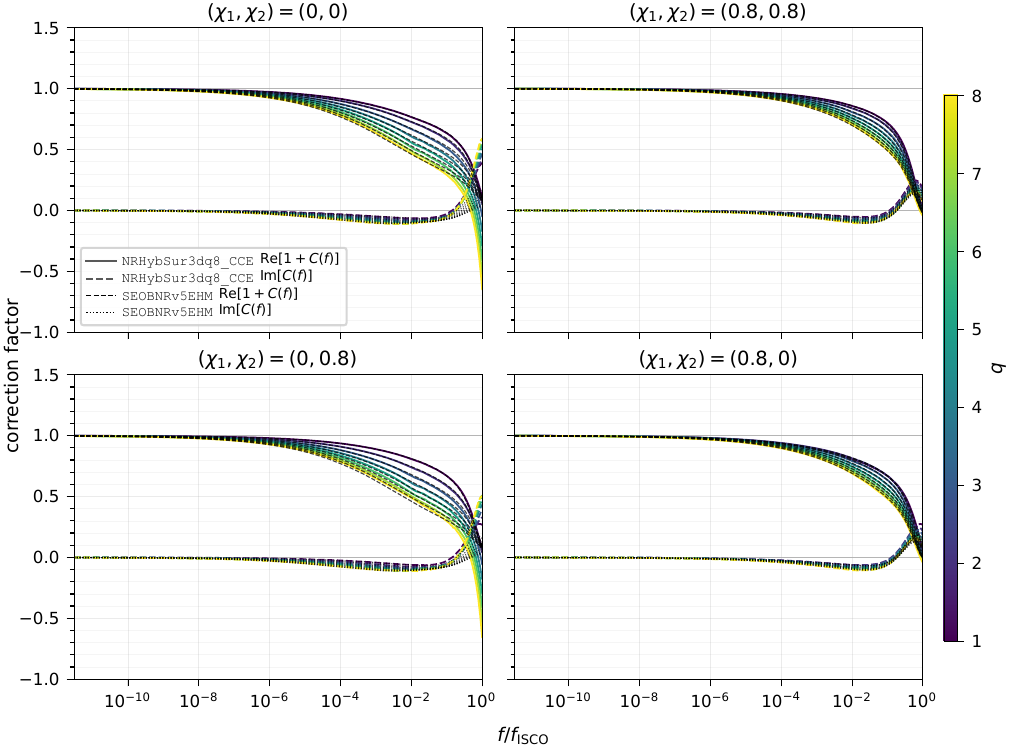}
		
\caption{Finite-frequency correction factor for the nonprecessing quasicircular binary memory mode $\tilde h_{2,0}$ for $q\in [1,8]$, in various aligned-spin configurations. The plotted quantities are $\operatorname{Re}[1+C(f)]$ and $\operatorname{Im}[C(f)]$. Note that $C(0)=0$, and $C(f)=[C(-f)]^*$ because the waveform is linearly polarized.}
\label{fig:circular_correction_q_spin}
\end{figure*}

\begin{figure}[hbt]
\centering
\includegraphics[width=0.48\textwidth]{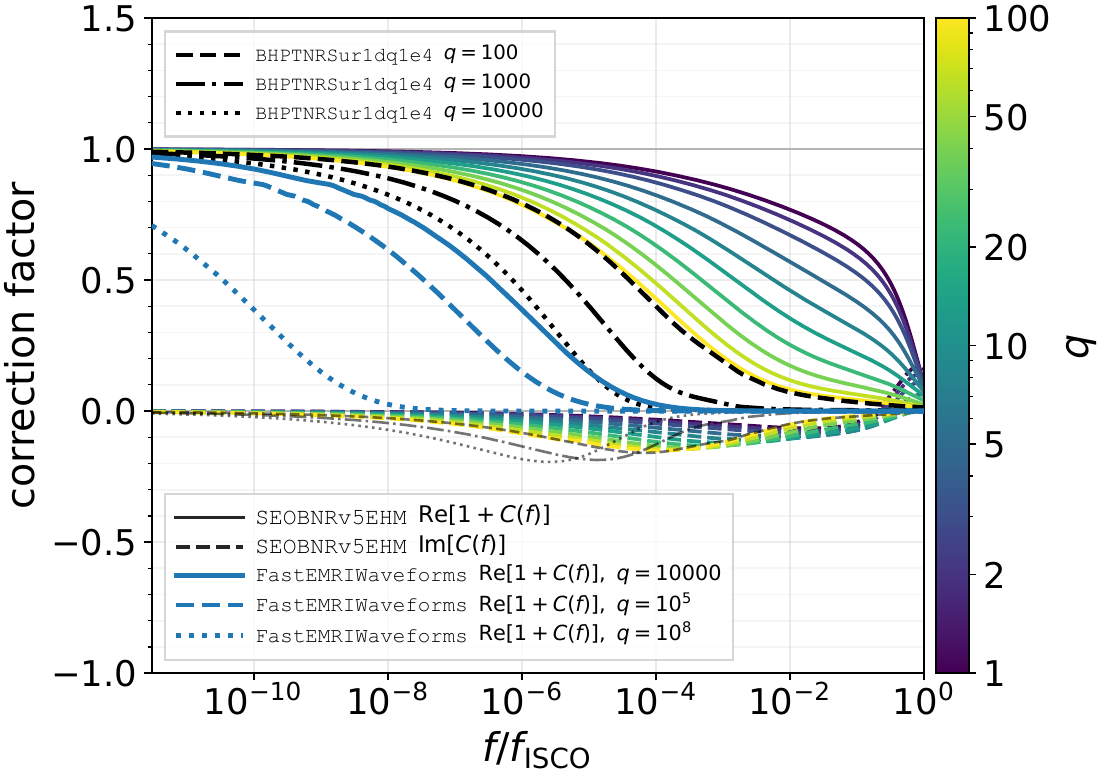}
\caption{Finite-frequency correction of the nonspinning, nonprecessing quasicircular binary memory mode $\tilde h_{2,0}$. The plotted quantities are $\operatorname{Re}[1+C(f)]$ and $\operatorname{Im}[C(f)]$. Colored lines show the \texttt{SEOBNRv5EHM}-based results for $q\in[1,100]$; black and blue lines show the large-$q$ results based on \texttt{BHPTNRSur1dq1e4} and \texttt{FastEMRIWaveforms} oscillatory modes.}
\label{fig:circular_correction_EOBq100}
\end{figure}

For the dominant memory mode $h_{2,0}$, we evaluate the finite-frequency correction factor $C(f)=2\pi f\,\tilde h_{2,0}(f)/(i\Delta h_{2,0})-1$, with $t=0$ set at the peak of $|h_{2,2}|$ and $f_{\rm ISCO}\equiv\sqrt{M/(6M)^3}/\pi$. For $q\in [1,8]$, \texttt{NRHybSur3dq8\_CCE} provides $h_{2,0}$ directly. We also use \texttt{SEOBNRv5EHM}~\cite{Gamboa2025SEOBNRv5EHM,Mihaylov:2023bkc}, \texttt{BHPTNRSur1dq1e4}~\cite{Islam:2022laz}, and \texttt{FastEMRIWaveforms}~\cite{Katz:2021yft,ChapmanBird:2025fewKerr} to reconstruct the null-memory $h_{2,0}$ mode from the oscillatory modes using Eq.~\eqref{null_memory_formula}. These models cover the comparable- and intermediate-mass-ratio, perturbative large-mass-ratio, and adiabatic extreme-mass-ratio regimes, respectively. For $q\le 8$, Fig.~\ref{fig:circular_correction_q_spin} compares the \texttt{NRHybSur3dq8\_CCE} results with those obtained from \texttt{SEOBNRv5EHM} oscillatory modes using\footnote{Such a calculation omits the ordinary part of the strain, which affects the merger-ringdown shape but is negligible in the late-time memory offset for such BBH mergers~\cite{Mitman_2024}.} Eq.~\eqref{null_memory_formula} and with an effective 0PN extrapolation; the results agree well at low frequencies. The nonspinning large-$q$ extension in Fig.~\ref{fig:circular_correction_EOBq100} is computed from \texttt{SEOBNRv5EHM}, \texttt{BHPTNRSur1dq1e4}, and \texttt{FastEMRIWaveforms} oscillatory modes, again using Eq.~\eqref{null_memory_formula} with an effective 0PN extrapolation. At low frequencies, the deviation from the soft limit increases with mass ratio at fixed $f/f_{\rm ISCO}$. In the extreme-mass-ratio, adiabatic limit, the correction factor depends primarily on the rescaled frequency $M_z f/\nu\approx qM_z f$. The corresponding memory amplitudes $\Delta h_{2,0}$ are summarized in Fig.~\ref{fig:delta_h20_scaled_vs_q}. Our nonspinning result is close to the NR-surrogate fit of Ref.~\cite{wzqk-62wc}. For \texttt{BHPTNRSur1dq1e4} waveforms whose initial frequencies are not sufficiently low, a more accurate effective PN extrapolation yields better agreement.
	
\subsection{Constraint on the source from the displacement-memory amplitude}
From Eq.~\eqref{null_memory_formula}, the offset of vacuum null displacement memory at future null infinity in an asymptotically flat spacetime can be expressed by $\Delta h(\Theta,\Phi)=\frac{2\pi}{R}\bar\eth^2 \mathfrak{D}^{-1}\frac{dE_\text{gw}}{d\Omega_2}$, where $\bar\eth$ is the spin-lowering operator, $\mathfrak{D}=D^2(D^2+2)/8$, and $D^2$ is the spherical Laplacian. Including the ordinary contribution, the result is~\cite{Mitman_2020} $\Delta h=\frac{1}{2}\bar\eth^2 \mathfrak{D}^{-1}\Delta Q$, with $\Delta Q=\left(\Delta m+4\pi\frac{dE_\text{gw}}{d\Omega_2}\right)\frac{1}{R}$, where $\Delta m$ is the change of Bondi mass aspect.\footnote{For example, for massive particles at timelike infinity, the Bondi mass aspect may be taken as $m(\Theta,\Phi)=\sum_I\frac{(1-|\mathbf{v}_I|^2)^{3/2}}{[1-\mathbf{v}_I\cdot\mathbf{n}(\Theta,\Phi)]^3}m_I$~\cite{PhysRevD.108.104052,He_2015}, with $|\mathbf{n}|=1$; it satisfies $\int\frac{d\Omega_2}{4\pi}m\left(\begin{smallmatrix}1\\ \mathbf{n}\end{smallmatrix}\right)=\sum_I {m_I\left(\begin{smallmatrix}1\\ \mathbf{v}_I\end{smallmatrix}\right)}/{\sqrt{1-|\mathbf{v}_I|^2}}$. Substituting $\Delta m$ in Eq.~\eqref{displacement_memory_spherical} gives Eqs.~\eqref{scattering_memory} and \eqref{scattering_memory_spherical}. In the massless limit, $m\to \sum_I 4\pi\omega_I \delta^2(\mathbf{n},\mathbf{n}_I)=4\pi(dE_\text{null}/d\Omega_2)$.} For $l\ge 2$, this gives
\begin{equation}\label{displacement_memory_spherical}
(\Delta h)_{l,m}=4\sqrt{\frac{(l-2)!}{(l+2)!}}(\Delta Q)_{l,m},
\end{equation}
where $(\Delta Q)_{l,m}$ and $(\Delta h)_{l,m}$ denote the coefficients of
$\Delta Q$ and $\Delta h$ in the $Y_{lm}$ and ${}_{-2}Y_{lm}$ bases,
respectively. A constraint on the source can be derived from $|\Delta h|$,
\begin{equation}
\begin{aligned}
\mathcal{Q} &\equiv \int d\Omega_2\,|\Delta Q|^2-\sum_{l=0,1}\sum_m|(\Delta Q)_{l,m}|^2
\\
&=\sum_{l\ge 2}\sum_m |(\Delta Q)_{l,m}|^2 \ge \frac{|\Delta h|^2}{\mathcal{I}(\Theta,\Phi)},
\end{aligned}
\end{equation}
with $\mathcal{I}=\sum_{l\ge 2}^\infty\sum_m 16\frac{(l-2)!}{(l+2)!}\left|_{-2}Y_{lm}\right|^2$. Since $\sum_m |_{-s}Y_{lm}|^2=(2l+1)/(4\pi)$, $\mathcal{I}=4/(3\pi)$, so $\mathcal{Q}\ge (3\pi/4)|\Delta h|^2$. Taking into account the cosmic expansion (see Sec.~\ref{Propagation of waveforms}), this bound becomes $\mathcal{Q}_{R\to d_L}\ge (3\pi/4)|\Delta h|^2/(1+z)^2$.
	
\section{Detectability Analysis}\label{sec:4}
	
\subsection{TDI response}\label{sec:TDI response}
\begin{figure*}[hbt]
\centering
\includegraphics[width=0.85\textwidth]{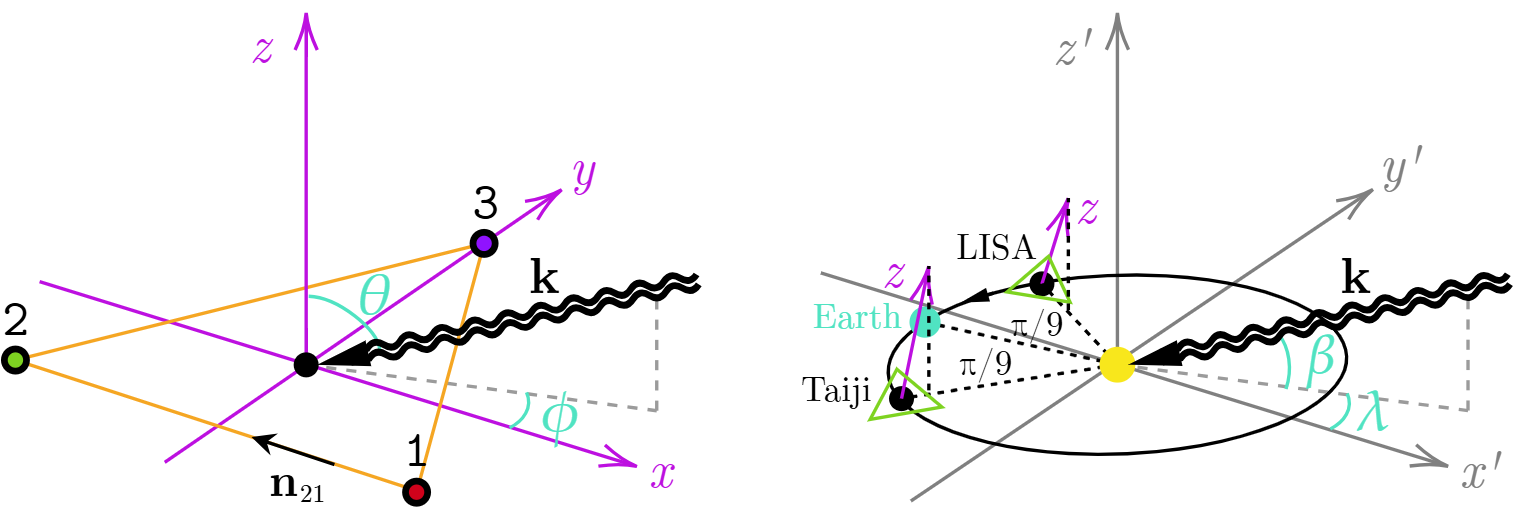}
\caption{\textbf{Left}: the detector configuration in the comoving and corotating ($xyz$) frame centered at the guiding center. SC are indexed 1, 2, 3 clockwise when looking down at their solar panels. SC 3 lies on the $y$-axis, and $\mathbf{n}_{21}$ points in the negative $x$-direction. For LISA, Taiji, and BBO Stage 1, the constellation rotates clockwise about the $z$-axis at an approximately constant angular velocity $2\pi/\text{yr}$ in the SSB frame, while the guiding center travels in a quasicircular heliocentric orbit on the ecliptic plane with radius $R\approx 1\,\text{AU}$. For TianQin, the SC follow geocentric orbits with a period of about 3.65 days~\cite{hu2024gravitationalwaveastrophysicstianqin}. \textbf{Right}: schematic of the considered guiding center orbits of LISA and Taiji in the SSB ($x'y'z'$) frame. The $z$-axis of the detector frame undergoes an approximate precession about the $z'$-axis with a constant angular velocity $2\pi/\text{yr}$ and a fixed inclination angle $\angle(\mathbf{e}_z,\mathbf{e}_{z'})=\pi/3$. The LISA detector trails the Earth by $\pi/9$, while the Taiji detector leads the Earth by $\pi/9$~\cite{Ruan_2020}.}\label{fig:detector}
\end{figure*}

In this section, we analyze soft memory signals at several planned space-based GW detectors. A LISA-like detector consists of three spacecraft (SC) forming an approximately equilateral triangle constellation, as depicted in Fig.~\ref{fig:detector}. Laser beams are exchanged among the SC to form multiple interferometers. Six laser beams connecting the local and distant test masses interfere with the local lasers, allowing measurement of the frequency modulation of light that carries information about passing GWs.
	
We denote the position of the $I$th SC by $\mathbf{x}_I(t)$. Consider light emitted from the $s$th SC at time $t_s$ and received by the $r$th SC at time $t$, i.e., $\mathbf{x}_r(t)=\mathbf{x}_s(t_s)+L_{rs}(t)\,\mathbf{n}_{rs}(t)$, with the unit vector $\mathbf{n}_{rs}(t)\equiv [\mathbf{x}_r(t)-\mathbf{x}_s(t_s)]/|\mathbf{x}_r(t)-\mathbf{x}_s(t_s)|$ along the light propagation and the effective armlength $L_{rs}(t)\equiv |\mathbf{x}_r(t)-\mathbf{x}_s(t_s)|$. Since the motion of SC during the light-travel time is negligible, we can make the approximation $\mathbf{n}_{rs}(t)\approx [\mathbf{x}_r(t)-\mathbf{x}_s(t)]/|\mathbf{x}_r(t)-\mathbf{x}_s(t)|$, $L_{rs}(t)\approx |\mathbf{x}_r(t)-\mathbf{x}_s(t)|$ and $t_s\approx t-L_{rs}(t)$.
	
The propagation of laser light between the SC can be described in the leading-order eikonal approximation, as summarized in Sec.~\ref{Propagation of waveforms} and Appendix~\ref{appendix_propagation_details}. The light frequency $\omega_s$ at $\mathbf{x}_s(t_s)$ is fixed by the laser source, while the light frequency $\omega_r$ measured locally at $\mathbf{x}_r(t)$ is affected by both the motion of $\mathbf{x}_{r,s}$ [see Eq.~\eqref{coupling_signal} in Appendix~\ref{appendix_B}] and the light propagation. The GW contribution to the one-way frequency shift $y_{rs}(t)\equiv [\omega_r(t)-\omega_s]/\omega_s$ can be conveniently analyzed in the TT coordinates, with the relevant metric given by Eq.~\eqref{GW_spacetime}. For light propagating between two static test masses, the relative frequency modulation comes solely from the GW-induced modification of the photon dispersion relation, which for a plane wave $h_{ij}(t,\mathbf{x})=h_{ij}(u=t-\hat{\mathbf{k}}\cdot\mathbf{x})$ is given by Eq.~\eqref{frequency_shift_in_GW}:
\begin{equation}\label{one-way}
y_{rs}(t)=\frac{1}{2}n_{rs}^in_{rs}^j\frac{h_{ij}(t-L_{rs}-\hat{\mathbf{k}}\cdot\mathbf{x}_s)-h_{ij}(t-\hat{\mathbf{k}}\cdot\mathbf{x}_r)}{1-\hat{\mathbf{k}}\cdot \mathbf{n}_{rs}},
\end{equation}
with $\hat{\mathbf{k}}$ denoting the unit vector in the direction of GW propagation.
	
In reality, the laser frequency is subject to random fluctuations. Such laser frequency noise, together with test-mass acceleration noise and optical metrology noise, constitutes the main instrumental noise limiting the detector's sensitivity to GWs. For LISA-like detectors, laser frequency noise would dominate the measurements and must be mitigated through suitable time-delayed linear combinations of the six one-way data streams $y_{rs}(t)$. This postprocessing procedure, called time-delay interferometry (TDI)~\cite{Tinto:2004wu}, synthesizes a virtual equal-arm interferometer and thus suppresses the laser frequency noise. For example, the Michelson $X$ combination in the first-generation TDI is
\begin{align}\label{X_1(t)}
X_1(t)=\left(y_{13}+D_{13}y_{31}+D_{131}y_{12}+D_{1312}y_{21}\right)-(3\leftrightarrow 2)
,
\end{align}
where the time-delay operators are defined by $D_{ij}x(t) \equiv x(t-L_{ij})$ and $	D_{i_1\cdots i_n}x(t) \equiv \prod_{k=1}^{n-1} D_{i_k i_{k+1}}x(t)$. For computing GW signals, it suffices to make the approximation $D_{i_1\cdots i_n}x(t)\approx x\left(t-\sum_{k=1}^{n-1}L_{i_k i_{k+1}}\right)$.
	
The TDI response is further simplified in the static equal-arm approximation (SEA), in which we neglect the motion of SC and use $L_{rs}\approx L=\text{const}$. The static-detector approximation is reasonable for a short-term observation, and the equal-arm approximation can also be sufficiently accurate in such cases, since the armlengths vary only at the percent level. Given the time-domain waveform $h_{ij} (t)=h_{ij}(t,\mathbf{0})$, the frequency spectrum of the time-domain TDI signals (with a sufficiently high sampling frequency) can be derived in this approximation by using Eq.~\eqref{one-way}. The result for \eqref{X_1(t)} is
\begin{align}\label{X_1}
\tilde X_1 &=
\frac{1}{2}\left[
n_{13}^in_{13}^j
\left( F_{13} + \text{e}^{2\pi i f L} F_{31} \right)
-(3\leftrightarrow 2)
\right]\tilde h_{ij},
\end{align}
with $\tilde h_{ij}(f)=\frac{1}{2}(e_{ij}^++ie_{ij}^\times)\tilde h(f)+\frac{1}{2}(e_{ij}^+-ie_{ij}^\times)[\tilde h(-f)]^*$ ($[\tilde h(-f)]^*$ being the Fourier transform of $h^*(t)$), and
\begin{equation}
F_{rs}\equiv \left(1-\text{e}^{4\pi i fL}\right)\frac{
\text{e}^{2\pi i f L(1-\hat{\mathbf{k}}\cdot \mathbf{n}_{rs})}-1}
{1-\hat{\mathbf{k}}\cdot \mathbf{n}_{rs}}
\text{e}^{2\pi i f\,\hat{\mathbf{k}}\cdot\mathbf{x}_r}.
\label{F_rs}
\end{equation}
Note that in the low-frequency limit ($fL\ll 1$),
\begin{align}\label{low-freq_limit}
\tilde X_1\approx 8(\pi f L)^2 \left(
n_{13}^in_{13}^j-n_{21}^in_{21}^j
\right)\tilde h_{ij}\,\text{e}^{2\pi i f\,\hat{\mathbf{k}}\cdot\mathbf{X} },
\end{align}
where $\mathbf{X}$ is the guiding center of the constellation.
	
The other two Michelson combinations $Y$ and $Z$ can be obtained from $X$ through cyclic permutations $(1,2,3) \to (2,3,1)$ and $(1,2,3) \to (3,1,2)$, respectively. For the signal analysis, we consider the optimal combinations $\{A,E,T\}$:
\begin{align}\label{AET}
\left(\begin{matrix}
A\\E\\T
\end{matrix}\right)
=
\left(\begin{matrix}
-\frac{1}{\sqrt{2}} & 0 &\frac{1}{\sqrt{2}}
\\
\frac{1}{\sqrt{6}} & -\frac{2}{\sqrt{6}} & \frac{1}{\sqrt{6}}
\\
\frac{1}{\sqrt{3}} & \frac{1}{\sqrt{3}} & \frac{1}{\sqrt{3}}
\end{matrix}\right)
\left(\begin{matrix}
X\\Y\\Z
\end{matrix}\right)
,
\end{align}
for which the noise correlation matrix is diagonalized in the idealized case~\cite{Prince_2002}. In the SEA, $\tilde T(f)$ is strongly suppressed for $fL\ll 1$ relative to Eq.~\eqref{low-freq_limit}, and the use of first-generation TDI is equivalent to the use of second-generation TDI. Since the SEA is accurate for the soft memory signals considered here, we consider only the first-generation TDI observables in this paper.
	
For a single LISA-like detector in the SEA, it is convenient to parametrize the triad $\{\mathbf{a,b,\hat{k}}\}$ in the detector frame defined in the left panel of Fig.~\ref{fig:detector}, where we choose $\mathbf{a}=-\mathbf{e}_\phi$, $\mathbf{b}=-\mathbf{e}_\theta$ and $\hat{\mathbf{k}}=-\mathbf{e}_r$. The positions of the SC ($I\in\{1,2,3\}$) in this detector frame are given by
\begin{equation}
\left(\begin{matrix}
x_I \\ y_I \\ z_I
\end{matrix}\right)=\frac{L}{\sqrt{3}}\left(\begin{matrix}
\cos 2\sigma_I \\ \sin 2\sigma_I \\ 0
\end{matrix}\right),
\quad
\sigma_I=\frac{\pi}{4}+I\frac{2\pi}{3},
\end{equation}
with the guiding center $\mathbf{X}=0$.
	
For long-term observations with a single detector or joint observations with multiple detectors, we parametrize $\{\mathbf{a,b,\hat{k}}\}$ in the solar-system-barycentric (SSB) ecliptic coordinate frame, as depicted in the right panel of Fig.~\ref{fig:detector}. In the equal-arm approximation, the positions of SC in the SSB frame can be obtained as $x_{Ii}'= X'_i(\eta)+R_{ij}(\eta)\,x_{Ij}$, with the guiding center $\mathbf{X}'=R(\cos\eta,\sin\eta,0)$. For the description of GW polarizations, we adopt the standard choice~\cite{Katz_2022}: $\mathbf{a}=-\mathbf{e}_{\phi'}$, $\mathbf{b}=-\mathbf{e}_{\theta'}$, and $\hat{\mathbf{k}}=-\mathbf{e}_{r'}$, with $(\theta',\phi')=(\pi/2-\beta,\lambda)$, where $\beta$ and $\lambda$ are the ecliptic latitude and longitude of the source direction. The explicit rotation matrix $R_{ij}(\eta)$ and the relation between the detector-frame angles and the SSB-frame angles are given in Appendix~\ref{appendix_tdi_coordinates}.
	
\subsection{Soft memory signals}\label{sec:Soft memory signals}
We now turn to the TDI response of a single LISA-like detector to soft memory events. We consider a signal over the interval $t\in [0,T]$. From Eqs.~\eqref{displacement_memory_waveform}--\eqref{integrated-displacement_memory_waveform}, the general parametrization of the frequency-domain waveform $\tilde h(f)=\tilde h_+-i\tilde h_\times$ is
\begin{equation}
\tilde h=\frac{iH\text{e}^{-2i\psi}}{2\pi f}\text{e}^{2\pi i ft_*}
\end{equation}
for the displacement memory,
\begin{equation}
\tilde h=-\frac{\dot H \text{e}^{-2i\psi}}{(2\pi f)^2}\text{e}^{2\pi i ft_*}
\end{equation}
for the velocity memory, and
\begin{equation}
\tilde h=O\,\text{e}^{-2i\psi}\text{e}^{2\pi i ft_*}
\end{equation}
for the integrated-displacement memory, where $H,\dot H,O>0$ are the memory amplitudes, $\psi\in[0,\pi)$ is the polarization angle, and $t_*\in(0,T)$ is the burst arrival time. Note that $\psi$ depends on the choice of polarization tensors \eqref{TT_polarization}. In the following, we denote the polarization angles in the detector and SSB frames by $\psi$ and $\Psi$, respectively. Additionally, the sky location of the source is given by $\{\theta,\phi\}$ in the detector frame, and $\{\beta,\lambda\}$ in the SSB frame. The former will be used in the following discussion of single-detector response.
	
\begin{figure}[hbt]
\centering
\includegraphics[width=0.48\textwidth]{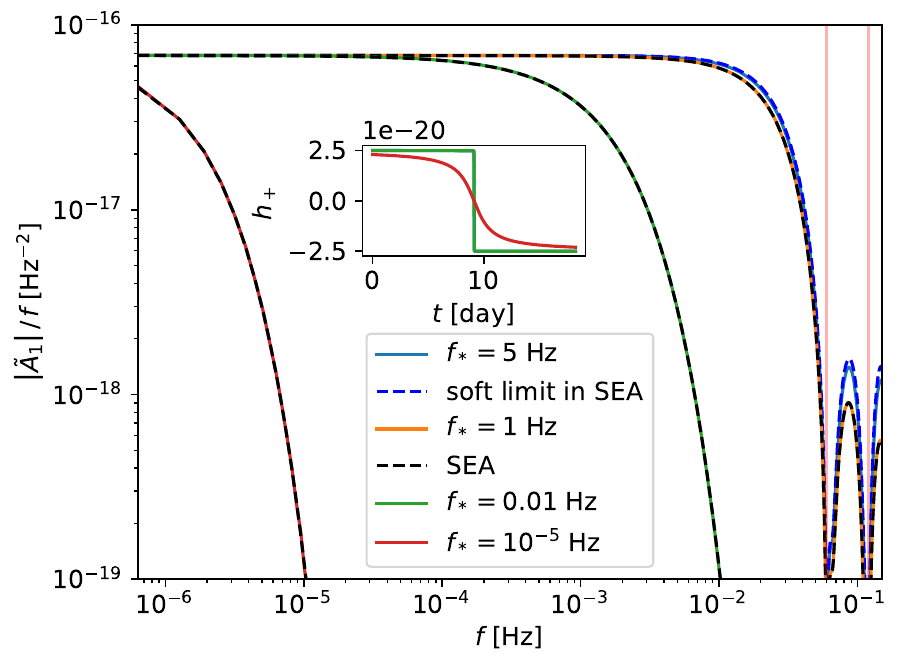}
\caption{Comparison of the analytic and numerical spectra of $A_1(t)$ for the arctan model \eqref{arctan_model} at LISA, with parameters $H=10^{-19}$, $T=0.05\,\text{yr}$, $t_*=0.5T$, $\{\beta,\lambda,\Psi\}=\{-1.4,2.3,\pi/3\}$, and $f_*\in\{10^{-5},0.01,1,5\}\,\text{Hz}$. The analytic spectrum (black dashed line) is obtained in the SEA, with the soft limit indicated by the blue dashed line. The numerical spectrum (colored solid line) is computed using \texttt{fastlisaresponse} with the equal-armlength orbit and a sampling frequency of $1\,\text{Hz}$. The vertical dashed lines correspond to integer multiples of $f=1/(2L)$. The time-domain waveforms are shown in the inset.}\label{fig:slow_jump}
\end{figure}
	
In the SEA, the frequency-domain signals $\{\tilde A_1,\tilde E_1,\tilde T_1\}$ can be calculated using Eqs.~\eqref{X_1} and \eqref{AET}. The signals are degenerate under the transformation from $\{\theta,\psi\}$ to $\{\theta',\psi'\}=\{\pi-\theta,\pi-\psi\}$, and change only by an overall sign under the transformation from $\psi$ to $\psi'=\psi\pm\pi/2$. In the low-frequency approximation \eqref{low-freq_limit}, there is an additional degeneracy under $\phi\leftrightarrow \phi+\pi$. Moreover, the signals in the SEA at $f\to 0$ depend only on four parameters, while a soft memory signal has five parameters in total. The full measurement of a soft memory event with a single detector is therefore possible precisely because the response function deviates from its low-frequency approximation. When this deviation is small, the signal suffers from a strong degeneracy.
	
\begin{figure}[hbt]
\centering
\includegraphics[width=0.48\textwidth]{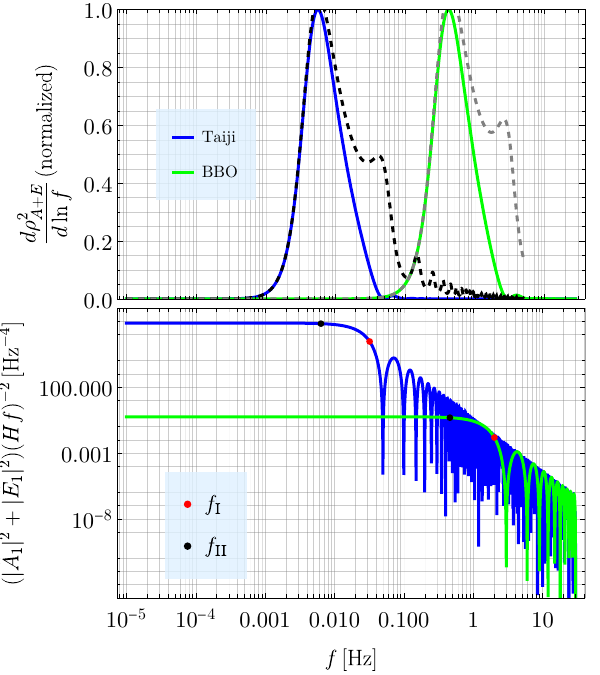}
\caption{Normalized logarithmic spectral density of $\rho_{A+E}^2$ (top panel) and the corresponding signal spectrum (bottom panel) at Taiji and BBO for a soft displacement-memory signal with $\{\phi,\theta,\psi\}=\{0,0,0\}$.}\label{fig:displacement_memory_SNR}
\end{figure}
	
\begin{figure}[hbt]
\centering
\includegraphics[width=0.48\textwidth]{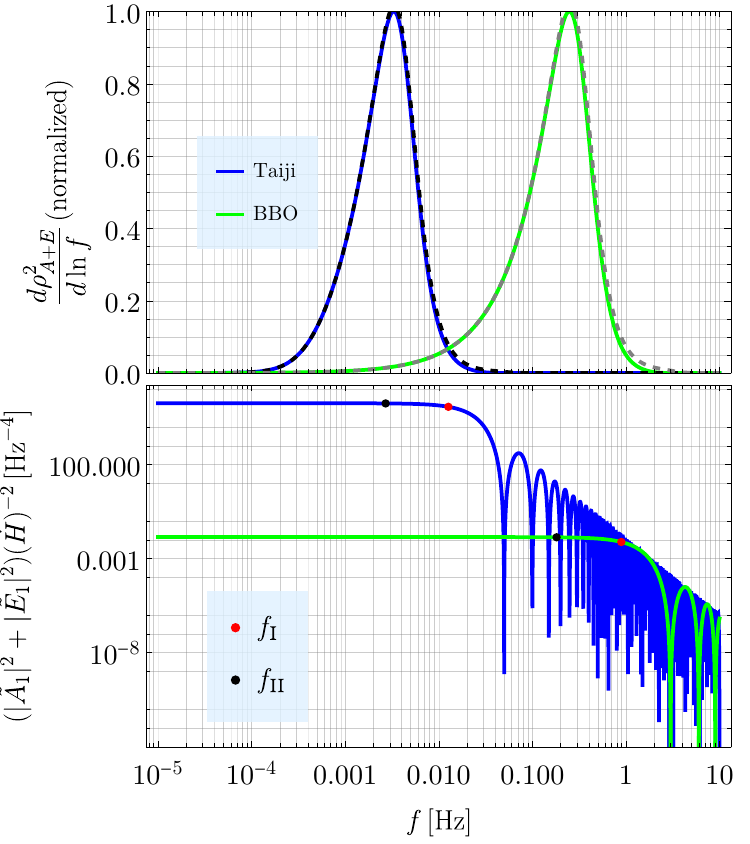}
\caption{Normalized logarithmic spectral density of $\rho_{A+E}^2$ (top panel) and the corresponding signal spectrum (bottom panel) at Taiji and BBO for a soft velocity-memory signal with $\{\phi,\theta,\psi\}=\{0,0,0\}$.}\label{fig:velocity memory_SNR}
\end{figure}
	
\begin{figure}[hbt]
\centering
\includegraphics[width=0.45\textwidth]{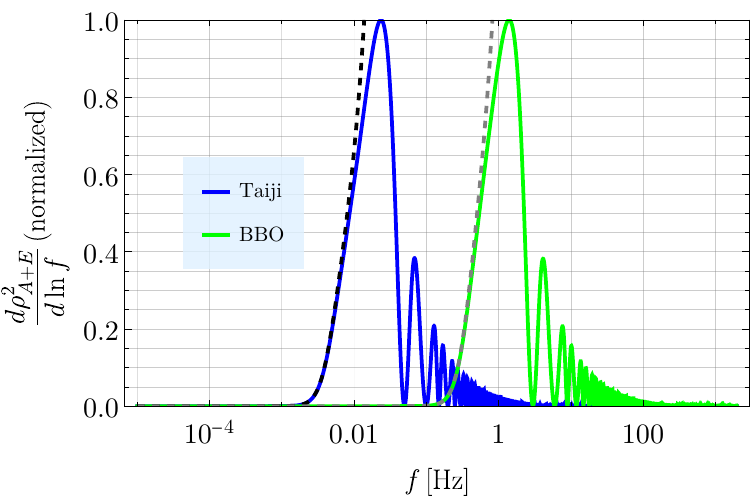}
\caption{Normalized logarithmic spectral density of $\rho_{A+E}^2$ at Taiji and BBO for a soft integrated-displacement-memory signal with $\{\phi,\theta,\psi\}=\{0,0,0\}$.}\label{fig:integrated-displacement memory_SNR}
\end{figure}
	
For the LISA detector, we verified that the analytic approximation matches the time-domain computation of \texttt{fastlisaresponse}~\cite{Katz_2022}, up to an overall sign. The static-detector approximation in fact remains good when the deviation from the soft limit is large. This is shown in Fig.~\ref{fig:slow_jump} for the arctan model \eqref{arctan_model} with different transition timescales $1/f_*$. For $f_*=1\,\text{Hz}$, the spectrum is already close to the soft waveform at $f<0.01\,\text{Hz}$. For $f_*=10^{-5}\,\text{Hz}$, the soft limit does not provide a good approximation at $f>10^{-6}\,\text{Hz}$, but the SEA remains accurate.
	
\subsection{Optimal signal-to-noise ratio}
	
We proceed to estimate the detectability of soft memory events through matched filtering. Since the $\{A,E,T\}$ channels are orthogonal in the SEA, the noise-weighted inner product between two time-domain signals $U_a(t)$ and $U_b(t)$ is
\begin{equation}
\langle a|b\rangle\equiv\sum_{U=A,E,T}\int_{f_\text{min}}^{f_\text{max}} df\frac{4\,\text{Re}\left[\tilde U_a(f)\,\tilde U_b^*(f)\right]}{S_U(f)},
\end{equation}
where $S_U(f)$ denotes the one-sided power spectral density (PSD) of the noise in the $U$th channel, which is assumed to be wide-sense stationary. For a signal with duration $T$, we take $f_\text{min}=1/T$. In practice, the continuous integral is replaced by a discrete sum with frequency step $\Delta f=1/T$, but their difference is small if $\Delta f$ is sufficiently small; in this section, we use the continuous integral. The detector noise model used in the SNR estimates is given in Appendix~\ref{appendix_tdi_coordinates}.
	
For a given signal $U_a(t)$, the optimal signal-to-noise ratio $\rho$ for matched filtering is given by
\begin{equation}\label{SNR}
\rho^2=\langle a|a\rangle=
\sum_U\int d\ln f\left[\frac{4f |\tilde U_a|^2}{S_U(f)}\right].
\end{equation}
In this paper, we do not include the T channel, which is much weaker than the A and E channels and is significantly affected by unequal armlengths at low frequencies; see Appendix~\ref{appendix_A} for further discussion. The resulting SNR is denoted by $\rho_{A+E}$, which is nearly independent of $\phi$. Note that $S_{A,E}=0$ at $f=1/(2L)$, but no divergence occurs because the factor $\left|1-\text{e}^{4\pi i fL}\right|^2=4\sin^2(2\pi fL)$ is canceled by the corresponding factor in the signal in Eq.~\eqref{F_rs}.
	
The logarithmic spectral density distributions of the squared SNR for the soft displacement-memory and velocity-memory signals with $\phi=\theta=\psi=0$ are shown in Figs.~\ref{fig:displacement_memory_SNR} and \ref{fig:velocity memory_SNR}, respectively (the squared SNR is proportional to the area under each curve). The spectral density is normalized such that the peak value equals unity. We also show results obtained using the low-frequency approximation of the response function as dashed lines; these correspond to the flat signal spectrum at the low-frequency end in the bottom panel. A significant portion of the spectrum cannot be described using the low-frequency approximation for displacement memory. By contrast, the low-frequency approximation is much better for Taiji for velocity memory, indicating a strong degeneracy of the signal.
	
To quantify the spectral range where the SNR receives most of its contribution, we introduce the frequencies $f_\text{I}$, $f_\text{II}$, and $f_0$ as follows. For $\theta=\psi=\phi=0$, $\rho^2_{A+E}(f_\text{min}=10^{-5}\,\text{Hz}, f_\text{max}=2f_\text{I})/\rho^2_{A+E}(f_\text{min}=10^{-5}\,\text{Hz}, f_\text{max}=f_\text{I})-1\approx 10^{-2}$, $\rho^2_{A+E}(f_\text{min}=10^{-5}\,\text{Hz}, f_\text{max}=f_\text{II})\approx \frac{1}{2}\rho^2_{A+E}(f_\text{min}=10^{-5}\,\text{Hz}, f_\text{max}=f_\text{I})$, and $\rho^2_{A+E}(f_\text{min}=f_0, f_\text{max}=f_\text{I})/\rho^2_{A+E}(f_\text{min}=2f_0, f_\text{max}=f_\text{I})-1\approx 10^{-2}$. Thus, the maximum SNR for $\phi=\theta=\psi=0$ is largely determined by the spectrum within $[f_\text{0}, f_\text{I}]$, and half of the maximum squared SNR comes from the frequency range $[f_\text{0}, f_\text{II}]$. According to this continuous-integral estimate, extending the signal duration beyond $1/f_0$ only adds a small low-frequency contribution to the SNR. We list the values of $f_\text{0,I,II}$ for LISA, Taiji, TianQin, and BBO in Table~\ref{detector_table_2}. The low-frequency part of the spectrum is more important for velocity memory. The situation is reversed for integrated-displacement memory, as shown in Fig.~\ref{fig:integrated-displacement memory_SNR}. In the following, we set $f_\text{min}=f_0$ and $f_\text{max}=f_\text{I}$, unless stated otherwise.
	
\begin{figure}[hbt]
\centering
\includegraphics[width=0.5\textwidth]{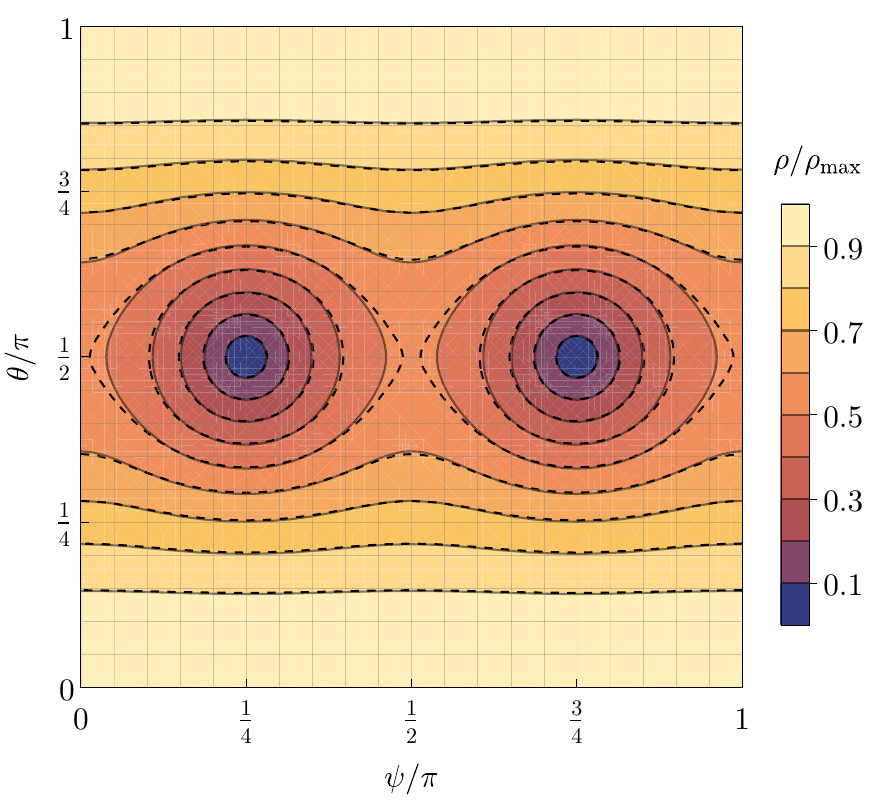}
\caption{Angular distribution of $\rho_{A+E}(\theta,\psi)/\rho_{A+E}(0,0)$ at Taiji for $\phi=0$. The solid (dashed) contours show displacement (velocity) memory at fixed $H$ ($\dot H$).}\label{fig:SNR}
\end{figure}

The angular dependence of the SNR exhibits the following symmetry: $\rho_{A+E}(\phi,\theta,\psi)=\rho_{A+E}(-\phi,\theta,\pi-\psi)=\rho_{A+E}(\phi,\pi-\theta,\pi-\psi)=\rho_{A+E}(\phi,\theta,\psi+\pi/2)=\rho_{A+E}(\phi+\pi,\theta,\psi)$. To a good approximation, we also have $\rho_{A+E}(\phi,\theta,\psi)\approx \rho_{A+E}(\theta,\psi)\approx \rho_{A+E}(\pi-\theta,\psi)$. Thus, the angular dependence of the SNR is almost fixed by $\rho^2_{A+E}(\theta,\psi)$ within $\psi\in [0,\pi/2]$ and $\theta\in[0,\pi/2]$. The result is roughly the same for all considered detectors and is shown for Taiji in Fig.~\ref{fig:SNR}. The results for displacement and velocity memory are also found to be similar. The SNR is maximized at $\theta\in\{0,\pi\}$, where it is independent of $\psi$. The SNR vanishes at $\theta=\pi/2$, $\psi\in\{\pi/4,3\pi/4\}$. For a given $\theta$, the SNR is maximized at $\psi\in\{0,\pi/2,\pi\}$ and minimized at $\psi\in\{\pi/4,3\pi/4\}$. For a given $\psi$, the SNR is maximized at $\theta\in\{0,\pi\}$ and minimized at $\theta=\pi/2$. In Table~\ref{detector_table_2}, we list the minimum memory amplitudes $H_{10}^\text{min}$ and $\dot H_{10}^\text{min}$ for $\rho_{A+E}=10$ at the considered detectors, which correspond to the events with $\theta=\psi=0$.
	
It is also useful to examine the SNR averaged over $\{\phi,\theta,\psi\}$, which provides a measure of the ``mean'' signal strength. Generically, a linearly polarized GW can be written as $h_{ij}(t)=\left(c_{2\psi}\,e_{ij}^+ + s_{2\psi}\,e_{ij}^\times\right)\mathcal{H}(t)$, and the signal in the static-detector approximation takes the form: $\tilde U(f) =\mathcal{F}^U_{ij}(f)\,\tilde h_{ij}(f)= \tilde U(\psi=0)\,c_{2\psi} + \tilde U(\psi=\pi/4)\,s_{2\psi}$. The averaged version of Eq.~\eqref{SNR} is then given by
\begin{equation}
\begin{aligned}
\left\langle\rho^2\right\rangle &\equiv\int \frac{\sin\theta\,d\theta\,d \phi\, d\psi}{4\pi^2}\langle a|a\rangle
\\
&= \int \frac{\sin\theta\,d\theta\,d \phi}{8\pi}
\left(
\langle a|a\rangle_{\psi=0}+\langle a|a\rangle_{\psi=\pi/4}
\right)
\\
&=
\sum_U\sum_{\lambda=+,\times}\int d f\,4\mathcal{F}^U_\lambda\,\left|\tilde{ \mathcal{H}}(f)\right|^2/S_U(f),
\end{aligned}
\end{equation}
where
\begin{equation}\label{F_function}
\mathcal{F}^U_\lambda(f)\equiv \int\frac{d\Omega_2}{8\pi}\,\left|\mathcal{F}^U_{ij}e_{ij}^\lambda\right|^2.
\end{equation}
Note that in the SEA, the low-frequency limit is $\lim_{f\to 0}\sum_\lambda \mathcal{F}_\lambda^{A,E}\to \frac{288}{5}(fL\pi)^4$~\cite{Hartwig_2023}. In Table~\ref{detector_table_2}, we list the averaged memory amplitudes $\langle H\rangle_{10}$ and $\langle\dot H\rangle_{10}$ for $\langle\rho_{A+E}^2\rangle^{1/2}=10$ at the considered detectors. We find that the values of $H_{10}$ and $\dot H_{10}$ are of the same order of magnitude for LISA, Taiji, and TianQin, whereas BBO can achieve significantly higher sensitivity to the soft displacement-memory signal, provided it occurs at $f\lesssim 1\,\text{Hz}$.
	
\begin{table}[htb!]
\renewcommand{\arraystretch}{1}
\setlength{\tabcolsep}{5.5pt}
\begin{center}
\begin{tabular}{ccccc}
\hline
\hline
& LISA & 
Taiji & TianQin & BBO \\
\hline
$H^\text{min}_{10}/10^{-20}$ & 1.3 & 0.69 & 2.0 & $6.3\times 10^{-4}$ \\
$\langle H \rangle_{10}/10^{-20}$ & 2.0 & 1.1 & 3.3 & $1.0\times 10^{-3}$\\
$f_0$ [mHz] & 0.56 & 0.71 & 1.1 & 51 \\
$f_0^{-1}/10^{-5}$ [yr] & 5.7 & 4.5 & 3.0 & 6.3$\times 10^{-2}$ \\
$f_\text{II}$ [mHz] & 5.1 & 6.4 & 11 & 462 \\
$f_\text{I}$ [Hz] & 0.035 & 0.032 & 0.25 & 2.0 \\
\noalign{\vskip 0.3em}
\hline
\noalign{\vskip 0.3em}
$\dot H^\text{min}_{10}/10^{-22}$\,[Hz] & 2.7 & 1.8 & 8.5 & 0.12 \\
$\langle\dot H\rangle_{10}/10^{-22}$\,[Hz] & 4.2 & 2.9 & 13 & 0.18\\
$f_0$ [mHz] & 0.13 & 0.16 & 0.19 & 3.5 \\
$f_0^{-1}/10^{-4}$ [yr] & 2.4 & 2.0 & 1.7 & 9.0$\times 10^{-2}$ \\
$f_\text{II}$ [mHz] & 2.1 & 2.7 & 3.9 & 182 \\
$f_\text{I}$ [Hz] & 0.010 & 0.013 & 0.019 & 0.89 \\
\hline
\hline
\end{tabular}
\end{center}
\caption{$f_\text{0,I,II}$ and the minimum and averaged memory amplitudes for $\rho_{A+E}=10$ at the considered detectors, for displacement (top rows) and velocity memory (bottom rows). The minimum memory amplitude threshold and $f_\text{0,I,II}$ are obtained at $\phi=\theta=\psi=0$, whereas the averaged memory amplitude threshold is obtained using the averaged response function.}
\label{detector_table_2}
\end{table}
	
\section{Parameter-estimation simulations}\label{sec:5}
Once a soft memory signal is identified by matched filtering, Bayesian inference can constrain its parameters. Given the data $d$ and a model $\Lambda$ with parameters $\boldsymbol{\xi}$, the posterior probability distribution, according to Bayes' rule, is $P(\boldsymbol{\xi}|d, \Lambda)
= {p(\boldsymbol{\xi}|\Lambda)\,\mathcal{L}}/{\mathcal{Z}}$, where $\mathcal{L} = p(d|\boldsymbol{\xi}, \Lambda)$ is the likelihood, 
$\mathcal{Z}=p(d|\Lambda)=\int d\boldsymbol{\xi}\,p(\boldsymbol{\xi}|\Lambda)\,\mathcal{L}$ is the evidence, and 
$p(\boldsymbol{\xi}|\Lambda)$ is the prior. The likelihood is taken to be $\mathcal{L} \propto \exp \left(-\frac{1}{2}\langle d-h \mid d-h\rangle\right)$. In the present case, the data stream containing the signal $s(t)$ in a TDI channel is given by $d(t)=s(t)+n(t)$, where $n(t)$ is the noise. Ideally, the noise is zero-mean, wide-sense stationary and described by the one-sided PSD in Eq.~\eqref{noise_PSD}.
	
\begin{figure}[hbt]
\centering
\includegraphics[width=0.48\textwidth]{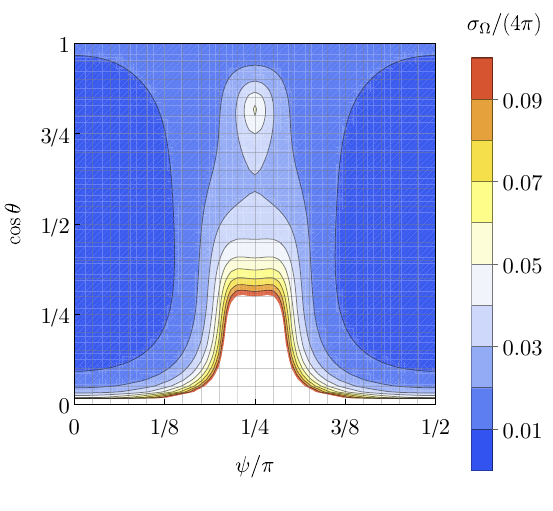}
\\
\includegraphics[width=0.48\textwidth]{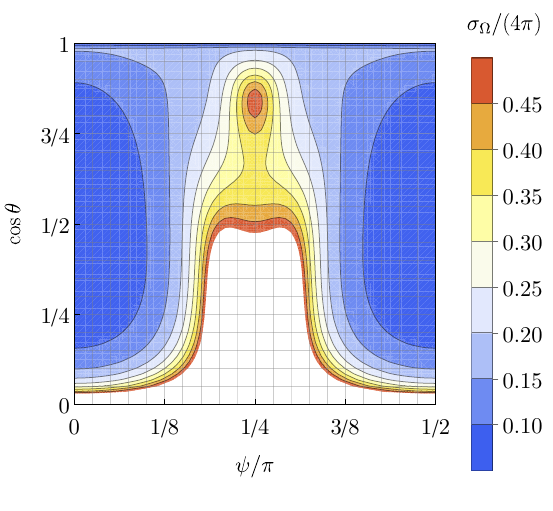}
\caption{FIM estimate of $\sigma_\Omega/(4\pi)$ for a soft memory signal at Taiji, for displacement (top panel) and velocity (bottom panel) memory. The memory amplitudes are set to $H=10^{-19}$ and $\dot H=2.6\times 10^{-21}\,\text{Hz}$, so that $\rho_{A+E}$ at $\theta=\psi=0$ is identical. The other parameters are $\phi=0$, $\tau_*=0.3$, and $T=10^{-3}\,\text{yr}$.}\label{fig:Fisher}
\end{figure}

In the limit of high SNR, the measurement precision can be estimated using the Fisher information matrix (FIM). For a model $h_{\boldsymbol{\xi}}(t)$ with parameters $\boldsymbol{\xi}=\{\xi_i\}_i$, the FIM evaluated at the true parameter values $\boldsymbol{\xi}=\boldsymbol{\hat \xi}$ is $\Gamma_{ij}\equiv\left\langle {\partial h}/{\partial \xi_i}\mid {\partial h}/{\partial \xi_j}\right\rangle_{\boldsymbol{\xi}=\boldsymbol{\hat \xi} }$, and the inferred covariance matrix of $\boldsymbol{\xi}$ is estimated to be $\Sigma_{ij}=(\Gamma^{-1})_{ij}$, with the root-mean-squared error of $\xi_i$ given by $\sigma_i=\sqrt{\Sigma_{ii}}$. Since $t_*\in (0,T)$, we use $\tau_*=t_*/T\in (0,1)$ as a parameter. We then consider a five-parameter model with $\boldsymbol{\xi}=\{\log_{10}H,\tau_*,\phi,\cos\theta,\psi\}$. The angular resolution of the source sky location can be estimated by the solid-angle measure,
\[
\sigma_\Omega =2\pi \sqrt{\Sigma_{c_\theta c_\theta}\Sigma_{\phi\phi} - \Sigma_{c_\theta\phi}^2}
=2\pi \sqrt{\Sigma_{s_\beta s_\beta}\Sigma_{\lambda\lambda} - \Sigma_{s_\beta\lambda}^2}
~,
\]
which is invariant under coordinate rotation. The result for $\phi=0$ is shown in Fig.~\ref{fig:Fisher}. This result is $\phi$-dependent, and the estimated errors are proportional to $1/H$, while $\sigma_{\tau_*}$ is also proportional to $1/T$. Because of the signal symmetry, we show only $\psi\in [0,\pi/2]$ and $\theta\in[0,\pi/2]$. Notably, the FIM estimate of the measurement error increases drastically as $\theta\to \pi/2$. The reason is that when $\theta=\pi/2$, only $h_+$ contributes to the signal, so $H$ and $\psi$ are degenerate; additionally, $\phi$ and $\psi$ are degenerate at $\theta\in\{0,\pi\}$. For a given $\theta$, the estimated error is minimized at $\psi\in\{0,\pi/2\}$. We also observe that the FIM estimate of the measurement error is larger for velocity memory at roughly the same SNR, consistent with the stronger degeneracy discussed in Sec.~\ref{sec:Soft memory signals}.
	
The FIM provides a local high-SNR estimate whose accuracy can be limited by the parameter degeneracies. In the following, we present results from Bayesian parameter-estimation simulations, focusing on displacement memory. For concreteness, we consider measurements with Taiji, the LISA-Taiji network (described in Fig.~\ref{fig:detector}), and BBO. For LISA-Taiji, we set the guiding-center phases to $\eta_\text{Taiji}=\eta_\text{LISA}+2\pi/9$ and $\eta_\text{LISA}=\pi/4$, and the constellation rotation phase to $\Phi_0=0$; the log-likelihood is taken as the sum of those of LISA and Taiji. As mentioned in Sec.~\ref{sec:TDI response}, the signals at a single detector are degenerate under the transformation $\{\theta,\psi\}\leftrightarrow\{\pi-\theta,\pi-\psi\}$; this degeneracy can be removed by restricting $\psi\in [0,\pi/2]$; it cannot be easily removed in the SSB-frame parametrization. We therefore adopt the parameter set $\boldsymbol{\xi}=\{\log_{10}H,\tau_*,\lambda,\sin\beta,\Psi\}$ for LISA-Taiji, and $\boldsymbol{\xi}=\{\log_{10}H,\tau_*,\phi,\cos\theta,\psi\}$ for a single detector. The priors are taken to be uniform, as listed in Table~\ref{prior_table}. The amplitude bounds are chosen relative to the injected amplitudes and span 4 orders of magnitude. For the noisy injections, the frequency-domain noise realizations are generated with \texttt{PyCBC}~\cite{Usman:2015kfa} on the discrete grid $\Delta f=1/T$. For paired comparisons, each realistic-waveform injection is accompanied by a soft-waveform injection with the same noise realization.
	
\begin{table}[htb!]
\renewcommand{\arraystretch}{1}
\setlength{\tabcolsep}{10pt}
\begin{center}
\begin{tabular}{ccc}
\hline
\hline
Parameter & 
Lower Bound & Upper Bound \\
\hline
$\log_{10}H$ & $\log_{10}\hat H-2$ & $\log_{10}\hat H+2$ \\
$\log_{10}(\dot H/\text{Hz})$ & $\log_{10}(\hat{\dot H}/\text{Hz})-2$ & $\log_{10}(\hat{\dot H}/\text{Hz})+2$ \\
$\tau_*$ & 0 & 1\\
\noalign{\vskip 0.3em}
\hline
\noalign{\vskip 0.3em}
$\psi$ & 0 & $\pi/2$\\
$\phi$ & 0 & $2\pi$
\\
$\cos\theta$ & $-1$ & 1
\\
\noalign{\vskip 0.3em}
\hline
\noalign{\vskip 0.3em}
$\Psi$ & 0 & $\pi$
\\
$\lambda$ & 0 & $2\pi$
\\
$\sin\beta$ & $-1$ & 1
\\
\hline
\hline
\end{tabular}
\end{center}
\caption{Lower and upper bounds of the uniform priors. The parameters $\{\psi,\phi,\cos\theta\}$ are defined in the detector frame for a single detector, while the parameters $\{\Psi,\lambda,\sin\beta\}$ are defined in the SSB frame (see Sec.~\ref{sec:TDI response}). $\psi$ and $\Psi$ denote the polarization angles in the detector and SSB frames, respectively. $\hat H$ and $\hat{\dot H}$ denote the injected true amplitudes.}
\label{prior_table}
\end{table}
	
\begin{figure*}[hbt]
\centering
\includegraphics[width=1\textwidth]{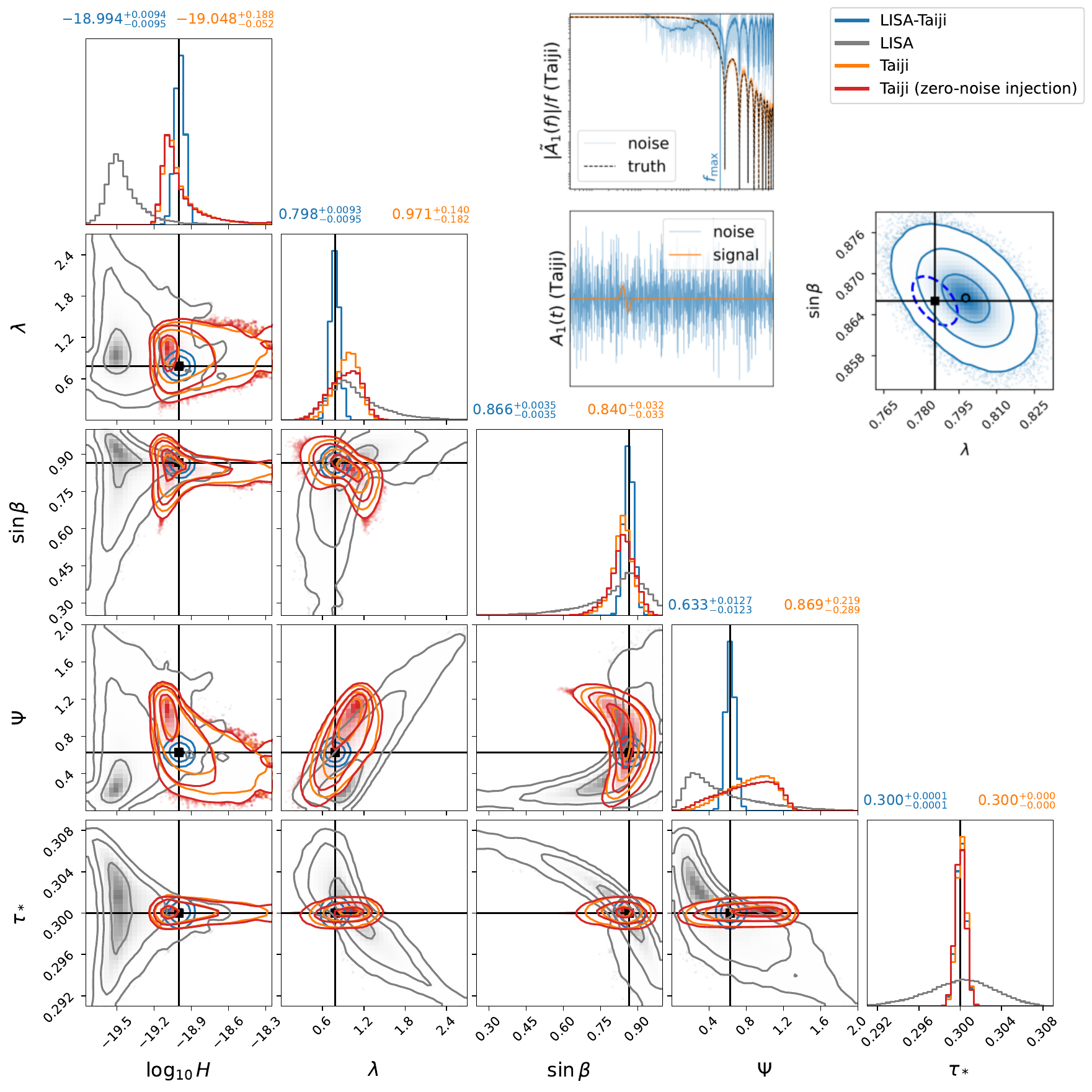}
\caption{Posterior distribution for a soft displacement-memory signal, with $T=10^{-3}\,\text{yr}\approx 9\,\text{h}\approx (3\times 10^{-5}\,\text{Hz})^{-1}$ and $f_\text{max}=0.04\,\text{Hz}$. The injected parameter values (black solid lines) are $\hat H=10^{-19}$, $\hat\lambda=\pi/4$, $\hat\beta=\pi/3$, $\hat\Psi=\pi/5$ (corresponding to $\hat\theta\approx 1.47$, $\hat\phi\approx 2.20$, $\hat\psi\approx 1.22$), and $\hat\tau_*=0.3$. The single-detector LISA and Taiji samples in the SSB-frame parametrization are obtained by transforming the corresponding detector-frame samples. The optimal SNRs are $12$ for LISA, $58$ for Taiji, and $60$ for the LISA-Taiji network. The contours mark the conventional $\{1,2,3\}\sigma$ isodensity levels of the 2D distributions, while the vertical dashed lines represent the $\{16\%,50\%,84\%\}$ quantiles of the marginalized distributions, i.e., the posterior median and central 68\% credible interval. The corresponding values for the LISA-Taiji and Taiji analyses are reported in the panel titles. The insets show the band-limited injected signal in the time domain and the frequency-domain signals, together with noise, in the $A$ channel at Taiji (for comparison, frequency-domain signals with parameters drawn from the ``Taiji'' samples are also shown), as well as a zoomed-in view of the posterior distribution of the source sky location. The blue dot marks the posterior median and the blue dashed contour indicates the FIM estimate of the $1\sigma$ level.}\label{fig:LISA-Taiji_4_plot}
\end{figure*}
	
\begin{figure*}[t]
\centering
\includegraphics[width=\textwidth]{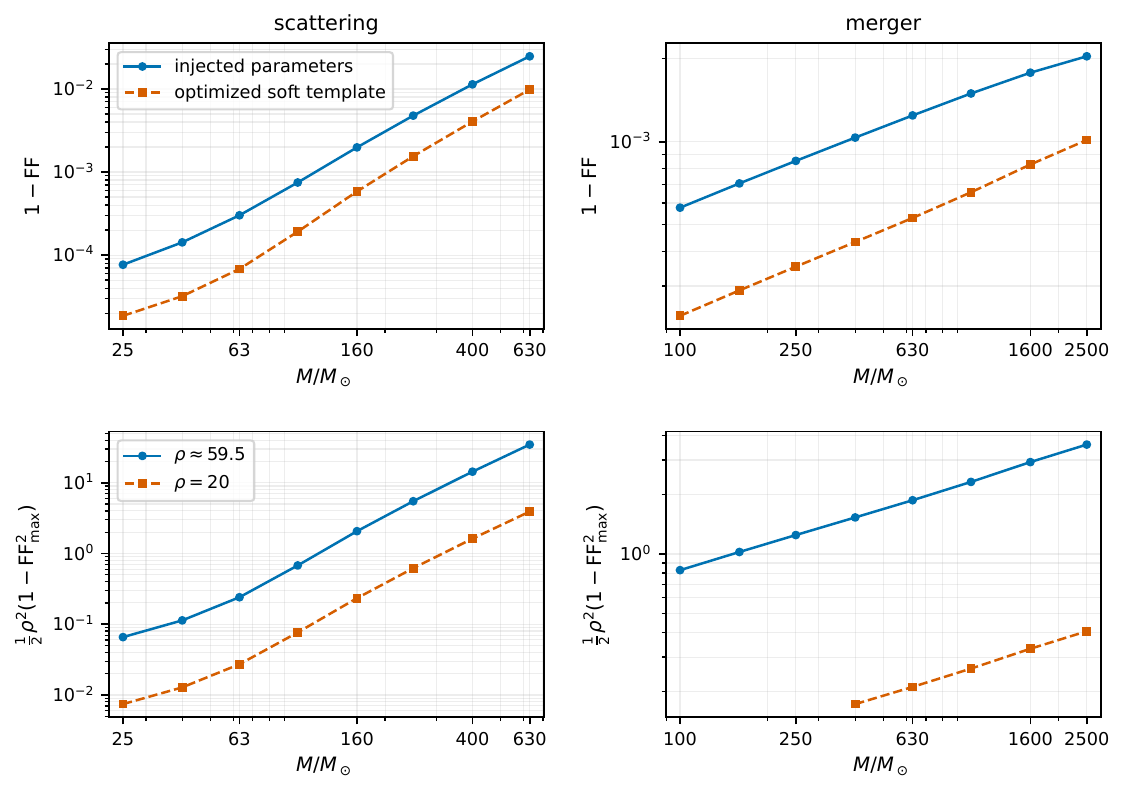}
\caption{Waveform-shape systematics for LISA-Taiji parameter estimation with scattering and merger memory waveforms, shown as functions of the source-frame total mass $M$. We inject realistic waveforms and use the soft template in the likelihood. All cases use $T=10^{-3}\,\mathrm{yr}$ and $f_\text{max}=0.04\,\mathrm{Hz}$. The fixed SSB-frame parameters are $\lambda=\pi/4$, $\beta=\pi/3$, $\Psi=\pi/5$, and $\tau_*=0.3$; for each mass, the amplitude is rescaled to the quoted SNR. The scattering time origin is the periastron crossing, and the merger time origin is the peak of $|h_{2,2}|$; this origin is placed at $t_*=\tau_*T$ in the injected signal. The scattering scan uses the 1PN compact-binary scattering waveform with $v_0=0.13$, $b/M=100$, $\nu=0.2$, source-frame viewing direction $\{\Theta,\Phi\}=\{\pi/6,\pi/3\}$, and $25\le M/M_\odot\le 630$. The merger scan uses nonspinning, quasicircular $q=1$ BBH memory waveforms with $\Theta=\pi/3$ over the mass ranges shown. The finite-frequency spectra are rescaled using the detector-frame mass $M_z=(1+z)M$. The top panels show the mismatch $1-\mathrm{FF}$ between the realistic waveform and the soft template, where $\mathrm{FF}\equiv\langle h_\mathrm{real}|h_\mathrm{soft}\rangle/(\langle h_\mathrm{real}|h_\mathrm{real}\rangle\langle h_\mathrm{soft}|h_\mathrm{soft}\rangle)^{1/2}$. For ``injected parameters,'' $\{\lambda,\beta,\Psi,\tau_*\}$ are fixed to their injected values. For ``optimized soft template,'' the overlap is maximized over these parameters, giving $\mathrm{FF}_{\rm max}$. The bottom panels show the corresponding optimized log-likelihood loss, $\frac{1}{2}\rho^2(1-\mathrm{FF}_{\rm max}^{2})$, for the high-SNR case with $\rho\approx 59.5$ and for $\rho=20$.}\label{fig:lisataiji_waveform_systematics_mismatch}
\end{figure*}
	
\begin{figure*}[t]
\centering
\includegraphics[width=\textwidth]{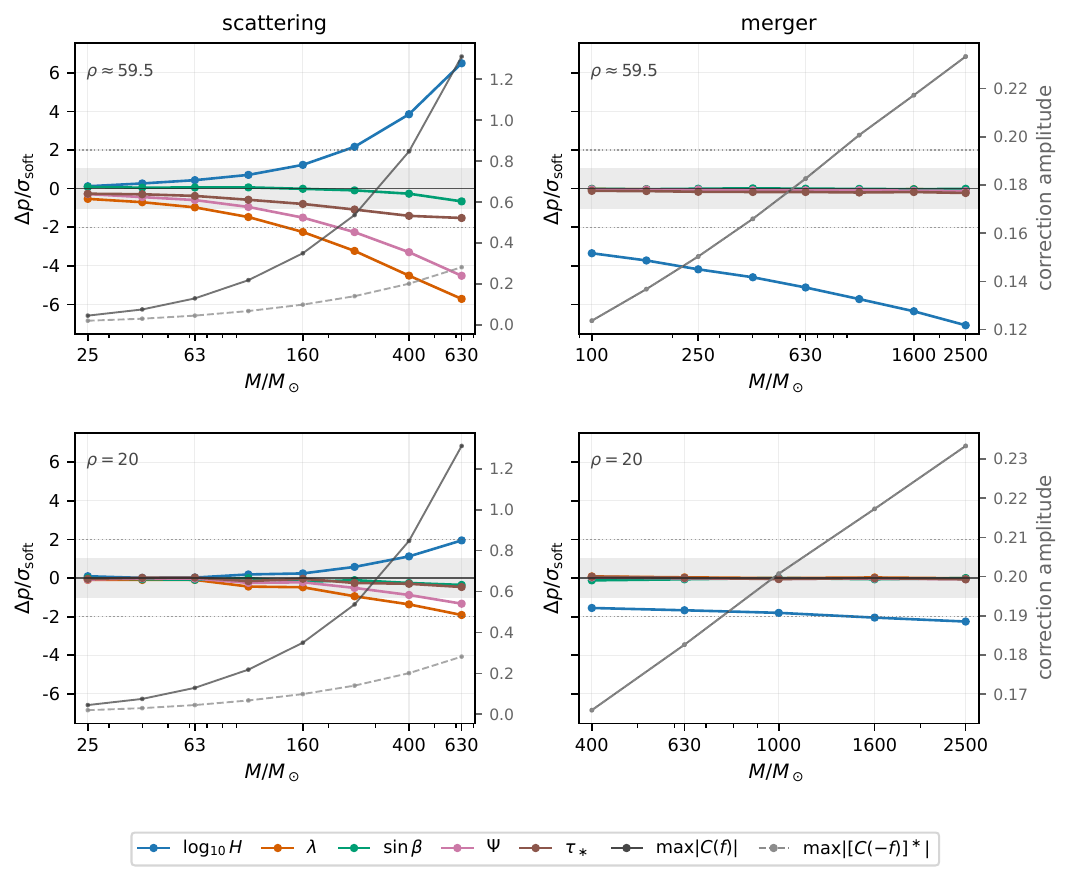}
\caption{ Posterior shifts caused by fitting realistic scattering and merger waveforms with the soft-template likelihood in LISA-Taiji, shown as functions of the source-frame total mass $M$ for the same source families, fixed angular parameters, and SNR rescalings as in Fig.~\ref{fig:lisataiji_waveform_systematics_mismatch}. The colored points show $\Delta p/\sigma_{\rm soft}$ for each parameter $p$, where $\Delta p\equiv (p_{\rm real}^{\rm med}-\hat p_{\rm real})-(p_{\rm soft}^{\rm med}-\hat p_{\rm soft})$. Here $p_{\rm soft}^{\rm med}$ is the paired soft-waveform posterior median and $\sigma_{\rm soft}\equiv(q_{84}-q_{16})/2$, with $q_{16}$ and $q_{84}$ being the corresponding marginalized quantiles. For the corresponding soft-waveform analyses, ordered as $(\log_{10}H,\lambda,\sin\beta,\Psi,\tau_*)$, $p_{\rm soft}^{\rm med}\approx(-19.00,0.795,0.863,0.635,0.3001)$, and $\sigma_{\rm soft}\approx(0.010,0.0086,0.0035,0.012,1.0\times10^{-4})$ for $\rho\approx59.5$, while $p_{\rm soft}^{\rm med}\approx(-19.48,0.810,0.858,0.653,0.3003)$ and $\sigma_{\rm soft}\approx(0.030,0.026,0.011,0.041,3.0\times10^{-4})$ for $\rho=20$. The shaded band and dotted lines mark $|\Delta p|<\sigma_{\rm soft}$ and $|\Delta p|=2\sigma_{\rm soft}$, respectively. The gray lines use the right axis and show the maximum correction amplitudes over the analysis band.}\label{fig:lisataiji_waveform_systematics_bias}
\end{figure*}
	
\begin{figure*}[t]
\centering
\includegraphics[width=\textwidth]{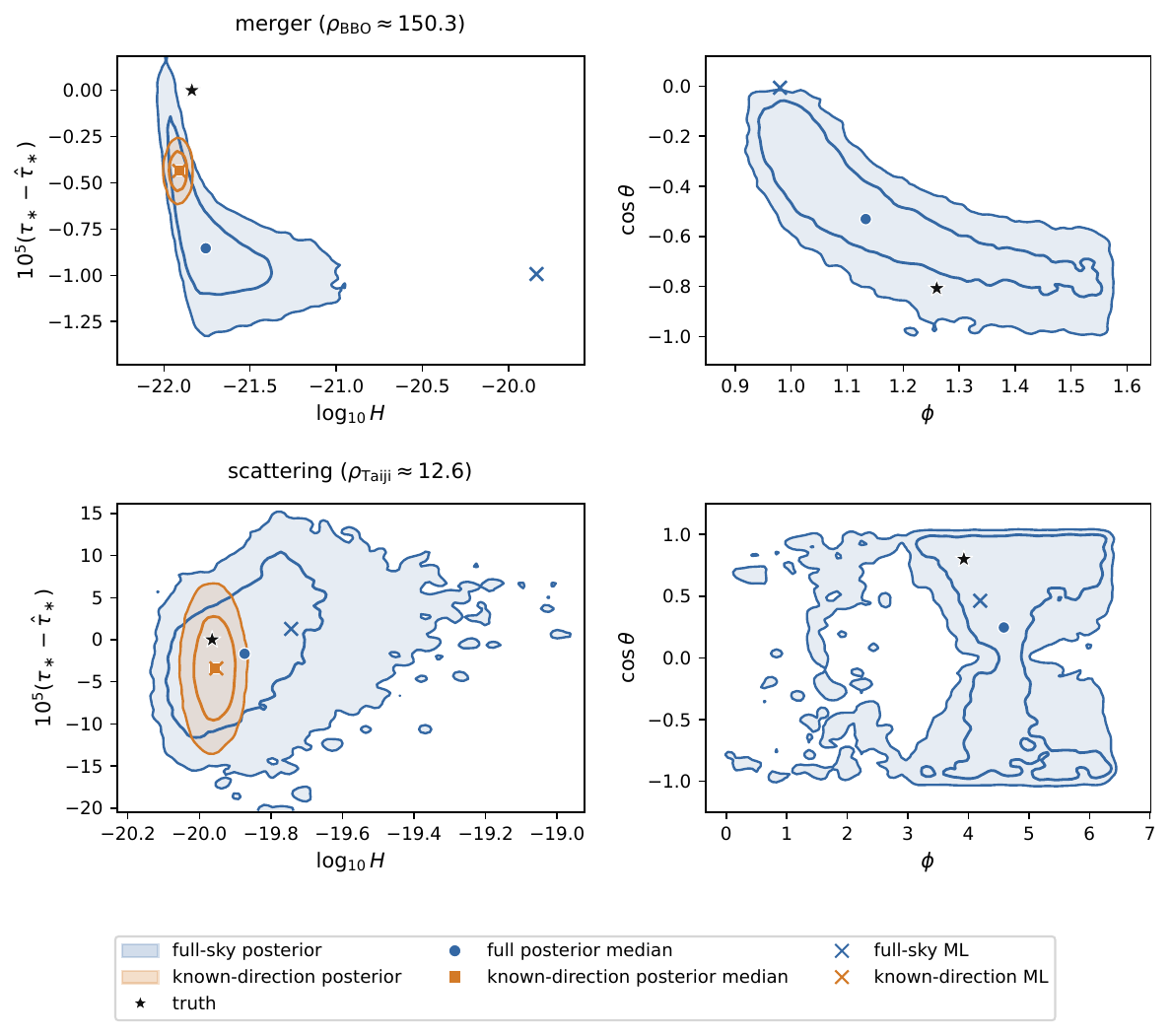}
\caption{ Single-detector parameter-estimation examples for realistic waveforms analyzed with the soft-template likelihood. The scattering waveform injection observed with Taiji uses $T=10^{-3}\,\mathrm{yr}$, $M=100\,M_\odot$, $d_L=5\,\mathrm{Mpc}$, $v_0=0.13$, $b/M=100$, $\nu=0.2$, $\{\Theta,\Phi\}=\{\pi/6,\pi/3\}$, and the 1PN scattering waveform; the injected detector-frame parameter values are $(\log_{10}\hat H,\hat\phi,\cos \hat\theta,\hat\psi,\hat\tau_*)\approx(-19.96,3.93,0.8,\pi/4,0.7)$. The merger waveform injection observed with BBO uses $T=10^{-4}\,\mathrm{yr}$, a nonspinning, quasicircular BBH memory waveform with $q=1$, $M=100\,M_\odot$, $d_L=1\,\mathrm{Gpc}$, and $\Theta=\pi/3$; the injected detector-frame parameter values are $(\log_{10}\hat H,\hat\phi,\cos \hat\theta,\hat\psi,\hat\tau_*)\approx(-21.84,1.26,-0.81,0.31,0.3)$. The likelihood inner product uses $1/T\le f<0.04\,\mathrm{Hz}$ for Taiji and $0.01<f<1\,\mathrm{Hz}$ for BBO. In each row, the left panel compares the full-sky and known-direction posteriors in the $(\log_{10}H,10^5(\tau_*-\hat{\tau}_*))$ plane, while the right panel shows the sky-location posterior for the full-sky fit. Stars mark the injected parameters, colored markers show posterior medians, and cross symbols mark the corresponding maximum-likelihood estimates.}\label{fig:single_detector_true_waveform_pe_main}
\end{figure*}

Figure~\ref{fig:LISA-Taiji_4_plot} shows a simulated LISA-Taiji measurement in the high-SNR regime, with $\hat H=10^{-19}$. Among the fitted parameters, the posterior constrains $\tau_*$ most tightly, partly because of the relatively long signal duration, and constrains the memory amplitude $\log_{10}H$ better than the angular parameters $\{\lambda,\sin\beta,\Psi\}$. Joint observations with LISA and Taiji greatly improve the measurement precision relative to that achieved with either detector alone. The plotted noise realization shifts the posterior medians but does not change these qualitative conclusions; the noise-free local FIM provides a reference for the posterior widths.
	
Appendix~\ref{appendix_PE_posteriors} presents posterior distributions for lower-SNR cases, known sky locations, and single-detector realistic-waveform tests. The lower-SNR regime is examined in Fig.~\ref{fig:LISA-Taiji_3_plot}. We find that even for $\rho\sim 10$, Taiji-only observations retain information on $\log_{10}H$ and $\tau_*$, while the angular parameters remain more strongly affected by degeneracy. If the sky location of the source is known \textit{a priori} from other observations, the remaining parameter constraints improve substantially. For the soft velocity-memory signal with the same SNR, the measurement errors are slightly larger. External sky localization removes the sky-position degeneracy and tightens the remaining parameters, while the measurable amplitude scale is still set mainly by the matched-filter SNR.
		
We assess waveform-shape systematics by injecting realistic scattering and merger memory waveforms and fitting them with the soft-template likelihood.\footnote{In this section, when quoting correction amplitudes over the analysis band, we use the band-referenced correction factor with $C(f_\text{min})=0$, so that $H$ is defined by the spectrum at the lower edge of the band. We use the 1PN scattering waveforms described in Sec.~\ref{sec:Displacement memory from compact binary scattering} and merger memory waveforms based on \texttt{NRHybSur3dq8\_CCE} (see Sec.~\ref{Sec:Displacement memory from quasi-circular BBH merger}).} In these LISA-Taiji scans, the SSB-frame parameters are fixed to $\lambda=\pi/4$, $\beta=\pi/3$, $\Psi=\pi/5$, and $\tau_*=0.3$; the source-frame total mass $M$ is varied, and the memory amplitude is rescaled to the quoted SNR. The scattering waveform uses the source family with $v_0=0.13$, $b/M=100$, $\nu=0.2$, and source-frame viewing direction $\{\Theta,\Phi\}=\{\pi/6,\pi/3\}$, aligned at periastron crossing; the merger waveform uses a nonspinning, quasicircular $q=1$ BBH memory waveform with viewing angle $\Theta=\pi/3$, aligned at the peak of $|h_{2,2}|$. For the merger scans, the mass ranges are chosen so that the LISA-Taiji analysis band remains below the onset of the dominant quadrupolar oscillatory mode, with the largest masses lying near this boundary. In both cases, this origin is placed at $t_*=\tau_*T$, and the finite-frequency scaling is evaluated with the detector-frame mass $M_z=(1+z)M$.
		
Figure~\ref{fig:lisataiji_waveform_systematics_mismatch} shows the template mismatch and the optimized likelihood loss. Figure~\ref{fig:lisataiji_waveform_systematics_bias} shows the difference between the posterior-median shifts from the injected values for realistic- and soft-waveform injections with the same noise realization. For scattering waveforms, the correction grows toward the large-mass end. In the high-SNR case, the maximum additional posterior-median shift reaches about $6.5\,\sigma_{\rm soft}$ at $M=630\,M_\odot$, and the optimized likelihood loss reaches about $34.6$. At $\rho=20$, the maxima are about $1.95\,\sigma_{\rm soft}$ and $3.91$, respectively. For merger waveforms, the maximum optimized losses are smaller, about $3.59$ in the high-SNR case and $0.406$ at $\rho=20$. The normalized merger shift is dominated by $\log_{10}H$ at the largest masses, whereas the angular and timing shifts remain much smaller; this reflects the narrow high-SNR amplitude posterior rather than a large waveform mismatch.
		
As single-detector realistic-waveform examples, we simulate a compact-binary scattering waveform observed with Taiji and a quasicircular stellar-mass BBH merger waveform observed with BBO, again using the soft-template likelihood. The scattering waveform injection observed with Taiji uses the same scattering source family with $M=100\,M_\odot$, $d_L=5\,\mathrm{Mpc}$ ($z\approx 10^{-3}$), and $\{\Theta,\Phi\}=\{\pi/6,\pi/3\}$; the merger waveform injection observed with BBO uses a nonspinning, quasicircular BBH memory waveform with $q=1$, $M=100\,M_\odot$, $d_L=1\,\mathrm{Gpc}$ ($z\approx 0.2$), and $\Theta=\pi/3$. For this merger example, the BBO analysis band lies below the onset of the dominant quadrupolar oscillatory mode. Figure~\ref{fig:single_detector_true_waveform_pe_main} shows the posterior constraints on the memory amplitude, arrival time, and sky location and lists the injected detector-frame parameter values. Figure~\ref{fig:single_detector_true_waveform_pe_appendix} shows the marginalized posterior densities for all sampled parameters. The BBO-observed merger example has $\rho\approx 150.3$, while the Taiji-observed scattering example has $\rho\approx 12.6$. The full-sky analyses show the broad sky degeneracies left by a single constellation, and the known-direction analyses show how external localization sharpens the remaining soft-template parameters. In the realistic-waveform tests, this sharpening also makes model mismatch more visible, especially for the high-SNR BBO-observed merger case.
		
We also test a supermassive black-hole binary (SMBHB) scenario in which the dominant oscillatory waveform is first fitted and subtracted before applying the soft-template likelihood. In this restricted fit, the intrinsic parameters and sky location are fixed to their injected values, while an overall amplitude, a time shift, the orbital phase, the polarization angle, and the inclination are varied. For the Taiji example shown in Fig.~\ref{fig:smbhb_residual_memory_pe}, the oscillatory waveform has $\rho\approx1.4\times10^4$, while the injected memory modes have $\rho\approx177.2$. After primary-waveform subtraction, a significant memory component remains measurable. For the same postsubtraction residual, $(\rho_\mathrm{res},H/\hat H)\approx(125.5,0.308^{+0.003}_{-0.003})$ at $f_\text{max}=40\,\mathrm{mHz}$ and $(18.0,0.724^{+0.046}_{-0.041})$ at $f_\text{max}=1\,\mathrm{mHz}$. The lower choice of $f_\text{max}$ reduces the amplitude bias but lowers the SNR and broadens the posterior.
	
\clearpage
	
\section{Stochastic soft-memory signals}\label{sec:6}
	
A data stream of sufficiently long duration may contain multiple events with memory, or even a large population of them. In the latter case, the detector effectively observes a stochastic GW background (SGWB) of soft-memory signals~\cite{Zhao:2021zlr}. Even if the individual events cannot be measured, the SGWB may still be detectable if the event rate is sufficiently large. This may occur, for example, for the SGWB from compact binary scattering~\cite{Garc_a_Bellido_2022,Kerachian_2024}. The SGWB from compact binary mergers also contains a memory contribution, although it is expected to be subleading in the frequency band of interest compared with the oscillatory GW contribution from sources in the same population; after subtraction of the oscillatory compact-binary foreground in primordial SGWB searches~\cite{Yamamoto2026}, however, the memory contribution can become relevant. In this section, we examine the signal PSD of an idealized SGWB of soft displacement-memory signals at the LISA-like detectors and assess its detectability. We also briefly comment on the SGWB of soft velocity-memory signals.
	
Generically, the GW can be written as
\[
h_{ij}(t,\mathbf{x})=\sum_{\lambda}\int_{-\infty}^\infty df\int d\Omega_2\,\tilde h_\lambda(f,\hat{\mathbf{k}})\,e_{ij}^\lambda(\hat{\mathbf{k}})\,\text{e}^{-2\pi i f(t-\hat{\mathbf{k}}\cdot\mathbf{x})}.
\]
If the SGWB is stationary, isotropic, and unpolarized (such an SGWB will henceforth be referred to as ``idealized''), the two-point statistics satisfy~\cite{romano2023searchesstochasticgravitationalwavebackgrounds}
\[
\left\langle \tilde h_\lambda(f,\hat{\mathbf{k}})\,\tilde h^*_{\lambda'}(f',\hat{\mathbf{k}}')\right\rangle
=\frac{S_h(f)}{16\pi}\,\delta(f-f')\,\delta^2(\hat{\mathbf{k}},\hat{\mathbf{k}}')\,\delta_{\lambda,\lambda'},
\]
where $S_h(f)$ is the strain power spectral density and $\delta^2(\hat{\mathbf{k}},\hat{\mathbf{k}}')$ is the Dirac delta distribution on the unit 2-sphere, i.e., $\int d\Omega_2(\hat{\mathbf{k}}')\,\delta^2(\hat{\mathbf{k}},\hat{\mathbf{k}}')\,F(\hat{\mathbf{k}}')=F(\hat{\mathbf{k}})$. Note that $S_h$ is related to the normalized GW energy density spectrum via
\begin{equation}
\Omega_\text{gw}(f)=\frac{1}{\rho_\text{crit}}\frac{d\rho_\text{gw}}{d \ln f}=\frac{2\pi^2}{3H_0^2}f^3S_h(f), \label{Omega_gw}
\end{equation}
where $H_0$ and $\rho_\text{crit}=3H_0^2/(8\pi)$ are the Hubble parameter and critical density today.
	
Let us consider the local waveform: $h_{ij}(t)=h_{ij}(t,\mathbf{0})$. The one-sided PSD $\mathcal{P}_{ij}(f)$ of $h_{ij}(t)$ is defined by $\langle \tilde h_{ij}(f)\,\tilde h_{ij}^*(f')\rangle=\frac{1}{2}\delta(f-f')\,\mathcal{P}_{ij}(f)$. It is useful to derive a relation between $\mathcal{P}_{ij}$ and $S_h$. Since $\tilde h_{ij}(f)=\sum_{\lambda} \int d\Omega_2\,\tilde h_\lambda(f,\hat{\mathbf{k}})\,e_{ij}^\lambda(\hat{\mathbf{k}})$, for the idealized SGWB,
\[
\begin{aligned}
\frac{\left\langle\tilde h_{ij}(f)\,\tilde h^*_{kl}(f')\right\rangle}{\delta(f-f')} &=\frac{S_h}{16\pi}\sum_{\lambda} \int d\Omega_2\,e_{ij}^\lambda(\hat{\mathbf{k}})\,e_{kl}^{\lambda *}(\hat{\mathbf{k}})
\\
&=\frac{S_h}{10}\left(\delta_{ik}\delta_{jl}+\delta_{il}\delta_{jk}-\frac{2}{3}\delta_{ij}\delta_{kl}\right)
,
\end{aligned}
\]
where we used the TT polarization sum $\sum_\lambda e_{ij}^\lambda\,e_{kl}^{\lambda *}=2\Lambda_{ij,kl}$, with $\Lambda_{ij,kl}$ defined in Eq.~\eqref{quad_formula}. Thus $\mathcal{P}_{ij}=\frac{1}{5}\left(1+\frac{1}{3}\delta_{ij}\right)S_h$, $\mathcal{P}_{ii}=\frac{4}{15}S_h\ \text{(no sum over $i$)}$, and
\begin{equation}\label{Sh_and_Pij}
S_h=\frac{1}{2}\sum_{i,j}\mathcal{P}_{ij}.
\end{equation}
For a LISA-like detector, the $U$-channel TDI response to a plane wave in the static-detector approximation is given by $\tilde U(f)=\mathcal{F}^U_{ij}(f)\,\tilde h_{ij}(f)$. The one-sided PSD of $U(t)$ is thus
\begin{equation}\label{P_U}
\begin{aligned}
P_U(f) =S_h \sum_\lambda\mathcal{F}^U_\lambda
,
\end{aligned}
\end{equation}
where $\mathcal{F}^U_\lambda(f)$ is the averaged response function defined in Eq.~\eqref{F_function}.
	
\begin{figure}[b]
\centering
\includegraphics[width=0.48\textwidth]{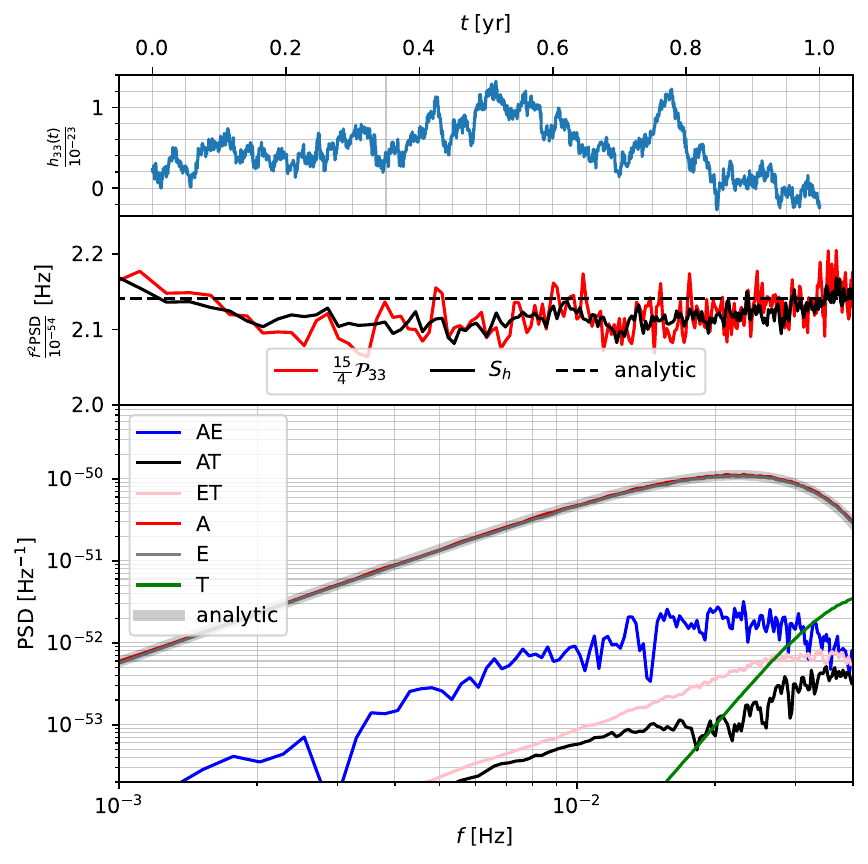}
\caption{Comparison of the simulated one-year SGWB of soft displacement-memory signals at LISA with the analytic prediction. The top panel shows the time-domain waveform $h_{33}(t)$, where $33$ denotes a Cartesian tensor component. The middle panel shows the SGWB spectrum, where the black solid line denotes the estimate of $S_h$ given by Eq.~\eqref{Sh_and_Pij}. The bottom panel shows the PSD and cross-spectral density (CSD) spectra of TDI signals, where the thick line denotes the signal PSD given by Eq.~\eqref{P_U}, using the analytic prediction \eqref{analytic_S_h} for $S_h$.}\label{fig:SGWB_1}
\end{figure}
	
\begin{table}[b]
\renewcommand{\arraystretch}{1}
\setlength{\tabcolsep}{7pt}
\begin{center}
\begin{tabular}{ccccc}
\hline
\hline
& \;LISA\; & 
\;Taiji\; & \;TianQin\; & \;BBO\; \\
\hline
$\frac{\Delta_{10}}{10^{-47}\text{Hz}}$ & 11 & 3.6 & 45 & $2.5\times 10^{-5}$ \\
\hline
\hline
\end{tabular}
\end{center}
\caption{Value of the displacement memory SGWB parameter $\Delta$ for $\rho_{A+E}=10$ and $T=1\,\text{yr}$ at the considered detectors. We use the frequency ranges
$[10^{-4},0.03]\,\mathrm{Hz}$ for LISA and Taiji,
$[10^{-3},0.25]\,\mathrm{Hz}$ for TianQin, and
$[0.1,1]\,\mathrm{Hz}$ for BBO.}
\label{detector_table_3}
\end{table}

\begin{figure}[hbt]
\centering
\includegraphics[width=0.48\textwidth]{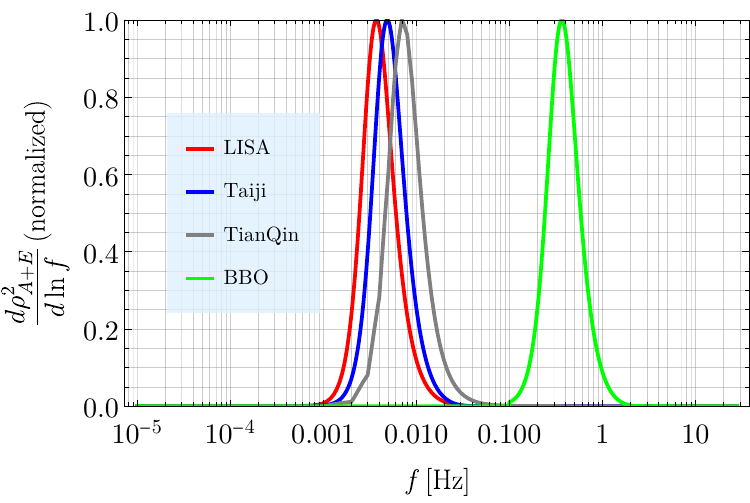}
\caption{Normalized logarithmic spectral density of $\rho_{A+E}^2$ at LISA, Taiji, TianQin, and BBO for an idealized SGWB of soft displacement-memory signals.}\label{fig:SGWB_SNR}
\end{figure}

\begin{figure}[hbt]
\centering
\includegraphics[width=0.48\textwidth]{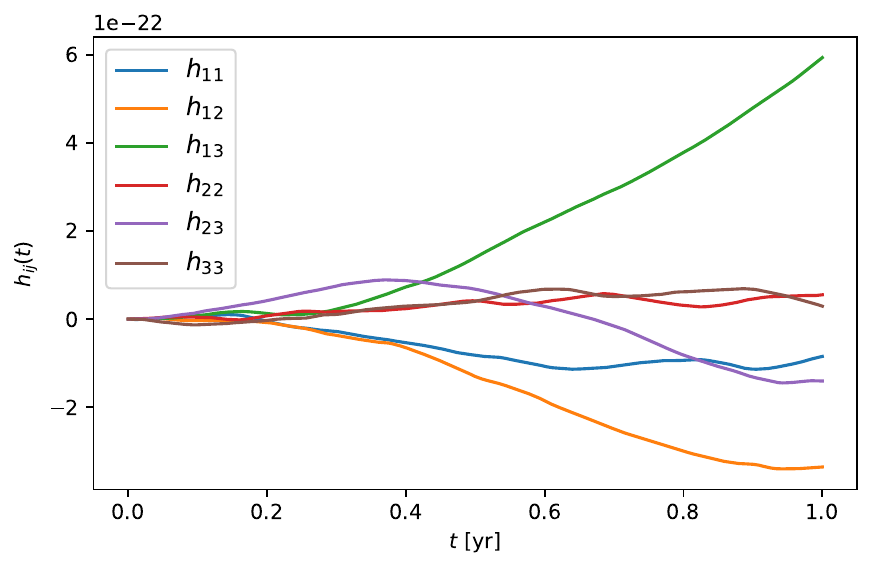}
\caption{Simulated one-year SGWB of soft velocity-memory signals. In this illustrative simulation, all event times are drawn within the observing interval and the prehistory is set to zero. The common initial value $h_{ij}(0)=0$ is therefore an artifact of this finite-window setup.}\label{fig:SGWB_PSD_relu}
\end{figure}
	
For an ensemble of soft displacement-memory signals in the interval $t\in (0,T)$, the resulting $h_{ij}(t)$ is a superposition of step functions with different step sizes $\mathcal{H}$ and step positions $t_*$. Such a time-domain waveform resembles Brownian motion and is not wide-sense stationary. Nevertheless, the Welch PSD estimator is
\begin{equation}\label{P_{ij}}
\mathcal{P}_{ij}\approx\frac{2}{T}\frac{\sum \mathcal{H}^2}{(2\pi f)^2}=\frac{N}{T}\frac{2\langle \mathcal{H}^2 \rangle}{(2\pi f)^2}.
\end{equation}
This is compatible with \cite{Zhao:2021zlr}. Here $N$ is the total number of steps, and we denote the mean and standard deviation of $X$ by $\langle X\rangle$ and $\sigma_X$, respectively.
	
We now consider an ensemble of mutually independent soft displacement-memory signals with a uniform distribution of $\{\lambda,\sin\beta,\Psi\}$. Here we refer to such an SGWB as ``idealized.'' A single event appears in $h_{33}(t)$ as a step with size $\mathcal{H}=\kappa H$, where $\kappa=-c^2_\beta\,c_{2\Psi}$. It follows that $\langle\mathcal{H}\rangle=0$, and $\langle\mathcal{H}^2\rangle=\sigma_\kappa^2\left(\langle H\rangle^2+\sigma_H^2\right)$, where $\sigma_\kappa^2=4/15$. Equations~\eqref{Sh_and_Pij} and \eqref{P_{ij}} then yield
\begin{equation}\label{analytic_S_h}
S_h=\frac{\Delta}{f^2},
\end{equation}
where $\Delta=\frac{2N}{T}\frac{\langle H\rangle^2+\sigma_H^2}{(2\pi)^2}$. The corresponding energy density $d\rho_\text{gw}/df$ [defined by Eq.~\eqref{Omega_gw}] is equivalent to a simple sum of the spectral energy flux density $\frac{1}{T}\frac{d E_\text{gw}}{R^2d\Omega_2df}=\frac{H^2}{8\pi T}$ associated with each memory jump. For example, if $H$ is distributed uniformly in $[0,H_\text{max}]$, then $\langle H\rangle^2+\sigma_H^2=\frac{1}{3}H_\text{max}^2$. To verify this description, a time-domain simulation of the SGWB is performed and the first-generation TDI signal at the LISA detector is computed using \texttt{fastlisaresponse} with the realistic ESA orbit. The SGWB is constructed from 4000 sources drawn from a uniform distribution of $\{H,t_*,\lambda,\sin\beta,\Psi\}$. We find that the result agrees well with the analytic prediction for $S_h$ given by Eq.~\eqref{analytic_S_h} and the signal PSD given by Eq.~\eqref{P_U}, as shown in Fig.~\ref{fig:SGWB_1}.
	
In principle, the signal PSD can be measured~\cite{Hartwig_2023,Wu_2026}, and the corresponding squared SNR is given by
\begin{equation}
\rho^2=T\sum_U\int_{f_\text{min}}^{f_\text{max}} df\,\left(\frac{P_U}{S_U}\right)^2.
\end{equation}
The SNR of the T channel is again found to be small and is neglected. In Table~\ref{detector_table_3} we list the values of the parameter $\Delta$ in Eq.~\eqref{analytic_S_h} for $\rho_{A+E}=10$ and $T=1\,\text{yr}$ at the considered detectors. The corresponding logarithmic spectral density of $\rho_{A+E}^2$ is shown in Fig.~\ref{fig:SGWB_SNR}. For the tabulated estimates in Table~\ref{detector_table_3}, we use the detector-specific frequency ranges listed in the caption.
	
In contrast to displacement memory, the SGWB of soft velocity-memory signals cannot be characterized by a PSD. Figure~\ref{fig:SGWB_PSD_relu} shows an example of such an SGWB, constructed from $10^4$ sources modeled as an ensemble of ramp waveforms with $\Delta \dot h_-=0$ (see Sec.~\ref{Sec:Velocity memory}), with a uniform distribution of $\{\dot H,t_*,\lambda,\sin\beta,\Psi\}$. The time-domain waveform resembles a low-frequency deterministic signal. Detecting such a signal component is expected to be particularly challenging.
	
\section{Summary and discussion}\label{sec:7}
As predicted by general relativity, oscillatory gravitational radiation is generally accompanied by a low-frequency component associated with GW memory. A measurement targeting only the high-frequency part is therefore incomplete. At sufficiently low frequencies, the waveforms approach clean, universal forms, with the memory amplitudes as the only intrinsic parameters. We have investigated the prospects for detecting such soft signals with future space-based interferometers, through observations of individual events and the stochastic background. In the optimal matched-filter estimate under the adopted instrumental-noise model, the SNR of a single event in LISA, Taiji, and TianQin can reach 10 for a displacement memory of amplitude $\sim 10^{-20}$. Such signals can arise from compact-binary scatterings, supernovae, gamma-ray bursts, or certain dark transients (including subsolar-mass BBH mergers). For the adopted BBO instrumental-noise model, a displacement-memory amplitude $\sim 10^{-23}$ gives an SNR of 10, a scale relevant for sufficiently loud stellar-mass compact-binary mergers. We examine the parameter-estimation precision for soft displacement-memory signals using simulated Bayesian analyses. These simulations show that the soft-template amplitude and arrival time are measurable by a single LISA-like detector at SNRs of order 10, while sky and polarization degeneracies are reduced by LISA-Taiji observations or by external sky localization. Sufficiently loud stellar-mass compact-binary mergers are within reach of single-detector soft-template measurements of null displacement memory. The deviation of the GW spectrum from an exact soft waveform can be described by a complex correction factor. We examine such corrections in realistic examples of compact binary scattering and quasicircular BBH mergers by injecting realistic waveforms and fitting them with the soft-template likelihood. The soft-template parameters remain statistically constrained; finite-frequency waveform-shape corrections can produce measurable posterior shifts, especially at high SNR, while the cases with $\rho=20$ show smaller but still quantifiable shifts. Many unresolvable soft signals can also form a stochastic gravitational-wave background. For displacement memory, the resulting background is expected to follow a strain power spectral density $S_h\propto f^{-2}$. We evaluate the corresponding SNR of the signal power spectrum for the considered detectors. These results establish soft signals from bursts with memory as concrete targets for space-based gravitational-wave detection.
	
This paper focuses on soft memory signals in general relativity; analogous soft signals may also arise in additional GW polarization modes~\cite{Tahura_2021,Hou_2021,Heisenberg_2023}. Similar signals can arise from massless or ultralight dark radiation, as discussed in Appendix~\ref{appendix_B}. The relevant templates can be readily constructed, and their distinguishability can be analyzed. Finally, soft memory signals that extend to sufficiently high frequencies can, in principle, be jointly observed by ground-based and space-based detectors, providing complementary information about the spectrum. Possible soft memory signals associated with a GW event observed by a ground-based detector can also be searched for with space-based detectors.
	
\section*{Acknowledgments}
We thank the anonymous referee for valuable comments and suggestions that helped improve the quality of the manuscript. Y. L. M. is supported in part by the National Key R \& D Program of China under Grant No. 2021YFC2202900, the National Natural Science Foundation of China (NSFC) under Grant No. 12547104, and Gusu Talent Innovation Program under Grant No. ZXL2024363. Y. T. is partly supported by NSFC under Grant No. 12547104.
	
\section*{Data Availability}
The numerical data and scripts supporting the LISA--Taiji parameter-estimation example are available at~\cite{lisataiji_soft_memory_pe_data}.

\appendix
	
\section{Eikonal approximation}\label{appendix_propagation_details}
	
The zeroth-order eikonal approximation of massless relativistic waves in a curved spacetime background is described by the massless Klein-Gordon equation. We consider a spatially flat Friedmann-Lemaitre-Robertson-Walker spacetime that is weakly perturbed: ${ds^2}/{a^2}=-(1+2\Psi)d\eta^2+2\Xi_id\eta dx^i+(\delta_{ij}+H_{ij})dx^idx^j$,
where $a(\eta)$ is the cosmological scale factor and $\eta$ is the conformal time. The massless Klein-Gordon equation in the coordinates $x^\mu=(\eta,\mathbf{x})$ reads ${ \partial_a\left(\sqrt{-g}\,g^{a b} \partial_b \phi\right)}/{\sqrt{-g}}=(W+B^a\partial_a)\phi=0$, where $W\equiv g^{ab}\partial_a\partial_b$ and $B^a\equiv\partial_b(\sqrt{-g}\,g^{ab})/\sqrt{-g}$. In the eikonal approximation~\cite{PhysRev.126.1899}, a Fourier mode of the complex scalar field is described by a rapidly varying real phase $S(\eta,\mathbf{x})$ and a slowly varying real amplitude $A(\eta,\mathbf{x})$ as $\phi = A\,\text{e}^{iS}$, with $\omega(\eta,\mathbf{x})\equiv-\partial_\eta S=2\pi f$ and $\mathbf{k}(\eta,\mathbf{x})\equiv \partial_\mathbf{x} S$. At zeroth order, $W\phi\approx 0$, so $k_\mu=\partial_\mu S=(-\omega,\mathbf{k})$ satisfies the local dispersion relation: $0=D(x,k)=-g^{ab}k_ak_b\approx a^{-2}[(1-2\Psi)\omega^2-|\mathbf{k}|^2+H_{ij}k_ik_j+2\Xi_i\omega k_i]$. $k_\mu(\eta,\mathbf{x}(\eta))$ evolves according to ${dk_\mu}/{d\eta}=(\partial_\mu D)/(\partial_\omega D)$, e.g.,
\begin{equation}\label{freq_evolution}
\frac{d\omega}{d\eta}\approx\frac{\omega^2\partial_\eta \Psi-\frac{1}{2}(\partial_\eta H_{ij})k_ik_j-(\partial_\eta \Xi_i)\omega k_i}{\omega}.
\end{equation}
Meanwhile, the amplitude evolves according to $\left[(\partial_{k_a}D)\partial_a+\left(\partial_{k_a}\partial_{k_b}D\right)(\partial_a k_b)/2-B^ak_a\right]A\approx 0$, i.e., $k^a\partial_a \ln A\approx -\nabla_ak^a/2$. The frequency measured by a timelike observer with four-velocity $u^\mu$ (normalized as $u^\mu u_\mu=-1$) is $\omega_\text{o}=-u^\mu k_\mu$. For example, for an observer (emitter) with $u_\text{o(e)}^\mu\approx a_\text{o(e)}^{-1}[1-\Psi_\text{o(e)}]\delta^\mu_0$ at the time $\eta_\text{o(e)}$ of observation (emission), $\omega_\text{o(e)}\approx a_\text{o(e)}^{-1}[1-\Psi_\text{o(e)}]\,\omega(\eta_\text{o(e)})$. In the absence of metric perturbations, $\omega(\eta_\text{o})=\omega(\eta_\text{e})$, so $\omega_\text{o} = \omega_\text{e}/(1+z)$, where $z\equiv a_\text{o}/a_\text{e}-1$ is the redshift of the source.
	
For an observer (or a source) with nonvanishing velocity $u^i$, there is an additional apparent frequency shift, which in a flat background is just the usual Doppler shift, given by $\varrho\approx 1-\hat{\mathbf{k}}\cdot\mathbf{v}$ for nonrelativistic motion, with $\mathbf{v}=\mathbf{v}_\text{o}-\mathbf{v}_\text{e}$. As an example of the effect of metric perturbations, consider the wave traveling in $H_{ij}(u=\eta-\mathbf{v}_g\cdot\mathbf{x})$, where $\mathbf{v}_g$ is a constant velocity of the background. Since $\frac{d}{d\eta}\approx\partial_\eta+\hat k_i\partial_i$ along the propagating direction $\hat{\mathbf{k}}$ of the wave, $\int_{\eta_1}^{\eta_2} d\eta\,\partial_\eta X(u)\approx \frac{\Delta X}{1-\mathbf{v}_g\cdot\hat{\mathbf{k}}}$,
where $\Delta X\equiv X (\eta_2,\mathbf{x}_2)-X (\eta_1,\mathbf{x}_1)$. From Eq.~\eqref{freq_evolution}, the accumulated frequency shift is therefore~\cite{EstabrookWahlquist1975}
\begin{equation}\label{frequency_shift_in_GW}
\varrho-1=\frac{\Delta\omega}{\omega}
=\frac{1}{\omega}\int d\eta\frac{d\omega}{d\eta}
\approx-\frac{\Delta \left(\frac{1}{2}\hat k_i \hat k_jH_{ij}\right)}{1-\mathbf{v}_g\cdot\hat{\mathbf{k}}}.
\end{equation}
	
\onecolumngrid
	
\section{Coordinate transformations and noise model}\label{appendix_tdi_coordinates}
	
\subsection{Detector-frame and SSB-frame coordinates}
	
In the equal-arm approximation, the positions of SC in the SSB frame can be obtained as $x_{Ii}'= X'_i(\eta)+R_{ij}(\eta)\,x_{Ij}$, with the guiding center $\mathbf{X}'=R(\cos\eta,\sin\eta,0)$ and $R_{ij}$ a rotation matrix that depends on a phase constant $\Phi_0$ (corresponding to $\eta_0+\xi_0$ in \cite{Krolak:2004xp}):
\begin{equation}
R_{ij}(\eta)=\frac{1}{8}\left(
\begin{matrix}
\sqrt{3} (3 s_{\Phi_0}+s_{2 \eta -\Phi_0})-c_{2 \eta -\Phi_0}+3c_{\Phi_0} & -s_{2 \eta -\Phi_0}-\sqrt{3} c_{2 \eta -\Phi_0}-3 s_{\Phi_0}+3 \sqrt{3} c_{\Phi_0} & -4 \sqrt{3} c_{\eta} \\
-s_{2 \eta -\Phi_0}-\sqrt{3} c_{2 \eta -\Phi_0}+3 s_{\Phi_0}-3 \sqrt{3} c_{\Phi_0} & \sqrt{3} (3 s_{\Phi_0}-s_{2 \eta -\Phi_0})+c_{2 \eta -\Phi_0}+3 c_{\Phi_0} & -4 \sqrt{3} s_{\eta} \\
2 \sqrt{3} c_{\eta -\Phi_0}-6 s_{\eta -\Phi_0} & 2 \sqrt{3} s_{\eta -\Phi_0}+6 c_{\eta -\Phi_0} & 4 \\
\end{matrix}
\right),
\end{equation}
where $c_z \equiv \cos z$ and $s_z \equiv \sin z$. Note that the coordinates of $\mathbf{e}_z$ in the SSB frame are $(-\sqrt{3}\cos\eta, -\sqrt{3}\sin\eta,1)/2$. The explicit relation between $(\theta,\phi)$ in the detector frame and $(\beta,\lambda)$ in the SSB frame is given by $\beta=\arcsin s_\beta$, $\lambda=\Theta(c_\lambda)\,\arcsin s_\lambda+\Theta(-c_\lambda)\,(\pi-\arcsin s_\lambda)$, with
	
\begin{equation}
\left(\begin{matrix}
c_\beta c_\lambda \\
c_\beta s_\lambda \\
s_\beta
\end{matrix}\right)
=
-\frac{1}{8}\left(\begin{matrix}
s_\theta \left[-\sqrt{3} (s_{2 \eta -\Phi_0-\phi}+3 s_{\Phi_0+\phi })+c_{2 \eta -\Phi_0-\phi}-3 c_{\Phi_0+\phi }\right]
+4 \sqrt{3} c_\eta c_\theta
\\
s_\theta \left(s_{2 \eta -\Phi_0-\phi}+\sqrt{3} c_{2 \eta -\Phi_0-\phi }-3 s_{\Phi_0+\phi}+3 \sqrt{3} c_{\Phi_0+\phi}\right)
+4 \sqrt{3} s_\eta c_\theta
\\
2 \left[s_\theta \left(3 s_{\eta -\Phi_0-\phi}-\sqrt{3} c_{\eta -\Phi_0-\phi }\right)-2 c_\theta\right]
\end{matrix}\right).
\end{equation}
	
\subsection{Noise model}
	
The one-sided noise PSD used in the SNR and parameter-estimation analyses takes the form~\cite{Prince_2002,babak2021lisasensitivitysnrcalculations,Hartwig_2023}:
\begin{equation}\label{noise_PSD}
\begin{aligned}
S_{A_1,E_1}=& \left\{16\left[3+2\cos (2\pi fL)+\cos(4\pi fL)\right]\,S_\text{acc}
+8\left[2+\cos (2\pi fL)\right]\,S_\text{oms}\right\}\sin^2(2\pi fL).
\end{aligned}
\end{equation}
For LISA, Taiji, and TianQin, we use the optical metrology noise $S_{\mathrm{oms}}(f)$ and test mass acceleration noise $S_\text{acc}(f)$ given by
\begin{align}
S_{\mathrm{oms}} &=\left(\frac{2\pi f s_\text{oms}}{c}\right)^2\left[1+\left(\frac{2\, \mathrm{mHz}}{f}\right)^4\right]\text{s},
\quad
S_{\mathrm{acc}} =\left(\frac{\text{Hz}^2s_\text{acc}}{2\pi fc}\right)^2\left[1+\left(\frac{4\, \mathrm{mHz}}{10f}\right)^2\right]\left[1+\left(\frac{f}{8\, \mathrm{mHz}}\right)^4\right]\text{s}
,
\end{align}
where $c$ denotes the speed of light. The values of $s_\text{oms}$ and $s_\text{acc}$ are listed in Table~\ref{detector_table_1}. For BBO~\cite{Crowder_2005,Corbin_2006}, we use
\begin{align}
S_{\mathrm{oms}} =
\frac{2\times 10^{-34}\,\text{s}}{(3L/\text{m})^2}\left(\frac{6\pi fL}{c}\right)^2,
\quad
S_{\mathrm{acc}} =
\frac{9\times 10^{-34}\,\text{s}}{(2\pi f/\text{Hz})^4(3L/\text{m})^2}\left(\frac{6\pi fL}{c}\right)^2
.
\end{align}
In its proposed design, the first stage of BBO comprises a single LISA-like constellation, and the second stage consists of 12 SC in four constellations. Here we consider the first stage of BBO. We also neglect the impact of Galactic confusion noise and other signal components in the data streams, which generally require global fitting.
	
\begin{table}[H]
\renewcommand{\arraystretch}{1}
\setlength{\tabcolsep}{4pt}
\begin{center}
\begin{tabular}{ccccc}
\hline
\hline
Detector & $s_\text{acc}$ [fm] & $s_\text{oms}$ [pm] & $L$ [Gm] & $1/(2L)$ [Hz] \\
\hline
LISA~\cite{amaroseoane2017laser,babak2021lisasensitivitysnrcalculations} & 3 & 15 & 2.5 & 0.06\\
Taiji~\cite{10.1093/nsr/nwx116} & 3 & 8 & 3 & 0.05\\
TianQin~\cite{Luo_2016} & 1 & 1 & 0.17 & 0.88\\
BBO~\cite{Corbin_2006}& $\times$ & $\times$ & 0.05 & 3\\
\hline
\hline
\end{tabular}
\end{center}
\caption{Detector-noise and armlength parameters used for the considered detectors.}
\label{detector_table_1}
\end{table}
	
\clearpage
\twocolumngrid
\section{Soft-memory signals in the T channel}\label{appendix_A}
	
In the main text, the signal in the T channel is neglected. In contrast to $A$ and $E$, the low-frequency spectrum of $T$ is sensitive to the armlength differences and deviates drastically from the result in the SEA. We therefore need to consider second-generation TDI variables, for example,
\begin{equation}
\begin{aligned}
X_2 =& X_1 +[(D_{13121}y_{12}+D_{131212}y_{21}
\\
&+D_{1312121}y_{13}+D_{13121213}y_{31})
-(3\leftrightarrow 2)],
\end{aligned}
\end{equation}
with $X_1$ given by Eq.~\eqref{X_1}. The noise PSD of $T_2$~\cite{Quang_Nam_2023} at low frequencies is also significantly enhanced relative to the spectrum $S_{T_2}=4\sin^2(4\pi fL)\,S_{T_1}$ in the SEA, where
\[
\frac{S_{T_1}}{\sin^2(2\pi fL)}=32\sin^2(\pi fL)\left[4\sin^2(\pi fL)\,S_\text{acc}+\,S_\text{oms}\right].
\]
	
\begin{figure}[hbt!]
\centering
\includegraphics[width=0.48\textwidth]{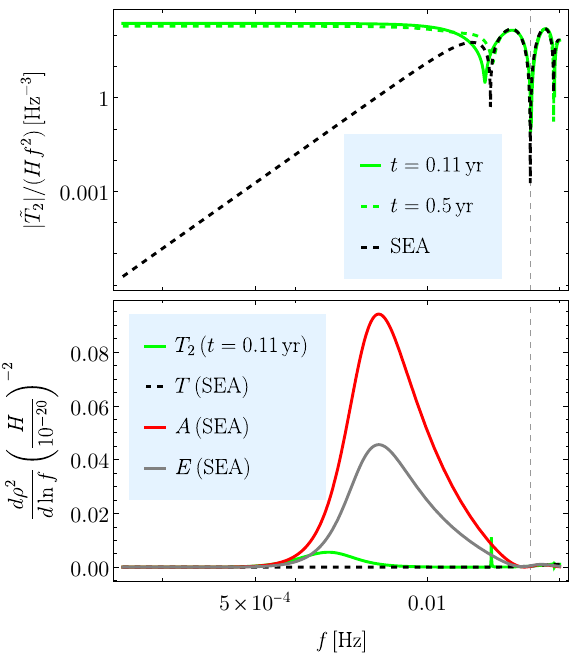}
\caption{Spectrum of $|\tilde T_2|/f^2$ (top panel) and the logarithmic spectral density of $\rho_{A,E,T}^2$ (bottom panel) for displacement memory with $\{\phi,\theta,\psi\}=\{0.32\pi,0.52\pi,0.25\pi\}$. The vertical dashed line corresponds to $f=1/(2L)$.}\label{fig:displacement_memory_T}
\end{figure}

For concreteness, we consider the soft displacement-memory signal at LISA, using the realistic orbital data provided by \texttt{LISAanalysistools}~\cite{michael_katz_2024_10930980} (still adopting the static-detector approximation). Figure~\ref{fig:displacement_memory_T} shows the result for $\{\phi,\theta,\psi\}=\{0.32\pi,0.52\pi,0.25\pi\}$ at $t=0.11\,\text{yr}$ and $t=0.5\,\text{yr}$ (in the time coordinate used for the orbital data), with $\{L_{23},L_{13},L_{12}\}\approx\{0.99,0.98,0.97\}L$ and $L=2.5\,\text{Gm}$. As in the A and E channels, $|\tilde T_2|/f^2$ also tends to a constant value at low frequencies, which is maximized at $t=0.11\,\text{yr}$ for the chosen values of $\{\phi,\theta,\psi\}$. To obtain the corresponding result in the SEA, we define an approximate detector frame by setting $\frac{1}{3}\sum_I\mathbf{x}_I=\mathbf{x}_3\times\mathbf{e}_y=\mathbf{0}$ and $\mathbf{n}_{12}\cdot\mathbf{e}_z=\mathbf{n}_{13}\cdot\mathbf{e}_z=0$. We find that the SNR of the T channel is significantly larger than that in the SEA, yet still much smaller than in the A and E channels. In particular, the inclusion of the T channel is not expected to break the degeneracy of the signal at low frequencies. Compared with the displacement memory signal with the same $\{\phi,\theta,\psi\}$ parameters, the SNR of the velocity-memory signal in the T channel exhibits a modest increase relative to the A and E channels, since the signal is shifted to lower frequencies while $S_T/S_A$ also decreases toward lower frequencies; nevertheless, the resulting SNR remains comparably low. We leave a thorough analysis using the full set of second-generation $\{A,E,T\}$ channels or other sets of TDI combinations for future work.

\section{Soft signals from dark radiation}\label{appendix_B}
\begin{figure}[b]
\centering
\includegraphics[width=0.48\textwidth]{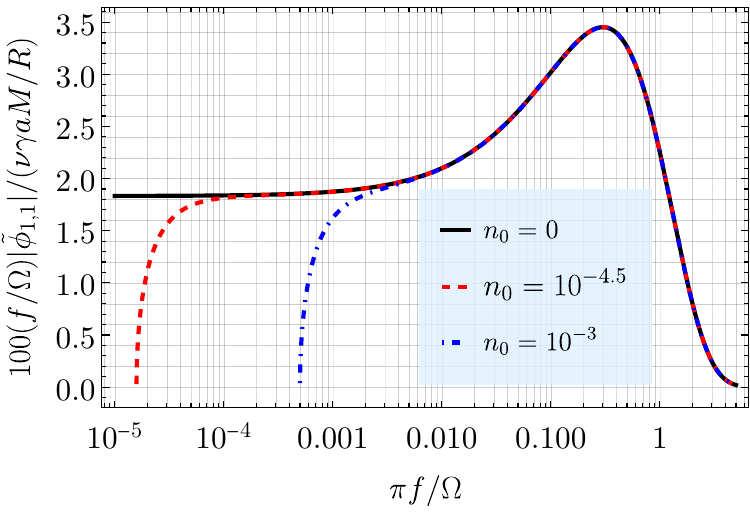}
\caption{Spectra of 0PN dipolar scalar waveforms from a charged binary on a Keplerian hyperbolic orbit, for different values of $n_0=m_\mathrm{b}/\Omega$.}\label{fig:dark}
\end{figure}

As in gravitational radiation, memory can also arise in massless scalar and vector radiation~\cite{Bieri_2013}. If the scalar or vector field couples to the test masses in a GW interferometer, this memory can also lead to soft signals (or their mimics for massive fields)\footnote{Note that even in the absence of nongravitational couplings, dark radiation can still produce GW memory.} as we discuss below. Consider a Klein-Gordon field $\phi$ and a Maxwell or Proca field $A_a$ sourced by point charges. The worldline action of a point particle with mass $m_\text{p}$, scalar charge $q_\phi$, and vector charge $q_A$ is given by
\begin{equation}
S=\int (-m_\text{p}+q_\phi\phi)\,d\tau+ \int q_A A_a dX^a, \label{coupling}
\end{equation}
where $\tau$ denotes the proper time. In flat spacetime, $\phi$ and $A_a$ (with $\partial_a A^a=\dot A^0+\nabla\cdot\mathbf{A}=0$) satisfy the wave equations:
\begin{equation}
(\partial_t^2-\nabla^2+m_\phi^2) \phi=n\equiv \sum_I\left[\frac{q_\phi}{u^0}\prod_i \delta(x^i-X^i)\right]_I, \notag
\end{equation}
and
\begin{equation}
(\partial_t^2-\nabla^2+m_A^2) A_a=j_a\equiv \sum_I\left[\frac{q_A u_a}{u^0}\prod_i \delta(x^i-X^i)\right]_I, \notag
\end{equation}
with $u^a=dx^a/d\tau$. For $m_\phi=m_A=0$, the memory in the direction $\hat{\mathbf{k}}$ produced in a scattering process, analogous to Eq.~\eqref{scattering_memory}, is given by $\Delta\phi =
\sum_I\frac{q_{I\phi}}{4\pi R} \Delta\left(\frac{\sqrt{1-|\mathbf{v}_I|^2}}{1-\mathbf{v}_I\cdot\hat{\mathbf{k}}}\right)$ and $\Delta A_i =P_{ij}\sum_I\frac{q_{IA}}{4\pi R} \Delta\left(\frac{v^j_I}{1-\mathbf{v}_I\cdot\hat{\mathbf{k}}}\right)$ (projected onto the radiation gauge by $P_{ij}=\delta_{ij}-\hat k_i \hat k_j$). As a concrete example, we consider a charged binary system (see, e.g., \cite{Cheng:2023qys,jq2j-d188}) on the same Keplerian hyperbolic orbit as in Sec.~\ref{sec:Displacement memory from compact binary scattering}. The scalar and vector waveforms can be expanded as
\[
\begin{aligned}
\phi &=  \int df\,\sum \tilde\phi_{l,m}(f)\, Y_{lm}\,\text{e}^{-2\pi i f u},
\\
A^0 & = \int df\sum \tilde A^0_{l,m}(f)\, Y_{lm}\,\text{e}^{-2\pi i f u},
\\
A^\text{L}&\equiv \mathbf{A}\cdot\hat{\mathbf{k}} = \int df\sum \tilde A_{l,m}^\text{L}(f)\, Y_{lm}\,\text{e}^{-2\pi i f u},
\\
A&\equiv \mathbf{A}\cdot(\mathbf{e}_\Theta-i \mathbf{e}_\Phi) = \int df\sum \tilde A_{l,m}(f) \,_{-1}Y_{lm}\,\text{e}^{-2\pi i f u},
\end{aligned}
\]
(see Sec.~\ref{sec:3} for the definition of $\{R,\Theta,\Phi\}$) where $u=t-v_gR$, with $v_g(f)=\sqrt{1-m_\text{b}^2/(2\pi f)^2}$ the group velocity ($\text{b}\in\{\phi,A\}$), and $\int_{-\infty}^{\infty} df$ runs only over $2\pi |f|\ge m_\text{b}$. The idealized positive-frequency spectra of the nonvanishing modes for the 0PN scalar dipole radiation are
\[
\frac{\tilde \phi_{1,\pm 1}(f)}{\frac{\nu \gamma a M}{R}} = \sqrt{1-\frac{n_0^2}{n^2}}\,\sqrt{\frac{\pi}{24}}\left(\pm i H_{in}'-\frac{\sqrt{e^2-1}}{e}H_{in}\right),
\]
where $n_0\equiv m_\text{b}/\Omega$, $n\equiv 2\pi f/\Omega$, $\gamma\equiv q_2/m_2-q_1/m_1$, and $H_{in}'=(H_{in-1}-H_{in+1})/2$. Note that $\tilde \phi_{1,1}(-f)=-[\tilde \phi_{1,-1}(f)]^*$. The corresponding dipolar scalar memory for $n_0=0$ is given by $4\pi R \Delta\phi=\nu M \gamma \hat{\mathbf{k}}\cdot\Delta \mathbf{v}$, with $\Delta \mathbf{v}=-(2a\Omega/e)\mathbf{e}_X$ for the hyperbolic orbit. The 0PN dipolar vector waveform can be obtained from the scalar waveform as $\tilde A^0_{1,\pm 1}=\tilde\phi_{1,\pm 1}$, $\tilde A^\text{L}_{1,\pm 1}=\tilde\phi_{1,\pm 1}/v_g$, and $\tilde A_{1,\pm 1}=\sqrt{2}\,\tilde\phi_{1,\pm 1}/v_g$, with the replacement $q_\phi\to q_A$. The spectra are shown in Fig.~\ref{fig:dark} for different values of $n_0$. Even for a massive scalar field, the spectrum can mimic a massless soft waveform over an intermediate-frequency band when $n_0$ is sufficiently small, before being suppressed near the mass threshold by the factor $\sqrt{1-n_0^2/n^2}$. Such a suppression is absent for scalar monopole radiation, which is generically generated beyond the Newtonian order during scattering even when the scalar charges are constant.
	
We now turn to the interferometer response to scalar and vector waves, assuming that the test masses are also described by the worldline action \eqref{coupling}, as considered, e.g., in \cite{Yu_2023}. The acceleration of a nonrelativistic test mass induced by weak scalar and vector waves is
\begin{equation}
\mathbf{F}_* 
\approx 
\frac{q_\phi}{m_\text{p}}\nabla\phi-\frac{q_A}{m_\text{p}}\left(\nabla A^0+\dot{\mathbf{A}}\right),
\end{equation}
where we neglect terms dependent on the test mass velocity $\dot{\mathbf{X}}$. For a monochromatic plane wave with frequency $\omega=2\pi f$ in the direction $\hat{\mathbf{k}}$, the induced velocity of the $I$th test mass at $\mathbf{x}_I$ is
\begin{equation}
\begin{aligned}
\mathbf{v}_I(t)&\equiv \int^t_{t_0} dt'\,\mathbf{F}_*(t',\mathbf{x}_I) 
\\
&\approx 
\Big[-\frac{q_\phi}{m_\text{p}}v_g\phi\,\hat{\mathbf{k}}+\frac{q_A}{m_\text{p}}\left(v_g A^0 \,\hat{\mathbf{k}}-\mathbf{A}\right)\Big]^t_{t_0},
\end{aligned}
\end{equation}
which contributes a Doppler shift to the one-way signal:
\begin{equation}
y_{rs}(t) \approx -\mathbf{n}_{rs}\cdot[\mathbf{v}_r(t)-\mathbf{v}_s(t-L_{rs})].
\label{coupling_signal}
\end{equation}
For a plane wave, the idealized spectrum $\tilde y_{rs}(f)$ in the static-detector approximation depends linearly on $\tilde \phi$ in the scalar case and on $\{\tilde A^0,\tilde A^\text{L},\tilde A\}$ in the vector case. Relativistic scalar or vector waves can therefore produce soft signals that differ from those due to nonrelativistic waves~\cite{Yu_2023,Yu:2024enm} and are, in principle, distinguishable from the soft signals of GW memory. The same holds for the one-way observable employed in Refs.~\cite{Gu__2025,PhysRevD.111.055031} to probe the axion-photon coupling in an axion field $\phi$, for which $y_{rs}\propto\frac{d}{dt}\left[\phi(t,\mathbf{x}_r(t))
-\phi(t-L_{rs},\mathbf{x}_s(t-L_{rs}))\right]$ if the light is circularly polarized. Self-interacting axion fields can also support nonperturbative structures such as domain walls~\cite{Sikivie:1982qv}, whose passage can induce memory-like field shifts.

\section{Additional posterior-distribution plots}\label{appendix_PE_posteriors}
	
This appendix collects posterior-distribution plots supporting the parameter-estimation discussion in Sec.~\ref{sec:5}.

\onecolumngrid

\begin{figure}[p]
\centering
\includegraphics[width=\textwidth]{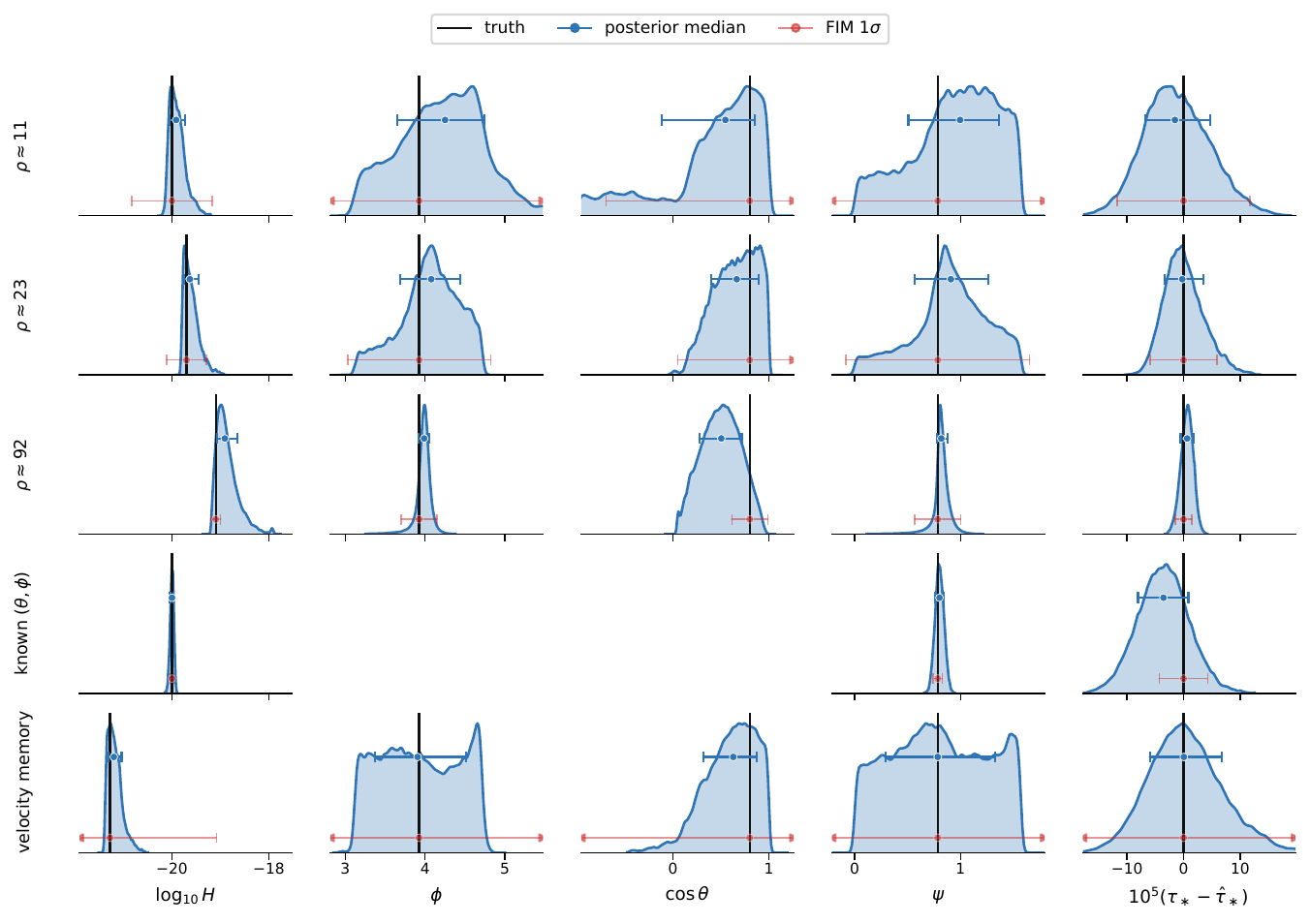}
\caption{Marginalized posterior densities for soft-memory signals measured at Taiji, with $T=10^{-3}\,\text{yr}$ and $f_\text{max}=0.04\,\text{Hz}$. The injected parameter values (black solid lines) are $\{\hat \tau_*,\hat\phi,\hat\theta,\hat\psi\}=\{0.7,3.93,0.64,0.79\}$. The upper three rows show displacement-memory results for different SNRs, with $\hat H=10^{-20},\,2\times 10^{-20},\,8\times 10^{-20}$, respectively. The fourth row shows the result for $\hat H=10^{-20}$ under the assumption of a known source sky location, so the fixed sky-location columns are omitted. The last row shows the result for a soft velocity-memory signal, where the first parameter is $\log_{10}(\dot H/\text{Hz})$ and $\hat{\dot H}=5.24\times 10^{-22}\,\text{Hz}$. The fifth column plots $10^5(\tau_*-\hat{\tau}_*)$, where $\hat{\tau}_*$ is the injected value. Blue points and intervals mark the posterior median and central 68\% credible interval. Red points and intervals show the local FIM $1\sigma$ estimates; intervals extending beyond the plotting range are clipped at the panel boundary, indicating poorly constrained parameters in the Fisher approximation.}\label{fig:LISA-Taiji_3_plot}
\end{figure}

\begin{figure}[p]
\centering
\includegraphics[width=0.90\textwidth]{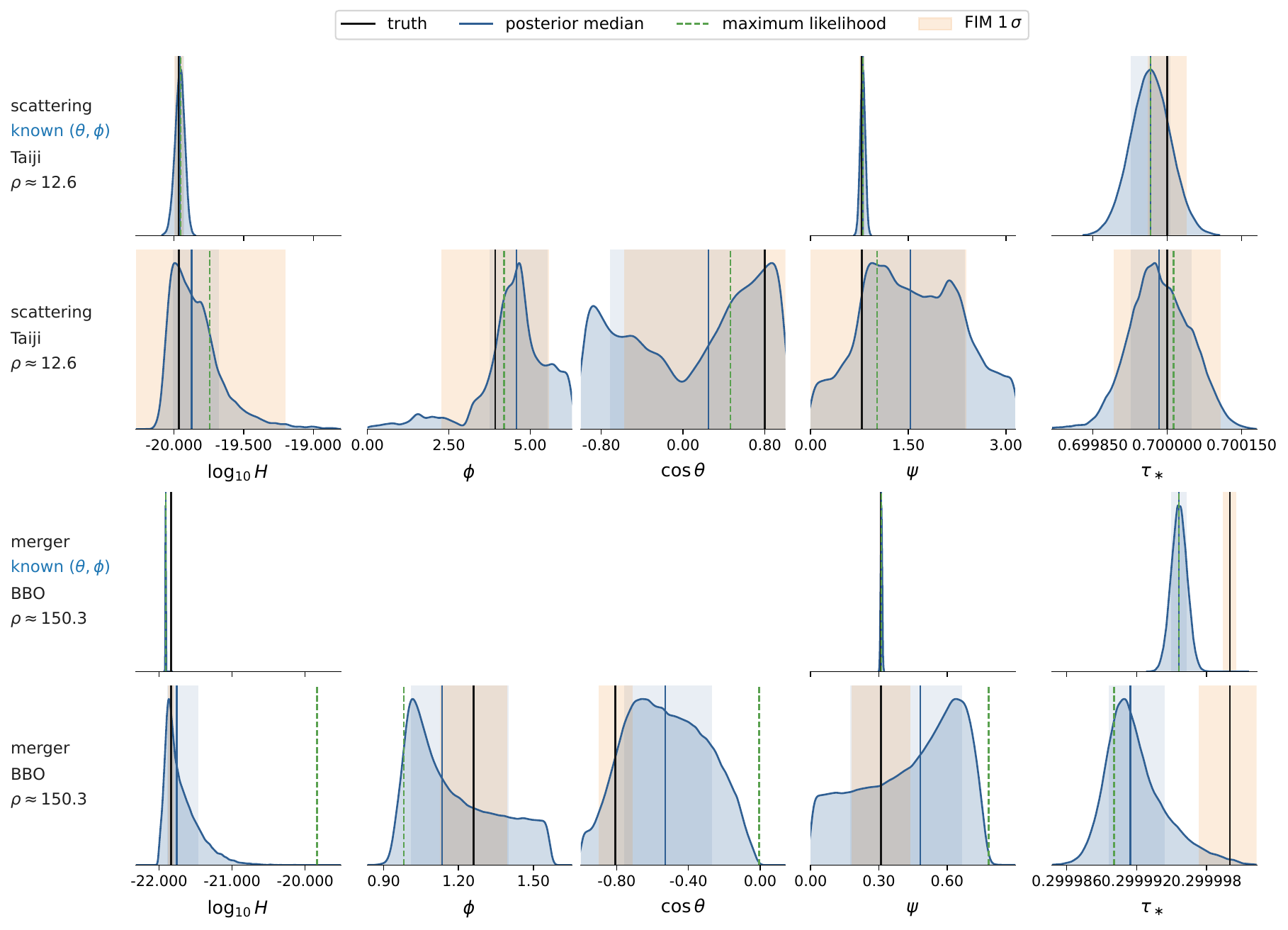}
\caption{Marginalized posterior densities for the single-detector realistic-waveform parameter-estimation examples summarized in Fig.~\ref{fig:single_detector_true_waveform_pe_main}, with the same source and injection parameters as stated there. The rows show the scattering waveform injection observed with Taiji and the merger waveform injection observed with BBO, with both full-sky and known-direction analyses. The injected signal is the realistic waveform and the likelihood uses the soft template. Vertical lines indicate the injected value, posterior median, and maximum-likelihood estimate; orange bands indicate the local FIM $1\sigma$ intervals where applicable.}\label{fig:single_detector_true_waveform_pe_appendix}
\vspace{2.0em}
\includegraphics[width=0.9\textwidth]{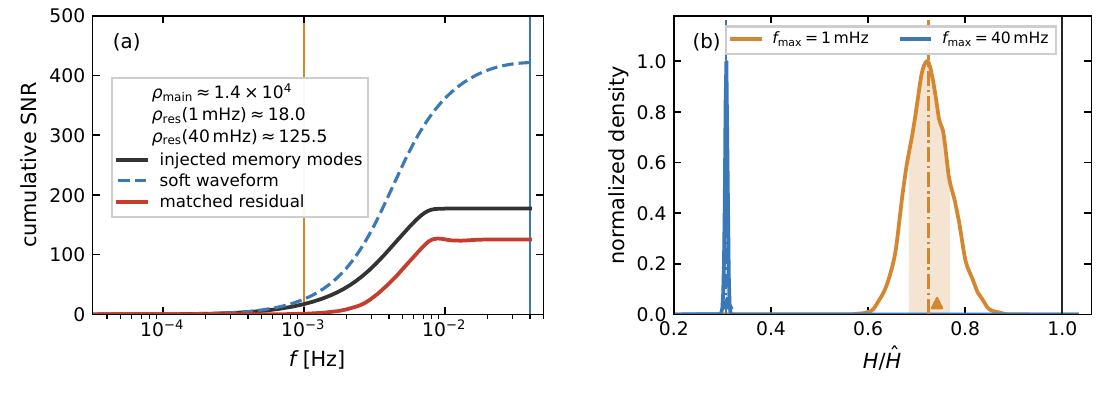}
\caption{Residual soft-memory measurement after subtracting the oscillatory SMBHB waveform. The Taiji injection uses $q=1$, source-frame total mass $M=10^6\,M_\odot$, $d_L=5\,\mathrm{Gpc}$ ($z\approx 0.8$), and $\Theta=\pi/3$. The reference soft waveform shown in panel (a) uses the same detector-frame parameters, with injected memory amplitude $\hat H$, $(\hat\phi,\cos\hat\theta,\hat\psi,\hat\tau_*)=(3.93,0.8,\pi/4,0.7)$, and $T=10^{-3}\,\mathrm{yr}$. The residual-memory PE uses uniform priors with ranges $\log_{10}H\in\log_{10}\hat H\pm1$, $\phi\in3.93\pm0.05$, $\cos\theta\in0.8\pm0.03$, $\psi\in\pi/4\pm0.05$, and $\tau_*\in0.7\pm0.005$. The oscillatory signal is generated with the available modes of \texttt{NRHybSur3dq8\_CCE}, while the injected memory is constructed from the corresponding $h_{2,0}$ and $h_{4,0}$ modes. The residual likelihood uses the soft template. Panel (a) compares the cumulative SNRs of the injected memory modes, the soft waveform, and the matched residual after primary-waveform subtraction. $\rho_{\rm main}$ denotes the optimal SNR of the oscillatory waveform before subtraction, and $\rho_{\rm res}$ denotes the matched soft-template SNR of the residual. Orange and blue denote $f_\text{max}=1\,\mathrm{mHz}$ and $40\,\mathrm{mHz}$, respectively, throughout. Panel (b) compares the corresponding $H/\hat H$ posteriors; shading marks the central $68\%$ credible intervals, dot-dashed lines the medians, black line the injected value, and triangles the maximum-likelihood estimates.}\label{fig:smbhb_residual_memory_pe}
\end{figure}
	
\clearpage
\twocolumngrid
\bibliography{paper}
\bibliographystyle{apsrev4-2}
\end{document}